\documentclass[12pt]{article}
\usepackage{epsfig}
\usepackage{axodraw}

\def\theequation{\thesection.\arabic{equation}}
\def\ra{\rightarrow}
\def\simlt{\stackrel{<}{{}_\sim}}

\newcommand{\boldsymbol}[1]{#1}

\newcommand{\bea}{\begin{eqnarray}}
\newcommand{\eea}{\end{eqnarray}}
\newcommand{\bd}{\begin{displaymath}}
\newcommand{\ed}{\end{displaymath}}
\newcommand{\be}{\begin{equation}}
\newcommand{\ee}{\end{equation}}
\newcommand{\ord}{{\cal{O}}}
\newcommand{\tb}{\tan\beta}

\def \Oi{{\mathcal O}}
\def\theequation{\thesection.\arabic{equation}}

\renewcommand{\baselinestretch}{1.2}
\begin{document}

\thispagestyle{empty}

{\normalsize\sf
\rightline {hep-ph/0210145}
\rightline{TUM-HEP-479/02}
\rightline{IFT-02/34}
\vskip 3mm
\rm\rightline{October 2002}
}

\vskip 5mm

\begin{center}
  
{\LARGE\bf $\mathbf{\Delta M_{d,s}}$, 
$\mathbf{B^0_{d,s}\rightarrow\mu^+\mu^-}$ 
and $\mathbf{B\rightarrow X_s}\boldsymbol{\gamma}$\\
in Supersymmetry at Large \boldmath{$\tan\beta$}\\
}

\vskip 10mm

{\large\bf Andrzej J.~Buras$^1$, Piotr H.~Chankowski$^2$,}\\
{\large\bf Janusz Rosiek$^{1,2}$ and {\L}ucja S{\l}awianowska$^2$} \\[5mm]

{\small $^1$ Physik Department, Technische Universit{\"a}t M{\"u}nchen,}\\
{\small D-85748 Garching, Germany}\\
{\small $^2$ Institute of Theoretical Physics, Warsaw University}\\
{\small Ho\.za 69, 00-681 Warsaw, Poland}

\end{center}

\vskip 5mm

\renewcommand{\baselinestretch}{1.1} 

\begin{abstract}
  We present an effective Lagrangian formalism for the calculation of
  flavour changing neutral and charged scalar currents in weak decays
  including $SU(2)\times U(1)$ symmetry breaking effects and the
  effects of the electroweak couplings $g_1$ and $g_2$. We apply this
  formalism to the MSSM with large $\tan\beta$ with the CKM matrix as
  the only source of flavour violation, heavy supersymmetric particles
  and light Higgs bosons. We give analytic formulae for the neutral
  and charged Higgs boson couplings to quarks including large
  $\tan\beta$ resummed corrections in the $SU(2)\times U(1)$ limit and
  demonstrate that these formulae can only be used for a
  semi-quantitative analysis. In particular they overestimate the
  effects of large $\tan\beta$ resummed corrections. We give also
  improved analytic formulae that reproduce the numerical results of
  the full approach within $5-10\%$. We present for the first time the
  predictions for the branching ratios $B^0_{s,d}\to \mu^+\mu^-$ and
  the $B^0_{d,s}-\bar B^0_{d,s}$ mass differences $\Delta M_{d,s}$
  that include simultaneously the resummed large $\tan\beta$
  corrections, $SU(2)\times U(1)$ breaking effects and the effects of
  the electroweak couplings. We perform an anatomy of the correlation
  between the {\it increase} of the rates of the decays $B^0_{s,d}\to
  \mu^+\mu^-$ and the {\it suppression} of $\Delta M_s$, that for
  large $\tan\beta$ are caused by the enhanced flavour changing
  neutral Higgs couplings to down quarks. We take into account the
  constraint from $B\to X_s \gamma$ clarifying some points in the
  calculation of the large $\tan\beta$ enhanced corrections to this
  decay.
\end{abstract}
\renewcommand{\baselinestretch}{1.2}

\newpage 
\setcounter{page}{1} 
\setcounter{footnote}{0}

\section{Introduction}
\label{sec:intro}
\setcounter{equation}{0} 

Supersymmetric theories are among the leading and most thoroughly
investigated candidates for extensions of the Standard Model (SM).
Even if supersymmetric particles are so heavy that they can be
produced only in very high energy collisions to be studied at
Tevatron, LHC and NLC, their virtual effects can be investigated 
also in low energy processes like particle-antiparticle mixing, 
CP violation and rare decays of hadrons and leptons. A number of
laboratories in Europe, USA and Japan will contribute to this
enterprise in an important manner in this decade.

Over the last fifteen years a vast number of papers discussing
supersymmetric effects in such low energy processes has been
published~\cite{MIPORO,GRNIRA,LS02}.  While $K^0$-$\bar K^0$ mixing,
$B_d^0$-$\bar B_d^0$ mixing, CP violation in $K\to \pi\pi$ decays 
and in particular the radiative decay $\bar B\to X_s\gamma$ played 
the leading role in these analyses so far, in view of forthcoming
experiments an even more important role will be played in the future
by the rare decays $\bar B\to X_s e^+e^-$, $B^0_{s,d}\to\mu^+\mu^-$
and $K\to\pi\nu\bar\nu$, the $B_s^0$-$\bar B_s^0$ mixing and various
CP violating transitions and asymmetries.

In the version of the supersymmetric theory, in which the
Cabibbo-Kobayashi-Maskawa (CKM) matrix remains the only source of
flavour and CP violation in the quark sector and $\tan\beta$ (the
ratio of the two vacuum expectation values $v_u/v_d$) is
between\footnote{For top squarks below 1 TeV $\tan\beta\simlt2$ is
most likely excluded by the unsuccessful Higgs boson search at LEP.}
2 and 10, the presence of supersymmetry does not drastically change
the SM predictions for processes in question.  While effects even as
high as $30\%$ could still be possible in $\Delta F=2$ transitions,
generally smaller effects are predicted in rare K- and B-meson decays.
Moreover, for supersymmetric particles heavier than $\sim500$ GeV
their effects become unmeasurably small.  Most recent analysis of this
scenario has been presented in~\cite{BUGAGOJASI}, where further
references to a rich literature can be found.  We will call this
scenario minimal flavour violation (MFV) at low $\tb$, deferring the
precise definition of our scenario to section 3.

Much larger effects are still possible in supersymmetric models with
new flavour violating interactions originating from the misalignment
of quark and squark mass
matrices~\cite{GAGAMASI,MIPORO,CHSL,CHRO,ISRE2}.  However, such models
contain many free parameters and their predictive power is small. The
same is also true for models with broken R-parity.

In this paper we investigate another very interesting scenario, namely
MFV in SUSY with large $\tb$.  It is theoretically appealing (e.g. it
is consistent with the approximate unification of top and bottom
Yukawa couplings at high energies predicted by some $SO(10)$ models)
and as predictive as MFV SUSY with low $\tb$ (the number of free
parameters is the same in both cases).  In this scenario very large
deviations from the SM predictions in certain low energy processes are
possible even for heavy supersymmetric particles but the calculations
are more technically involved due to the need for resummation of
large $\tb$ enhanced contributions.  One of the aims of the present
work is to formulate a general framework allowing to handle these
complications.

First discussions of large $\tan\beta$ scenarios in multi-Higgs
models and supergravity models can be found in~\cite{BRBUGE} and
\cite{OLPO}, respectively. The importance of supersymmetric large
$\tb$ effects in processes involving charged Higgs particles $H^\pm$
and flavour conserving processes involving neutral scalars has been
also known for some time \cite{HARASA,COSO,BAKO2,GUHOPE,HAHE}. 
They originate in the $\tb$ enhanced modifications of the standard
relations between the original Lagrangian down-quark mass parameters
$m_d$, $m_s$, $m_b$ (which determine the relevant Yukawa couplings)
and the running (``measured'') quark masses ${\overline m_d}$,
${\overline m_s}$, ${\overline m_b}$.  For example, in the case of the
$b$-quark one has
\begin{eqnarray}
m_b=\frac{\overline m_b}{1+\epsilon_b\tb}.
\label{FIRST}
\end{eqnarray}
The correction $\epsilon_b$ results from supersymmetric QCD and
electroweak one-loop corrections.  As $|\epsilon_b|$ can be of order
$0.01$, substantial enhancements or suppressions (depending on the
sign of $\epsilon_b$) of the Yukawa coupling are possible for
$\tb=\ord(50)$.  It has also been demonstrated~\cite{CAGANIWA1} that
expressing $m_b$ through ${\overline m_b}$ by means of eq.
(\ref{FIRST}) in the tree level couplings $\bar t H^+ b$, $\tilde
t\chi^+ b$ and in the flavour conserving neutral Higgs couplings
resumms for large values of $\tb$ dominant supersymmetric corrections
to all orders of perturbation theory. Such a resummation is necessary
for obtaining reliable predictions for measurable quantities
\cite{GUHOPE,CUHE}. 

Even more important for phenomenology is the fact that for large
values of $\tb$ flavour changing neutral currents in the down-quark
sector mediated by Higgs scalars can be significantly enhanced. In the
SM such currents are also induced at one loop by $tW^\pm$ penguin-like
diagrams with $Z^0$ replaced by the neutral Higgs boson but contribute
negligibly to the rare processes.  In supersymmetry with large
$\tan\beta$ neutral Higgs boson penguin-like diagrams with charginos
and stop-quarks in the loop can be more important than the standard
$Z^0$ penguins thanks to $\tb$ enhancement of the down-quark Yukawa
couplings. They have been first considered in \cite{HUYA} in 
connection with the $B\rightarrow X_sl^+l^-$ decay and in \cite{HAPOTO} 
and subsequently found to {\it increase} by orders of magnitude the
branching ratios of the rare decays
$B^0_{s,d}\to\mu^+\mu^-$~\cite{BAKO,CHSL,HULIYAZH,BOEWKRUR} and to
{\it decrease} significantly the $B^0_s$-$\bar B^0_s$ mass difference
$\Delta M_s$~\cite{BUCHROSL1} relative to the expectations based on
the SM.

Two additional related aspects of large supersymmetric corrections in
the down-quark sector are the following.  Elements of the physical CKM
matrix, to be called $V_{JI}^{\rm eff}$ in what follows, differ from
$V_{JI}$ present in the original Lagrangian by calculable flavour
dependent corrections $f_{JI}$~\cite{BLPORA,HAPOTO,BAKO}:
\begin{eqnarray}
V_{JI}=V_{JI}^{\rm eff} f_{JI}
\label{SECOND}
\end{eqnarray}
($J,I$ denote flavour indices).  $f_{JI}$ can substantially deviate
from unity for large $\tb$ and can in principle affect all processes
in which the CKM matrix is relevant.

Finally, as has been shown in~\cite{DEGAGI} in the context of the
analysis of $\bar B\to X_s\gamma$ decay, there are additional large
$\tb$ enhanced corrections in the charged Higgs couplings, that in
ref.~\cite{DEGAGI} have been parametrized by $\epsilon^\prime_b$ and
$\epsilon^\prime_t$.

It is desirable to calculate all the four effects, that is:

1) Large $\tb$ effects related to (\ref{FIRST}),

2) Enhanced neutral scalar-penguin diagrams, 

3) Large $\tb$ effects related to (\ref{SECOND}),

4) Enhanced corrections to charged Higgs vertex diagrams

\noindent in a self-consistent framework and to include them in a 
phenomenological analysis correlating predictions for $\bar B\to
X_s\gamma$, $B^0_{s,d}\to\mu^+\mu^-$ rates and for $B^0_{s,d}$-$\bar
B^0_{s,d}$ mass differences $\Delta M_{s,d}$.  It is also important to
formulate general rules allowing to include large $\tb$ effects that
can be used in other processes not considered in our paper.

Several steps towards this goal have been already made.  In
refs.~\cite{CAGANIWA1,DEGAGI,CAGANIWA2} large $\tb$ effects related to
1) and 4) have been discussed in the context of the $\bar B\to
X_s\gamma$ decay and recipes for including them in the charged Higgs
($H^\pm$) and charged Goldstone boson ($G^\pm$) vertices have been
formulated for specific quark flavours. In another elegant
analysis~\cite{ISRE1} the effects 1)-3) in the case of
$B^0_{s,d}\to\mu^+\mu^-$ decays and $B^0_{s,d}$-$\bar B^0_{s,d}$
mixing have been calculated in the $SU(2)\times U(1)$ symmetry limit,
neglecting the electroweak couplings $g_2$ and $g_1$ in comparison
with $\alpha_s$, $y_t$ and $y_b$. These authors confirmed sizeable
enhancements of $BR(B^0_{s,d}\to\mu^+\mu^-)$ and the suppression of 
$B^0_s$-$\bar B^0_s$
mixing pointed out previously in
refs.~\cite{BAKO,CHSL,HULIYAZH,BOEWKRUR} and \cite{BUCHROSL1},
respectively.

During the final stages of completion of our paper a model independent
analysis of rare processes in theories with the CKM matrix as the
unique source of flavour and CP violation has been presented
in~\cite{AMGIISST}.  While those authors also investigated large
$\tan\beta$ effects in $BR(B^0_{s,d}\to\mu^+\mu^-)$, $\Delta M_s$ and
$\bar B\rightarrow X_s\gamma$, their analysis was performed in the
same $SU(2)\times U(1)$ symmetry limit assuming the dominance of
$\alpha_s$, $y_t$ and $y_b$.

In the present paper we go beyond these analyses by
\begin{itemize}
\item Calculating all the four effects in an effective Lagrangian
  approach that goes beyond the $SU(2)\times U(1)$ symmetry limit
  considered in~\cite{ISRE1,AMGIISST} and includes the effects of the
  electroweak gauge couplings.
\item Calculating all the four effects in the $SU(2)\times U(1)$
  symmetry limit with vanishing electroweak gauge couplings thereby
  confirming and, in certain cases, correcting and generalizing the
  analytical rules for the inclusion of large $\tb$ effects presented
  in refs.~\cite{DEGAGI,CAGANIWA2}. Our results agree with those of
  \cite{ISRE1,AMGIISST} if the factors denoted here by $\epsilon_0$,
  $\epsilon_Y$, $\epsilon^\prime_0$ and $\epsilon_Y^\prime$ are
  assumed to be flavour independent.
\item Analysing numerically the  validity of the above approximation
  and giving simple recipes which improve it significantly.
\item Including all these effects in an analysis of the branching
  ratios $BR(B^0_{d,s}\rightarrow \mu^+\mu^-)$, $B^0_s$-$\bar B^0_s$
  mass difference $\Delta M_s$ and of correlations between them taking
  into account the constraint from $\bar B\rightarrow X_s\gamma$.
\end{itemize}

To our knowledge no complete analysis of all these topics has been
presented in the literature so far. While the recent analysis in
\cite{AMGIISST} considered many of these issues and went beyond
supersymmetry, it was performed in the $SU(2)\times U(1)$ symmetry
limit with $g_1=g_2=0$, the approximation the accuracy of which we
want to investigate here. In particular it is interesting to see how
big effects can still be expected in $B^0_{d,s}\rightarrow\mu^+\mu^-$
given the constraints on the magnitude of the scalar flavour changing
neutral currents stemming from the $B^0_s$-$\bar B^0_s$
mixing~\cite{BUCHROSL1}. This should also allow to assess the claim
made on the basis of approximate formulae by the authors of
ref.~\cite{ISRE1} that the effects found previously in
\cite{CHSL,BOEWKRUR} and~\cite{BUCHROSL1} can be decreased by a factor
as large as 5 when the large $\tb$ effects described by (\ref{FIRST})
and (\ref{SECOND}) are properly taken into account.

In a recent letter \cite{BUCHROSL2} we have presented some of the
formulae resulting from our analysis and in particular we have
analyzed the correlation between the increase of the rates of the
decays $B^0_{s,d}\rightarrow \mu^+\mu^-$ and the suppression of
$\Delta M_s$. In this paper we present the details of our formalism,
derive all results and extend the phenomenological analysis.

Our paper is organized as follows. In section 2 we present general
formalism for calculating the Higgs flavour changing and flavour
diagonal vertices based on the effective Lagrangian approach valid
beyond the $SU(2)\times U(1)$ symmetry limit.  In section 3 we relate
our approach to the one of~\cite{ISRE1,AMGIISST} and present explicit
expressions for flavour violating neutral scalar ($S^0$) and charged
scalar ($H^\pm,G^\pm$) couplings in the $SU(2)\times U(1)$ symmetry
limit. Subsequently we generalize these expressions to allow for
flavour dependence in the parameters $\epsilon_0$, $\epsilon_Y$,
$\epsilon^\prime_0$ and $\epsilon_Y^\prime$. We compare our formulae
with those present in the literature in section 4. In section 5 we
asses numerically the importance of the corrections calculated in
sections 2 and 3. We also investigate the effects of $SU(2)\times
U(1)$ breaking and of the effects of non-vanishing electroweak
couplings comparing our results with those present in the literature.
This includes in particular the parameters $\epsilon$ and $\epsilon'$
of refs. \cite{DEGAGI,CAGANIWA2}. Finally using the flavour dependent
formulation of section 3 we find analytic formulae for various
couplings that within $5-10\%$ reproduce the numerical results of the
full approach presented in section 2. In section 6 we collect the
basic formulae for $\Delta M_{d,s}$, $BR(B^0_{d,s}\rightarrow
\mu^+\mu^-)$ and $BR(\bar B\rightarrow X_s\gamma)$ in supersymmetry at
large $\tan\beta$.  Section 7 is devoted to the numerical analysis of
these quantities and the investigation of possible correlations
mentioned above. We conclude in section 8.


\section{General formalism}
\label{sec:genfor}
\setcounter{equation}{0} 

\subsection{Effective Lagrangian}

The most efficient way of handling heavy particle effects in low
energy processes involving quarks, is the effective Lagrangian
technique.  Let us assume that a sector of heavy fields in a theory
(e.g.  sfermions, gluinos, charginos and neutralinos in the MSSM) can
be integrated out without violating gauge invariance.  We consider
processes occurring at energies low enough so that the amplitudes
cannot have imaginary parts related by unitarity to particles
belonging to the heavy sector. Renormalized in the $\overline{\rm MS}$ 
scheme with some renormalization scale $Q$, virtual effects of
the decoupled particles ($Q$ should be chosen to be of the order of
their masses) corresponding to self-energy diagrams in
Fig.~\ref{fig:Sigcorr} can be summarized by the effective kinetic and
mass terms of light fermions:
\begin{eqnarray}
&&{\cal L}_{\rm eff}^{\rm kin}
=\overline{(d_J)_L}\left(1-\Sigma^d_{VL}(0)\right)^{JI}
i\not\!\partial(d_I)_L + 
\overline{(d_J)_R}\left(1-\Sigma^d_{VR}(0)\right)^{JI}
i\not\!\partial(d_I)_R\nonumber\\
&&\phantom{aaa}
-\overline{(d_J)_R}\left(m_d+\Sigma^d_{mL}(0)\right)^{JI}(d_I)_L
-\overline{(d_J)_L}\left(m_d+\Sigma^d_{mR}(0)\right)^{JI}(d_I)_R
\end{eqnarray}
plus similar terms for the up-type quarks and leptons.  Here $I,J$ 
are flavour indices with $d_1\equiv d$, $d_2\equiv s$, $d_3\equiv b$.
Analogous notation will be used for the up-type quarks. Exact 
expressions for quark self-energies are listed in the Appendix A.2.

In order to make the formulae more compact we denote (in some of them)
the diagonal quark mass matrices simply by $m_d\equiv{\rm
  diag}(m_{d_1},m_{d_2},m_{d_3})\equiv{\rm diag}(m_d,m_s,m_b)$,
$m_u\equiv{\rm diag}(m_{u_1},m_{u_2},m_{u_3})\equiv{\rm
  diag}(m_u,m_c,m_t)$.  Hermiticity of the effective Lagrangian
ensures that $(\Sigma^q_{VL})^\dagger=\Sigma^q_{VL}$,
$(\Sigma^q_{VR})^\dagger=\Sigma^q_{VR}$ and
$(\Sigma^q_{mL})^\dagger=\Sigma^q_{mR}$.

\begin{figure}[htbp]
\begin{center}
\begin{picture}(400,100)(0,0)
\Vertex(70,40){13}
\ArrowLine(60,40)(20,40)
\ArrowLine(120,40)(80,40)
\Text(30,30)[]{$J$}
\Text(110,30)[]{$I$}
\Text(260,40)[]{$=-i\left(\Sigma^q_{VL}\not\!pP_L 
+\Sigma_{VR}^q\not\!pP_R+\Sigma^q_{mL}P_L+\Sigma^q_{mR}P_R\right)^{JI}$}
\end{picture}
\end{center}
\caption{One-loop threshold corrections to fermion propagators.
  $P_L\equiv{1-\gamma^5\over2}$, $P_R\equiv{1+\gamma^5\over2}$ and
  $q\equiv d$ or $u$.}
\label{fig:Sigcorr}
\end{figure}

Similarly, virtual effects of the decoupled particles corresponding to
fermion-gauge boson vertices of the type shown schematically in
Fig.~\ref{fig:Vertcorr} can be summarised as
\begin{eqnarray}\label{LINT}
&&{\cal L}^{\rm int}_{\rm eff}=
-\overline{(d_J)_L}\gamma^\mu
 \left(F_L^{Zd}+\Delta F^{Zd}_L\right)^{JI}(d_I)_L Z_\mu^0
-\overline{(d_J)_R}\gamma^\mu
 \left(F_R^{Zd}+\Delta F^{Zd}_R\right)^{JI}(d_I)_R Z_\mu^0\nonumber\\
&&\phantom{aaaa}
-\overline{(u_J)_L}\gamma^\mu
 \left(F_L^{Zu}+\Delta F^{Zu}_L\right)^{JI}(u_I)_L Z_\mu^0
-\overline{(u_J)_R}\gamma^\mu
 \left(F_R^{Zu}+\Delta F^{Zu}_R\right)^{JI}(u_I)_R Z_\mu^0\\
&&\phantom{aaaa}
-\overline{(u_J)_L}\gamma^\mu
 \left(F_L^W+\Delta F^W_L\right)^{JI}(d_I)_L W_\mu^+
-\overline{(u_J)_R}\gamma^\mu
 \left(F_R^W+\Delta F^W_R\right)^{JI}(d_I)_R W_\mu^++{\rm H.c.}\nonumber
\end{eqnarray}
where $(F^{Zq}_L)^{JI}\propto\delta^{JI}$,
$(F_L^W)^{JI}=(g/\sqrt2)V_{JI}$ and, of course, $F_R^W=0$.

\begin{figure}[htbp]
\begin{center}
\begin{picture}(300,100)(0,0)
\Vertex(45,40){13}
\ArrowLine(40,45)(10,80)
\ArrowLine(10,0)(40,35)
\Photon(50,40)(110,40){3}{8}
\Text(10,70)[]{$J$}
\Text(10,10)[]{$I$}
\Text(200,40)[]{$=-i\gamma^\mu\left(\Delta F_L^{JI}P_L + 
\Delta F_R^{JI}P_R\right)$}
\end{picture}
\end{center}
\caption{One-loop threshold corrections to fermion-gauge boson vertices.}
\label{fig:Vertcorr}
\end{figure}

Finally, other virtual effects of the decoupled heavy particles can be
represented in the effective Lagrangian by various effective vertices
which are operators of dimension higher than four (non-renormalizable
operators) such as four-fermion operators or dimension 5
(chromo)magnetic operators etc.

\subsection{Diagonalization}

The next step is to rescale the quark fields so to render their
kinetic terms canonical:
\begin{eqnarray}
(d_I)_L\rightarrow\left(1+{1\over2}\Sigma_{VL}^d\right)^{IJ}(d_J)_L,
\phantom{aaaa}
(d_I)_R\rightarrow\left(1+{1\over2}\Sigma_{VR}^d\right)^{IJ}(d_J)_R,
\label{eqn:rescaling}
\end{eqnarray}
and the same for the up-type quarks.  After this operation the quark
mass terms become
\begin{eqnarray}
&&{\cal L}_{\rm mass}=
-\overline{(d_J)_R}\left(m_d+\Delta m_d\right)^{JI}(d_I)_L
-\overline{(d_J)_L}\left(m_d+\Delta m_d^\dagger\right)^{JI}(d_I)_R
\label{eqn:effmassterm}
\end{eqnarray}
where
\begin{eqnarray}
\Delta m_d\equiv\Sigma_{mL}^d(0)
+{1\over2}\Sigma_{VR}^d(0)  m_d+{1\over2}m_d \Sigma_{VL}^d(0)~.
\label{eqn:dmd}
\end{eqnarray}
Similar expressions are obtained for the up-type quarks.

The mass terms (\ref{eqn:effmassterm}) can be diagonalized by two
independent rotations:
\begin{eqnarray}
d_L\rightarrow\mathbf{D}_Ld_L,\phantom{aaa}
d_R\rightarrow\mathbf{D}_Rd_R
\label{eqn:grotations}
\end{eqnarray}
so that
\be\label{barmd}
\left[\mathbf{D_R^\dagger}(m_d+\Delta m_d)\mathbf{D_L}\right]^{JI}
=\overline m_{d_J} \delta^{JI}
\ee
with $\overline m_{d_J}$ denoting the corrected mass eigenvalues.  To
one-loop accuracy the transformations (\ref{eqn:grotations}) read:
\begin{eqnarray}
d_L\rightarrow\left(\mathbf{1+\Delta D}_L\right)d_L,\phantom{aaa}
d_R\rightarrow\left(\mathbf{1+\Delta D}_R\right)d_R.
\label{eqn:rotations}
\end{eqnarray}
The unitarity of $\mathbf{D}_{L(R)}$ requires that (up to higher order
terms) $\mathbf{\Delta D}^\dagger_{L(R)}=-\mathbf{\Delta D}_{L(R)}$
and $\mathbf{\Delta D}^{II}_{L(R)}=0$.  In this approximation the
corrected mass eigenvalues are given by\footnote{If the diagonal
  entries of the matrices $\Sigma^d_{mL}$, $\Sigma^d_{mR}$ are
  complex, the rotations (\ref{eqn:rotations}) have to be supplemented
  by additional chiral rotations of the quark fields in order to get
  real and positive quark masses.}
\begin{eqnarray}
{\overline m}_{d_J}= m_{d_J}\left(1+\kappa_{d_J}\right).
\label{eqn:corrmass}
\end{eqnarray}
where
\begin{eqnarray}
\kappa_{d_J}\equiv{\Sigma^{dJJ}_{mL}(0)\over m_{d_J}}
+{1\over2}\Sigma_{VR}^{dJJ}(0)+{1\over2}\Sigma_{VL}^{dJJ}(0)
\equiv{\left(\Delta m_d\right)^{JJ}\over m_{d_J}}.
\label{eqn:kappadef}
\end{eqnarray}
Formula (\ref{eqn:corrmass}) generalizes (\ref{FIRST}) to arbitrary
down-quark flavours.

In order to find the matrices $\mathbf{\Delta D}_{L(R)}$ we follow
ref.~\cite{ISRE1} and decompose $\Delta m_d$ explicitly into its
diagonal and off-diagonal parts
\begin{eqnarray}
\left(\Delta m_d\right)^{JI} = m_{d_J}\kappa_{d_J}\delta^{JI}
+\left(\Delta^\prime m_d\right)^{JI}
\label{eqn:dmdprime}
\end{eqnarray}
where by definition $(\Delta^\prime m_d)^{JI}=(\Delta m_d)^{JI}$ for
$J\neq I$ and $(\Delta^\prime m_d)^{JJ}=0$.

The condition determining the $\mathbf{\Delta D}_{L(R)}$ takes now the
form
\begin{eqnarray}
\left(\mathbf{\Delta D}^\dagger_R{\overline m_d}+
{\overline m_d}\mathbf{\Delta D}_L+\Delta^\prime m_d\right)^{JI}=0
\label{eqn:diagcond}
\end{eqnarray}
for $J\neq I$.  Note that following ref.~\cite{ISRE1} we multiply by
$\mathbf{\Delta D}_{L,R}$ also the $\tan\beta$ enhanced diagonal
correction to the down-type quark masses, i.e.  include terms which
are formally of higher order.  The non-zero entries of $\mathbf{\Delta
  D}^\dagger_{L(R)}$ are then
\begin{eqnarray}
\mathbf{\Delta D}^{JI}_L=-{{\overline m}_{d_J}\Delta m_d^{JI}
+(\Delta m_d^\dagger)^{JI}{\overline m}_{d_I}\over
{\overline m}_{d_J}^2-{\overline m}_{d_I}^2} ~~~J\neq I
\label{eqn:DLsol}
\end{eqnarray}
\begin{eqnarray}
\mathbf{\Delta D}^{JI}_R=-{{\overline m}_{d_J}(\Delta m_d^\dagger)^{JI} 
+ \Delta m_d^{JI}{\overline m}_{d_I}\over
{\overline m}_{d_J}^2-{\overline m}_{d_I}^2} ~~~J\neq I.
\label{eqn:DRsol}
\end{eqnarray}
Using the hierarchy of quark masses: $m_{d_3}\gg m_{d_2}\gg m_{d_1}$
(${\overline m}_{d_3}\gg{\overline m}_{d_2}\gg{\overline m}_{d_1}$)
and the fact that $\Delta m_d^{JI}\propto m_{d_J}$, $(\Delta
m_d^\dagger)^{JI}\propto m_{d_I}$ (see section 3 for explicit
expressions) we find for $J>I$
\begin{eqnarray}
&&\mathbf{\Delta D}^{JI}_L=
-{1\over{\overline m}_{d_J}}\Delta m_d^{JI}+{\cal O}
\left({{\overline m}^2_{d_I}\over{\overline m}^2_{d_J}}\right),\nonumber\\
&&\mathbf{\Delta D}^{JI}_R=
-{1\over{\overline m}_{d_J}}(\Delta m^\dagger_d)^{JI}
-{{\overline m}_{d_I}\over{\overline m}^2_{d_J}}\Delta m_d^{JI}
\sim{\cal O}\left({{\overline m}_{d_I}\over{\overline m}_{d_J}}\right)
\label{eqn:DJI1}
\end{eqnarray}
and for $J<I$
\begin{eqnarray}
&&\mathbf{\Delta D}^{JI}_L=
{1\over{\overline m}_{d_I}}(\Delta m^\dagger_d)^{JI}+{\cal O}
\left({{\overline m}^2_{d_J}\over{\overline m}^2_{d_I}}\right),\nonumber\\
&&\mathbf{\Delta D}^{JI}_R=
{1\over{\overline m}_{d_I}}\Delta m_d^{JI}
+{{\overline m}_{d_J}\over{\overline m}^2_{d_I}}(\Delta m^\dagger_d)^{JI}
\sim{\cal O}\left({{\overline m}_{d_J}\over{\overline m}_{d_I}}\right)
\label{eqn:DJI2}
\end{eqnarray}
Note that $\mathbf{\Delta D}_L$ is ${\cal O}(1)$ with respect to the
masses but $\mathbf{\Delta D}_R$ is always suppressed.  This
suppression can be occasionally removed by a large mass multiplying
$\mathbf{\Delta D}_R$.

The same formulae (with appropriate replacements $d\rightarrow u$,
$\mathbf{D}\rightarrow\mathbf{U}$) hold also for the up-type quarks.
In particular we have
\be\label{barmu}
\left[\mathbf{U^\dagger_R}(m_u+\Delta m_u)\mathbf{U_L}\right]^{JI}
=\overline m_{u_J} \delta^{JI}
\ee
with $\overline m_{u_J}$ denoting the corrected mass eigenvalues.
However the correction $\Delta m_u$ is not enhanced for large
$\tan\beta$ (see the discussion following (\ref{eqn:umasscorr})) in
contrast to $\Delta m_d$ that is enhanced in this limit.  Consequently
one can set ${\overline m}_{u_J}\approx m_{u_J}$ and the condition
analogous to (\ref{eqn:diagcond}) reads
\begin{eqnarray}
\left(\mathbf{\Delta U}^\dagger_R m_u+
m_u\mathbf{\Delta U}_L+\Delta^\prime m_u\right)^{JI}=0
\label{eqn:udiagcond}
\end{eqnarray}
for $J\neq I$.  The explicit expressions for $\mathbf{\Delta
  U}^{JI}_{L(R)}$ are then obtained directly from
(\ref{eqn:DLsol})-(\ref{eqn:DJI2}) by making appropriate replacements
($\overline m_{d_J}\to m_{u_J}$, etc).

Obviously, these are the masses ${\overline m}_{d_J}$, and
${\overline m}_{u_J}\approx m_{u_J}$ that have to be identified with
the running mass parameters of the low energy effective theory valid
for scales $\sim M_Z$.  Parameters ${\overline m}_{d_J}$, 
${\overline m}_{u_J}$ are therefore directly related through the known QCD
renormalization group running to the low energy quark mass parameters.

\subsection{Effective CKM matrix}

Rotations (\ref{eqn:rotations}) have the effect of renormalizing the
CKM matrix.  After the operations (\ref{eqn:rescaling}) and
(\ref{eqn:rotations}) the $W$-boson vertices become
\begin{eqnarray}
{\cal L}_{\rm eff}^{\rm W}=
-\overline{(u_J)_L}\gamma^\mu
 \left(\tilde F_L^W+\Delta\hat F^W_L\right)^{JI}(d_I)_L W_\mu^+
-\overline{(u_J)_R}\gamma^\mu
 \left(\Delta\hat F^W_R\right)^{JI}(d_I)_R W_\mu^++{\rm H.c.}
\end{eqnarray}
with
\begin{eqnarray}
\Delta\hat F^W_L\equiv\Delta F^W_L+{1\over2}\Sigma^u_{VL} F_L^W
+{1\over2} F_L^W \Sigma^d_{VL}, ~~~\Delta\hat F^W_R\equiv\Delta F^W_R
\end{eqnarray}
and 
\begin{eqnarray}
(\tilde F_L^W)^{JI}={g_2\over\sqrt2}V^{\rm eff}_{JI}\equiv
{g_2\over\sqrt2}\left(\mathbf{U}_L^\dagger V
\mathbf{D}_L\right)^{JI}\approx
{g_2\over\sqrt2}\left(V+\mathbf{\Delta U}_L^\dagger  V+
V \mathbf{\Delta D}_L\right)^{JI}~.
\label{eqn:CKMeff}
\end{eqnarray}

Since the corrections $\Delta\hat F_{L,R}^W$ are negligible (for example, 
for all $JI$ the gluino exchange computed in \cite{CIDEGAGI2} gives 
$\Delta\hat F_L^W/\tilde F_L^W\simlt3\times10^{-3}$), it is the matrix 
$V_{JI}^{\rm eff}$ that has to be identified with the CKM matrix whose 
elements are determined from the low energy processes. The matrix $V_{JI}$ 
is the CKM matrix of the MSSM which for a given spectrum of sparticles 
is related in a calculable way to the measured matrix $V_{JI}^{\rm eff}$.  
Thus, the formula (\ref{eqn:CKMeff}) is the realization of the relation 
(\ref{SECOND}). The correct procedure in phenomenological applications 
is then to calculate first all relevant quantities in terms of $V_{JI}$ 
and subsequently express the latter in terms of $V_{JI}^{\rm eff}$ that
should be determined from the measured branching ratios.

It is known~\cite{BLPORA,BAKO}, that for large $\tan\beta$ the ratios of 
the elements $V_{3I}^{\rm eff}/V_{3I}\approx V_{I3}^{\rm eff}/V_{I3}$ can, 
for $I\neq3$, substantially deviate from unity. Simple approximate (but 
very accurate) expressions for $f_{JI}\equiv V_{JI}/V^{\rm eff}_{JI}$ 
defined in eq. (\ref{SECOND}) will be given in section \ref{sec:su2xu1} 
(see eq. (\ref{eqn:fCKMcorr})) and the corresponding numerical analysis 
in section \ref{sec:comp}. 

The elements $|V_{ub}|$ and $|V_{cb}|$, that are affected by these
corrections are usually determined from tree level decays under the
assumption that new physics contributions to the relevant branching
ratios can be neglected. This assumption is clearly violated in the
case of supersymmetry at large $\tan\beta$. However, in this case 
the most important $\tan\beta$ enhanced loop corrections to tree 
level decays, can be absorbed in the $V^{\rm eff}$. Therefore, the 
quantities which experimentalists extract from these processes can 
be identified with $|V^{\rm eff}_{ub}|$ and $|V^{\rm eff}_{cb}|$ 
defined in (\ref{eqn:CKMeff}).

On the other hand, as we will see the ratio $|V_{ub}|/|V_{cb}|$ is to
an excellent approximation not affected by these corrections and
$|V_{ub}|/|V_{cb}|\approx|V^{\rm eff}_{ub}|/|V^{\rm eff}_{cb}|$.  It
should also be emphasized that
\be 
V^{\rm eff}(V^{\rm eff})^\dagger=\hat 1
\ee
so that all unitarity relations, in particular the formulae related to
the unitarity triangle, expressed usually in terms of $V_{JI}$, are
also valid for $V^{\rm eff}_{JI}$.

For the $Z^0$ couplings the rescalings (\ref{eqn:rescaling}) amount to
the replacements
\be
\Delta F_{L(R)}^{Zq}\rightarrow\Delta\hat F_{L(R)}^{Zq} \equiv\Delta
F_{L(R)}^{Zq}+{1\over2}\Sigma_{VL(R)}^q F_{L(R)}^{Zq}
+{1\over2}F_{L(R)}^{Zq}\Sigma_{VL(R)}^q.
\ee
To one-loop accuracy, the subsequent unitary rotations
(\ref{eqn:rotations}) do not affect the $Z^0$ vertices.
Flavour changing introduced in higher orders can be treated 
along the lines discussed in subsection \ref{subsec:cqvert}.

\subsection{Neutral Higgs-fermion vertices}
\label{subsec:hfvert}

The one-loop corrections to scalar ($S^0$) neutral Higgs boson
vertices induced by the decoupled particles can be written as
\begin{eqnarray}
{\cal L}^{S^0}_{\rm eff}=-
\overline{(d_J)_R}\left(F_L^{dS}+\Delta F^{dS}_L\right)^{JI}(d_I)_L S^0 
-\overline{(d_J)_L}\left(F_R^{dS}+\Delta F^{dS}_R\right)^{JI}(d_I)_R S^0
\label{eqn:Sverts}
\end{eqnarray}
and similarly for the up-type quarks (see Appendix A.3 for the 
expressions for $\Delta F^{dS}_{L,R}$).  For neutral Higgs and
Goldstone bosons we have

\begin{figure}[htbp]
\begin{center}
\begin{picture}(300,100)(0,0)
\Vertex(45,40){13}
\ArrowLine(40,45)(10,80)
\ArrowLine(10,0)(40,35)
\DashLine(50,40)(100,40){4}
\Text(5,70)[]{$d_J$}
\Text(5,10)[]{$d_I$}
\Text(95,50)[]{$S^0$}
\Text(220,40)[]{$=-i\left(\left(\Delta F_L^{dS}\right)^{JI}P_L 
+ \left(\Delta F_R^{dS}\right)^{JI}P_R\right)$}
\end{picture}
\end{center}
\caption{One-loop corrections to neutral Higgs-quark vertices.}
\label{fig:NHiggscorr}
\end{figure}

\begin{eqnarray}
(F_{L(R)}^{dS})^{JI}=\eta^{dS}_{L(R)} m_{d_J}\delta^{JI}
\equiv (F_{L(R)}^{dS})^J\delta^{JI}.
\end{eqnarray}
with explicit expressions for $\eta^{dS}_{L(R)}$ given in (\ref{FdS})
and (\ref{xS}).  After the operations (\ref{eqn:rescaling}) and
(\ref{eqn:rotations}) the expressions inside the brackets in
eq.~(\ref{eqn:Sverts}) become respectively
\begin{eqnarray}
F_L^{dS}+\Delta F^{dS}_L
+{1\over2}\Sigma^d_{VR}  F_L^{dS{\rm corr}}
+{1\over2}F_L^{dS{\rm corr}} \Sigma^d_{VL}
+\mathbf{\Delta D}^\dagger_R 
F_L^{dS{\rm corr}}+F_L^{dS{\rm corr}}
\mathbf{\Delta D}_L
\label{eqn:Sffvert}
\end{eqnarray}
\begin{eqnarray}
F_R^{dS}+\Delta F^{dS}_R
+{1\over2}\Sigma^d_{VL}  F_R^{dS{\rm corr}}
+{1\over2}F_R^{dS{\rm corr}} \Sigma^d_{VR}
+\mathbf{\Delta D}^\dagger_L 
F_R^{dS{\rm corr}}+F_R^{dS{\rm corr}}
\mathbf{\Delta D}_R
\label{eqn:Sffvert1}
\end{eqnarray}
where we have defined
\begin{eqnarray}
\left(F_{L(R)}^{dS{\rm corr}}\right)^J\equiv\eta^{dS}_{L(R)} m_{d_J}
+\left(\Delta F^{dS}_{L(R)}\right)^{JJ}.
\end{eqnarray}
Of course, in the phenomenological applications one has to express
$F_L^{dS}$ in terms of the ``observable'' ${\overline m}_{d_J}$ by
using eq.~(\ref{eqn:corrmass}):
\begin{eqnarray}
\left(F_L^{dS}\right)^{JI} = 
\eta^{dS}_L {{\overline m}_{d_J}\over
1+\kappa_{d_J}}\delta^{JI}\approx
\eta^{dS}_L {{\overline m}_{d_J}\over1+\epsilon_{d_J}}\delta^{JI}
\left[1 - {1\over2}\left(\Sigma^d_{VL}+\Sigma^d_{VR}\right)^{JJ}\right]
\label{eqn:f_resumed}
\end{eqnarray}
with an analogous formula for $\left(F_R^{dS}\right)^{JI}$.  In
writing the second expression in (\ref{eqn:f_resumed}) we have assumed
that the vector parts of the fermion self energies do not contain
large contributions and neglected terms of order
$\epsilon_d\left(\Sigma^d_{VL}+\Sigma^d_{VR}\right)$ and smaller.  We
have also defined the quantity $\epsilon_{d_J}$
\begin{eqnarray}
\epsilon_{d_J}\equiv
{\left(\Sigma_{mL}^d\right)^{JJ}\over m_{d_J}\tan\beta}
\label{eqn:eps_def}
\end{eqnarray}
which for $J=3$ corresponds to $\epsilon_b$ of ref.  \cite{DEGAGI}.
It is also instructive to make contact with the standard perturbative
(on-shell) calculation of the one loop $\bar dS^0d$ vertex which
gives~\cite{CHSL,BOEWKRUR}
\begin{eqnarray}
\eta^{dS}_L{\overline m}_{d_J}\delta^{JI} + 
\left(\Delta F_L^{dS}\right)^{JI} - \eta^{dS}_L \Sigma^{dJI}_{mL}~.
\end{eqnarray}
By expanding (\ref{eqn:Sffvert}) strictly to one loop (neglecting the
difference between $m_d$ and ${\overline m}_d$ in all 1-loop terms)
and using the explicit expressions (\ref{eqn:DLsol}),
(\ref{eqn:DRsol}) as well as (\ref{eqn:corrmass}) and
(\ref{eqn:kappadef}) it is easy to see that the two approaches indeed
coincide.  In particular, to 1-loop accuracy the vertex
(\ref{eqn:Sffvert}) is independent of the vector self energies, both
for the diagonal and off-diagonal transitions.

On the other hand, keeping $\epsilon_{d_J}$ in the denominator in
(\ref{eqn:f_resumed}) resumms dominant corrections of order
$\alpha_s(\mu/m_{\tilde g})\tan\beta$ to all orders of the
perturbation expansion~\cite{CAGANIWA1} which are important for
neutral Higgs decays into $b\bar b$ or $\tau^+\tau^-$ pairs
\cite{BAKO2,GUHOPE,CUHE}. 
Thus, the diagonal couplings of scalars to the down-type quarks are
\begin{eqnarray}
{\cal L}^{\rm diag}_{\rm eff}=-\overline{(d_J)_R}
\left(\eta_L^{dS}{{\overline m}_{d_J}\over1+\epsilon_{d_J}\tan\beta}
+\Delta F^{dS}_L\right)^{JJ}(d_J)_L S^0  - {\rm H.c.}
\label{eqn:Sffdiag}
\end{eqnarray}

The expression for the flavour non-diagonal vertex can be simplified
by taking into account the hierarchy of masses $m_{d_J}$: $m_{d_3}\gg
m_{d_2}\gg m_{d_1}$ (${\overline m}_{d_3}\gg{\overline
  m}_{d_2}\gg{\overline m}_{d_1}$) and using the expressions
(\ref{eqn:DJI1}) and (\ref{eqn:DJI2}) for the $\mathbf{\Delta
  D}_{L(R)}$ matrices.  One has also to take into account that (as we
will see shortly) in the limit of $SU(2)\times U(1)$ symmetry $(\Delta
m_d)^{JI}\propto m_{d_J}$ and $(\Delta m_d^\dagger)^{JI}\propto
m_{d_I}$.  In general one gets
\begin{eqnarray}\label{LNEUTRAL}
{\cal L}^{\rm off-diag}_{\rm eff}=
-\overline{(d_J)_R}\left[X^S_{RL}\right]^{JI}(d_I)_L S^0
-\overline{(d_J)_L}\left[X^S_{LR}\right]^{JI}(d_I)_R S^0
\end{eqnarray}
where for $J>I$
\begin{eqnarray}
\left[X^S_{RL}\right]^{JI}=\left(\Delta F^{dS}_L
+{1\over2}\Sigma^d_{VR}  F_L^{dS{\rm corr}}
+{1\over2}F_L^{dS{\rm corr}} \Sigma^d_{VL}\right)^{JI}
-{\left(F_L^{dS{\rm corr}}\right)^J\over{\overline m}_{d_J}}
\left(\Delta m_d\right)^{JI}
\label{eqn:XRL}
\end{eqnarray}
\begin{eqnarray}
&&\left[X^S_{LR}\right]^{JI}=\left(\Delta F^{dS}_R
+{1\over2}\Sigma^d_{VL}  F_R^{dS{\rm corr}}
+{1\over2}F_R^{dS{\rm corr}} \Sigma^d_{VR}\right)^{JI}
\label{eqn:XLR}\\
&&\phantom{aaaaaa}
-{\left(F_R^{dS{\rm corr}}\right)^J\over{\overline m}_{d_J}}
\left(\Delta m_d^\dagger\right)^{JI}
+\left({\left(F_R^{dS{\rm corr}}\right)^I\over{\overline m}_{d_J}}
-{\left(F_R^{dS{\rm corr}}\right)^J{\overline m}_{d_I}
\over{\overline m}_{d_J}^2}\right)(\Delta m_d)^{JI}\phantom{aaa}\nonumber
\end{eqnarray}
and for $J<I$ the expressions for $(X^S_{RL})^{JI}$ and
$(X^S_{LR})^{JI}$ can be obtained by using the rules (remember that
$(F^{dS{\rm corr}}_L)^{J\ast}=(F^{dS{\rm corr}}_R)^J$):
\begin{eqnarray}
\left[X^S_{RL}\right]^{JI} = 
\left[X^{S\dagger}_{LR}\right]^{JI}=\left[X^S_{LR}\right]^{IJ\ast}, 
\phantom{aa}
\left[X^S_{LR}\right]^{JI} = 
\left[X^{S\dagger}_{RL}\right]^{JI}=\left[X^S_{RL}\right]^{IJ\ast}
\label{eqn:rules}
\end{eqnarray}
Explicitly for $J<I$  one gets
\begin{eqnarray}
&&\left[X^S_{RL}\right]^{JI}=\left(\Delta F^{dS}_L
+{1\over2}\Sigma^d_{VR}  F_L^{dS{\rm corr}}
+{1\over2}F_L^{dS{\rm corr}} \Sigma^d_{VL}\right)^{JI}\label{eqn:XLRbis}\\
&&\phantom{aaaaaa}
-\left(\Delta m_d\right)^{JI}
{\left(F_L^{dS{\rm corr}}\right)^I\over{\overline m}_{d_I}}
+\left({\left(F_L^{dS{\rm corr}}\right)^J\over{\overline m}_{d_I}}
-{\left(F_L^{dS{\rm corr}}\right)^I{\overline m}_{d_J}
\over{\overline m}_{d_I}^2}\right)(\Delta m^\dagger_d)^{JI}\phantom{aaa}
\nonumber
\end{eqnarray}
\begin{eqnarray}
\left[X^S_{LR}\right]^{JI}=\left(\Delta F^{dS}_R
+{1\over2}\Sigma^d_{VL}  F_R^{dS{\rm corr}}
+{1\over2}F_R^{dS{\rm corr}} \Sigma^d_{VR}\right)^{JI}
-{\left(F_R^{dS{\rm corr}}\right)^I\over{\overline m}_{d_I}}
\left(\Delta m_d^\dagger\right)^{JI}~.
\label{eqn:XRLbis}
\end{eqnarray}

We observe that for $J>I$ $(I>J)$ the effects of the rediagonalization
in $X_{LR}$ $(X_{RL})$ are suppressed by a factor $\sim
m_{d_I}/m_{d_J}({m}_{d_J}/{m}_{d_I})$ compared to its effects in
$X_{RL}$ $(X_{LR})$.  It has to be remembered that $m_d$ and $V$
entering the calculation of $\Delta F^{dS}_{L(R)}$, $\Delta m_d$ etc.
have to be ultimately expressed in terms of ${\overline m}_d$ and
$V^{\rm eff}$.

\subsection{Charged Higgs-fermion vertices}
\label{subsec:chfvert}

In the same manner the operations (\ref{eqn:rescaling}) and
(\ref{eqn:rotations}) imply the following charged scalar ($H^\pm$) and
charged Goldstone boson ($G^\pm$) couplings
\be\label{LH}
{\cal L}^{H^+}_{\rm eff}=
\overline{(u_J)_R}\left[P^H_{RL}\right]^{JI}(d_I)_L H^+ +
\overline{(u_J)_L}\left[P^H_{LR}\right]^{JI}(d_I)_R H^+ +{\rm H.c.}
\ee
\be\label{LG}
{\cal L}^{G^+}_{\rm eff}=
\overline{(u_J)_R}\left[P^G_{RL}\right]^{JI}(d_I)_L G^+ +
\overline{(u_J)_L}\left[P^G_{LR}\right]^{JI}(d_I)_R G^+ + {\rm H.c.}
\ee
where ($i=H^+,G^+$)
\be\label{PRL}
\left[P^i_{RL}\right]=\eta^i_L m_u V^{\rm eff}+ 
\eta^i_L(\mathbf{\Delta U}^\dagger_R m_u-
m_u \mathbf{\Delta U}^\dagger_L) V^{\rm eff} +
\mathbf{U}^\dagger_R\Delta\hat F^i_L\mathbf{D}_L
\ee
\be\label{PLR}
\left[P^i_{LR}\right]=\eta^i_R  V^{\rm eff}m_d + 
\eta^i_R V^{\rm eff} (m_d \mathbf{\Delta D}_R -
\mathbf{\Delta D}_L m_d) 
+\mathbf{U}^\dagger_L\Delta\hat F^i_R\mathbf{D}_R
\ee

\begin{figure}[htbp]
\begin{center}
\begin{picture}(300,100)(0,0)
\Vertex(45,40){13}
\ArrowLine(40,45)(10,80)
\ArrowLine(10,0)(40,35)
\DashArrowLine(50,40)(110,40){4}
\Text(5,70)[]{$u_J$}
\Text(5,10)[]{$d_I$}
\Text(105,50)[]{$H_k^-$}
\Text(220,40)[]{$=i\left(\left(\Delta F_L^k\right)^{JI}P_L 
+ \left(\Delta F_R^k\right)^{JI}P_R\right)$}
\end{picture}
\end{center}
\caption{One-loop corrections to charged Higgs-quark vertices.}
\label{fig:CHiggscorr}
\end{figure}
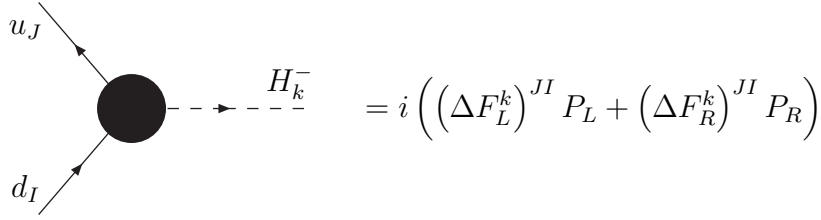

\noindent Here
\be 
\Delta\hat F^k_L=\Delta F^k_L +{1\over2}\Sigma_{VR}^u
F_L^k+{1\over2}F_L^k \Sigma_{VL}^d \label{PRLform}
\ee
\be
\Delta\hat F^k_R= \Delta F^k_R +{1\over2}\Sigma_{VL}^u
F_R^k+{1\over2}F_R^k \Sigma_{VR}^d \label{PLRform}
\ee
with 
\be
F^k_L=\eta^k_L m_u V, \qquad F^k_R=\eta^k_R V m_d
\ee
and $\Delta F^k_{L(R)}$ defined in analogy with (\ref{LINT}) (see
Appendix A.3 for the explicit expressions).

Finally
\begin{eqnarray}
\eta_L^H={g_2\over\sqrt2M_W}\cot\beta,\phantom{aaa} 
\eta_R^H={g_2\over\sqrt2M_W}\tan\beta
\end{eqnarray}
and
\begin{eqnarray}
\eta_L^G={g_2\over\sqrt2M_W},\phantom{aaa} \eta_R^G=-{g_2\over\sqrt2M_W}.
\end{eqnarray}

In writing (\ref{PRL}) and (\ref{PLR}) we have used (\ref{eqn:CKMeff})
in order to convert the CKM matrix in the tree level vertices
$F^i_{L(R)}$ into the effective CKM matrix $V^{\rm eff}$.

In what follows it will be useful to define the parameters
$\epsilon^{HL}_{JI}$, $\epsilon^{HR}_{JI}$, $\epsilon^{GL}_{JI}$ and
$\epsilon^{GR}_{JI}$ through
\be\label{PHIGGS}
\left[P^H_{RL}\right]^{JI}=\eta^H_L \overline m_{u_J} V^{\rm eff}_{JI}
(1-\epsilon^{HL}_{JI}), \qquad
\left[P^H_{LR}\right]^{JI}=\eta^H_R V^{\rm eff}_{JI} \overline m_{d_I} 
(1-\epsilon^{HR}_{JI}),
\ee
\be\label{PGOLD}
\left[P^G_{RL}\right]^{JI}=\eta^G_L \overline m_{u_J} V^{\rm eff}_{JI}
(1+\epsilon^{GL}_{JI}), \qquad
\left[P^G_{LR}\right]^{JI}=\eta^G_R V^{\rm eff}_{JI} \overline m_{d_I} 
(1+\epsilon^{GR}_{JI}).
\ee

This parametrization of the $H^+$ and $G^+$ couplings differs from the
parametrization in terms of $\epsilon^\prime$ factors introduced in
ref.~\cite{DEGAGI} but is more natural for our purposes.  In
particular it emphasizes the fact that ultimately all vertices should
be expressed in terms of $\overline m_i$ and $V_{JI}^{\rm eff}$.

The numerical analysis of the parameters $\epsilon^{HL}_{JI}$,
$\epsilon^{HR}_{JI}$, $\epsilon^{GL}_{JI}$ and $\epsilon^{GR}_{JI}$ is
presented in section~\ref{sec:comp}. While the general formulae for
these parameters are rather involved, they simplify considerably in
the $SU(2)\times U(1)$ symmetry limit with the gauge couplings $g_1$
and $g_2$ set to zero as we will see in section \ref{sec:su2xu1}.  In
particular the parameters $\epsilon^{GL}_{JI}$ and
$\epsilon^{GR}_{JI}$ vanish in this limit.  Therefore
$\epsilon^{GL}_{JI}$ and $\epsilon^{GR}_{JI}$ should be generally
smaller than $\epsilon^{HL}_{JI}$, and $\epsilon^{HR}_{JI}$.  This
expectation is confirmed by our numerical analysis presented in
section 5.

\subsection{Chargino-Quark-Squark Vertices}
\label{subsec:cqvert}

In the approach formulated above charginos and squarks are integrated
out and do not appear as explicit degrees of freedom in the low energy
effective theory.  Their virtual effects are accounted for by
corrections $\Delta m_{d,u}$, various formfactors $\Delta F$ and, as
already mentioned, by higher dimension operators.  Let us consider for
definiteness one of several such non-renormalizable operators
\begin{eqnarray}\label{NONREOP}
\Delta{\cal L}^{\rm eff}=-C_{JIMN}^{VLR} 
\left(\overline{(d_J)_L}\gamma^\mu(d_I)_L\right)
\left(\overline{(d_M)_R}\gamma_\mu(d_N)_R\right)
\end{eqnarray}
which is generated by chargino-stop box diagrams in the course of
integrating out these particles.  $C_{JIMN}^{VLR}$ are the Wilson
coefficients that depend on flavour indices $J,I,M,N$ and can be
calculated by using the original MSSM vertices
\begin{eqnarray}\label{CHL}
{\cal L}^{\rm MSSM}_\chi=-\sum_{i=1,2}\overline{\chi_i^-}
\tilde u^\dagger\left(c_L^i d_L + c_R^i d_R\right)+{\rm H.c.}
\end{eqnarray}
which (in the matrix notation) have the following structure
\cite{ROS}:
\begin{eqnarray}\label{RLi}
c_L^i=\left(a^i+b^i\cdot{m_u\over\sin\beta}\right)\cdot V,\phantom{aaaaa}
c_R^i={h^i\over\cos\beta}\cdot V\cdot m_d
\end{eqnarray}
Here $a^i$, $b^i$ and $h^i$ are $6\times3$ matrices that depend on
chargino compositions and left-right mixing of squarks.  The
coefficients like $C_{JIMN}^{VLR}$ corresponding to definite
chiralities of external quark fields are given as sums over squark and
chargino indices of appropriate products of $c^i_L$, $c^i_R$,
$(c^i_L)^\dagger$ and $(c^i_R)^\dagger$ weighted by some functions of
chargino and up-type squark masses.

The operations (\ref{eqn:rescaling}) and (\ref{eqn:rotations})
performed on the quark fields in the effective Lagrangian affect also
the coefficient $C_{JIMN}^{VLR}$.  Neglecting rescalings
(\ref{eqn:rescaling}) which are not enhanced by $\tan\beta$ it is easy
to see that the rotations (\ref{eqn:rotations}) amount simply to the
multiplication from the right of each $c^i_L$ and $c^i_R$ entering
$C_{JIMN}^{VLR}$ by the matrix $\mbox{\boldmath$D$}_L$ and
$\mbox{\boldmath$D$}_R$, respectively.  From (\ref{eqn:CKMeff}) it
follows that in the approximation $\mbox{\boldmath$\Delta
  U$}_{L(R)}\approx0$ this has the effect of replacing $V$ by $V^{\rm
  eff}$ in all $c^i_L$.  For the factors $c^i_R$ the effect of the
rotation is more complicated:
\begin{eqnarray}
V\cdot m_d\cdot \mbox{\boldmath$D$}_R \approx 
V^{\rm eff}\cdot\mbox{\boldmath$D$}_L^\dagger \cdot m_d
\cdot \mbox{\boldmath$D$}_R\approx 
V^{\rm eff}\cdot m_d + V^{\rm eff}\cdot\left(m_d\cdot
\mbox{\boldmath$\Delta D$}_R-\mbox{\boldmath$\Delta D$}_L\cdot m_d\right)
\nonumber
\end{eqnarray}

Thus, the most important effects of the rotations
(\ref{eqn:rotations}) in the non-renormalizable operators can be
simply taken into account by calculating their Wilson coefficients
using the modified chargino-squark couplings:
\begin{eqnarray}\label{RLiNew}
c^i_L\rightarrow\left(a^i+b^i\cdot{m_u\over\sin\beta}\right)
\cdot V^{\rm eff},
\end{eqnarray}
\begin{eqnarray}\label{RRiNew}
c_R^i\rightarrow{h^i\over\cos\beta}\cdot\left[V^{\rm eff}\cdot 
m_d +V^{\rm eff}\cdot\left(m_d \cdot \mbox{\boldmath$\Delta D$}_R-
\mbox{\boldmath$\Delta D$}_L\cdot m_d)\right)\right]
\label{eqn:ctd_mod}
\end{eqnarray}
where $m_{u_J}\approx{\overline m}_{u_J}$ and 
$m_{d_I}={\overline m}_{d_I}/(1+\epsilon_{d_I}\tan\beta)$.

\vskip0.3cm

The effective vertices of the effective Lagrangian calculated as
described in this section do depend on the renormalization scale
$\mu_R$.  According to the standard reasoning this scale should be of
the order of the mass scale of the particles which are integrated out.
The couplings and masses of the effective Lagrangian should be then
treated as running parameters. For example, the contribution of the
Higgs boson exchanges to the Wilson coefficients of the effective
Hamiltonians discussed in sec. 6 should be calculated at the scale
$\mu_{R^\prime}\sim M_{\rm Higgs}\sim M_{H^+}$. If the $M_{\rm
  SUSY}\gg M_{\rm Higgs}\sim M_{H^+}$, the couplings of the effective
Lagrangian should be evolved with the effective theory RGEs from
$\mu_R\sim M_{\rm SUSY}$ down $\mu_{R^\prime}\sim M_{\rm Higgs}\sim
M_{H^+}$.  In our numerical results we neglect this running as it
would not change the results drastically. It should be however taken
into account to predict physical quantities in the MSSM with higher
accuracy.

Finally let us stress that although in what follows we will restrict
our attention to the version of the MSSM in which flavour and CP
violation are ruled by the CKM matrix, the effective Lagrangian
approach described in this section is more general and can be used
\cite{CHSL,CHRO} also in the case of flavour and CP violation
originating in the squark sector.


\section{The limit of unbroken \boldmath{$SU(2)\times U(1)$} symmetry}
\label{sec:su2xu1}

\setcounter{equation}{0}

In this section we will discuss the approach of ref.~\cite{ISRE1} that
can be considered as a special limit of the more general approach
presented in section 2. We will also compare our results with the
recent analysis presented in \cite{AMGIISST} that goes beyond
supersymmetry but similarly to \cite{ISRE1} is based on the same
approximation. In order to relate these approaches to our approach we
consider decoupling of sparticles in the limit of unbroken
$SU(2)\times U(1)$ symmetry. The electroweak symmetry breaking is then
taken into account after sparticles are integrated out. The main goal
of this section is the derivation of explicit expressions for the
flavour changing neutral scalar couplings $\left[X^S_{RL}\right]^{JI}$
and $\left[X^S_{LR}\right]^{JI}$ and the charged scalar couplings
$\left[P^H_{RL}\right]^{JI}$, $\left[P^H_{LR}\right]^{JI}$,
$\left[P^G_{RL}\right]^{JI}$ and $\left[P^G_{LR}\right]^{JI}$ at large
$\tan\beta$ in the $SU(2)\times U(1)$ symmetry limit.  This
approximation should be valid if the sparticle mass scale is larger
than the mass scale of the Higgs boson sector (set by $M_{H^+}$). It
should be emphasized that the absence of vacuum expectation values
before decoupling implies neglecting the left-right mixing of
squarks even for non-vanishing $A_{u,d}$ and/or $\mu$ parameters.

In this section following the practice in the existing literature we
will assume also the dominance of $\alpha_s$ and top and bottom Yukawa
couplings neglecting the contributions of the electroweak couplings
$g_2$ and $g_1$. We will assess the validity of this approximation in
section \ref{sec:comp}.

\subsection{Minimal Flavour Violation and Flavour Dependence}

The large $\tan\beta$ enhanced corrections in the approach of
\cite{ISRE1,AMGIISST} are governed in the MSSM by four universal
parameters that we will denote by $\epsilon_0$, $\epsilon_Y$,
$\epsilon_0^\prime$ and $\epsilon_Y^\prime$. These parameters as
defined in \cite{ISRE1,AMGIISST} are flavour independent. Therefore
flavour dependence enters the Higgs and chargino couplings only
through the CKM matrix $V_{\rm eff}$.  This scenario can only be
realized if the soft SUSY breaking mass squared matrix $m^2_Q$ is
strictly proportional to the unit matrix. This corresponds to the 
Minimal Flavour Violation as defined in ref. \cite{AMGIISST} and 
to the scenario A in \cite{BOEWKRUR}.

However, in more realistic situations the diagonal entries
$m^2_{Q_3}$, $m^2_{U_3}$ and $m^2_{D_3}$ of the matrices $m^2_Q$,
$m^2_U$ and $m^2_D$ that are related to third generation squarks
differ from the entries corresponding to the first two generations to
be denoted collectively by $m^2$. As pointed out in \cite{MIPORO} in
this situation some flavour violation (still ruled by the CKM matrix)
unavoidably appears in the up- or down-type (or in both) squark mass
squared matrices. For $m^2\neq m^2_{Q_3}$, at the phenomenological
level, it becomes a matter of choice in which basis $m^2_Q$ is assumed
to be diagonal. The two simplest choices are
\begin{eqnarray}\label{FLAVOUR1}
({\cal M}^2_D)_{\rm LL}=m^2_Q+\dots \phantom{aaa}{\rm and}\phantom{aaa}
({\cal M}^2_U)_{\rm LL}= V m^2_QV^\dagger+\dots
\end{eqnarray}
or 
\begin{eqnarray}\label{FLAVOUR2}
({\cal M}^2_D)_{\rm LL}= V^\dagger m^2_QV+\dots \phantom{aaa}
{\rm and}\phantom{aaa}({\cal M}^2_U)_{\rm LL}=m^2_Q+\dots
\end{eqnarray}
where $m^2_Q$ is diagonal and the ellipses stand for other diagonal
terms.  In the classification in \cite{BOEWKRUR}, the settings in
(\ref{FLAVOUR1}) and (\ref{FLAVOUR2}) correspond to scenarios B and C,
respectively.

In our calculations in the approach of section 2 we choose the soft
SUSY breaking mass parameter $m^2_Q$ as in (\ref{FLAVOUR1}). The
scenario (\ref{FLAVOUR2}) with flavour violation in the down-type
squark mass matrix would require the inclusion of box and Higgs
penguin diagrams with gluinos and is beyond the scope of this paper.
Note that in SUSY
breaking scenarios like minimal SUGRA or gauge mediation, in which
proportionality of $m^2_Q$ to the unit matrix is assumed at some high
scale, the renormalization group evolution produces calculable off
diagonal terms both in $({\cal M}^2_U)_{\rm LL}$ and in $({\cal
M}^2_D)_{\rm LL}$. This is however a secondary effect as the
evolution has to generate first the splitting of the diagonal entries
of the squark mass matrices. The magnitude of the off diagonal entries
is then proportional to this splitting. Due to this fact the effects
of the splitting are always more important in flavour changing
amplitudes. Our framework should therefore constitute a good
approximation to predictions of the SUSY breaking scenarios in which
at some high scale the matrices $m^2_Q$, $m^2_U$ and $m^2_D$ are
proportional to the unit matrix and the trilinear soft parameters
$A_t$ and $A_b$ are real.

Allowing for different values of the parameters $m^2_{Q_3}$,
$m^2_{U_3}$, $m^2_{D_3}$ and $m^2$ introduces some flavour dependence
into $\epsilon_0$, $\epsilon_Y$, $\epsilon_0^\prime$,
$\epsilon_Y^\prime$ and the formulae for various couplings presented
in \cite{ISRE1,AMGIISST} have to be generalized to account for it.
However, for the sake of comparison with \cite{ISRE1,AMGIISST} we will
for the most part of this section treat $\epsilon_0$, $\epsilon_Y$,
$\epsilon_0^\prime$ and $\epsilon_Y^\prime$ as universal flavour
independent quantities and only at the end, in subsection 3.6, we will
provide formulae for various couplings that are also valid for flavour
dependent $\epsilon_0$, $\epsilon_Y$, $\epsilon_0^\prime$ and
$\epsilon_Y^\prime$. As we will see in sec. \ref{sec:comp}, obtaining
a numerically reliable approximation to the results of the complete
calculation done as in sec. 2 will require extending the approach of
refs. \cite{ISRE1,AMGIISST} by taking into account both the flavour
dependence of $\epsilon_0$, $\epsilon_Y$, $\epsilon_0^\prime$ and
$\epsilon_Y^\prime$ and the effects of the electroweak couplings $g_2$
and $g_1$. The latter will be automatically included in the formulae
presented in sec. \ref{sec:su2xu1flav} by defining appropriately
flavour dependent factors $\epsilon_0$, $\epsilon_Y$,
$\epsilon_0^\prime$ and $\epsilon_Y^\prime$ as discussed in section
\ref{sec:comp}.  Thus, the analytic expressions of sec.
\ref{sec:su2xu1flav} are the basis of the approximation that within a
few percent reproduces the numerical results of the full approach of
section 2 which includes automatically both the $SU(2)\times U(1)$
breaking effects and the effects of electroweak couplings.

\subsection{Effective Lagrangians}

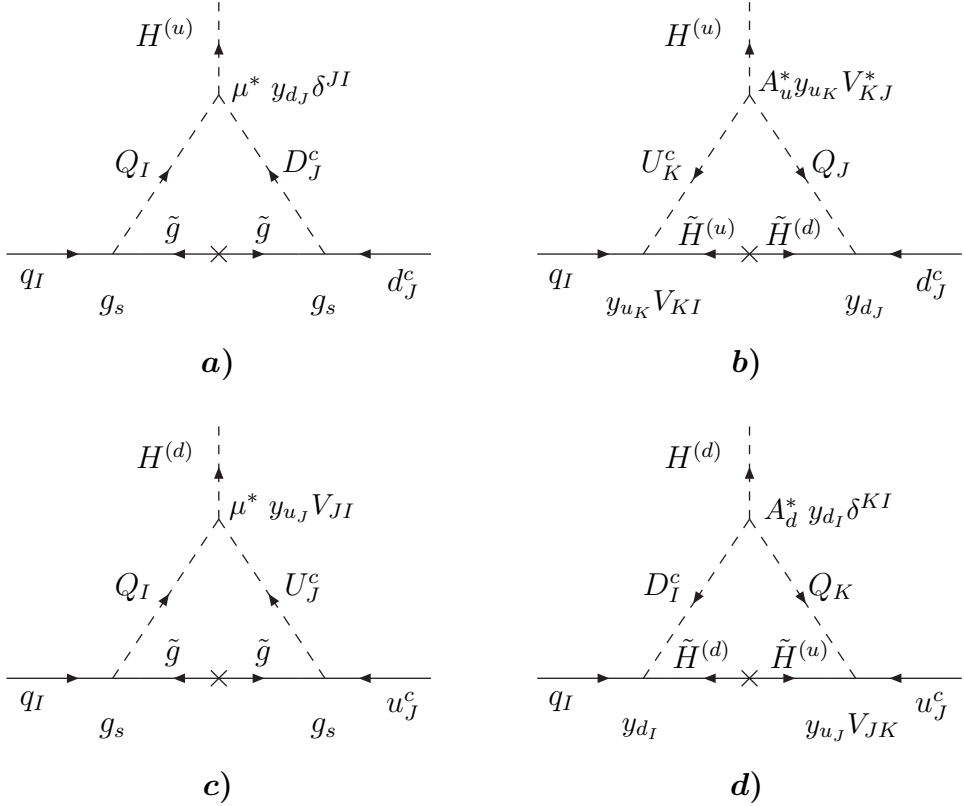
\begin{figure}[htbp] 
\begin{center}
\begin{picture}(400,310)(0,0)
\ArrowLine(90,200)(120,200)
\ArrowLine(90,200)(60,200)
\ArrowLine(10,200)(60,200)
\ArrowLine(170,200)(120,200)
\Line(87,203)(93,197)
\Line(87,197)(93,203)
\Text(20,195)[t]{$q_I$}
\Text(160,195)[t]{$d^c_J$}
\Text(73,215)[t]{$\tilde{g}$}
\Text(107,215)[t]{$\tilde{g}$}
\Text(58,240)[t]{$Q_I$}
\Text(122,240)[t]{$D^c_J$}
\Text(70,290)[t]{$H^{(u)}$}
\Text(50,185)[t]{$g_s$}
\Text(130,185)[t]{$g_s$}
\Text(118,270)[t]{\small $\mu^\ast ~y_{d_J}\delta^{JI}$}
\DashArrowLine(50,200)(90,260){4}
\DashArrowLine(130,200)(90,260){4}
\DashArrowLine(90,260)(90,295){4}
\Text(90,165)[t]{$\mbox{\boldmath$a)$}$}
\ArrowLine(290,200)(320,200)
\ArrowLine(290,200)(260,200)
\ArrowLine(210,200)(260,200)
\ArrowLine(370,200)(320,200)
\Line(287,203)(293,197)
\Line(287,197)(293,203)
\Text(220,195)[t]{$q_I$}
\Text(360,195)[t]{$d^c_J$}
\Text(275,215)[t]{$\tilde H^{(u)}$}
\Text(308,215)[t]{$\tilde H^{(d)}$}
\Text(258,240)[t]{$U^c_K$}
\Text(322,240)[t]{$Q_J$}
\Text(270,290)[t]{$H^{(u)}$}
\Text(255,187)[t]{\small $y_{u_K}V_{KI}$}
\Text(335,185)[t]{$y_{d_J}$}
\Text(320,270)[t]{\small $A_u^\ast y_{u_K}V^\ast_{KJ}$}
\DashArrowLine(290,260)(250,200){4}
\DashArrowLine(290,260)(330,200){4}
\DashArrowLine(290,260)(290,295){4}
\Text(290,165)[t]{$\mbox{\boldmath$b)$}$}
\ArrowLine(90,40)(120,40)
\ArrowLine(90,40)(60,40)
\ArrowLine(10,40)(60,40)
\ArrowLine(170,40)(120,40)
\Line(87,43)(93,37)
\Line(87,37)(93,43)
\Text(20,35)[t]{$q_I$}
\Text(160,35)[t]{$u^c_J$}
\Text(50,25)[t]{$g_s$}
\Text(130,25)[t]{$g_s$}
\Text(73,55)[t]{$\tilde{g}$}
\Text(107,55)[t]{$\tilde{g}$}
\Text(58,80)[t]{$Q_I$}
\Text(122,80)[t]{$U^c_J$}
\Text(118,110)[t]{\small $\mu^\ast ~y_{u_J}V_{JI}$}
\Text(70,130)[t]{$H^{(d)}$}
\DashArrowLine(50,40)(90,100){4}
\DashArrowLine(130,40)(90,100){4}
\DashArrowLine(90,100)(90,135){4}
\Text(90,5)[t]{$\mbox{\boldmath$c)$}$}
\ArrowLine(290,40)(260,40)
\ArrowLine(290,40)(320,40)
\ArrowLine(210,40)(260,40)
\ArrowLine(370,40)(320,40)
\Line(287,43)(293,37)
\Line(287,37)(293,43)
\Text(220,35)[t]{$q_I$}
\Text(360,35)[t]{$u^c_J$}
\Text(273,55)[t]{$\tilde H^{(d)}$}
\Text(310,55)[t]{$\tilde H^{(u)}$}
\Text(258,80)[t]{$D^c_I$}
\Text(322,80)[t]{$Q_K$}
\Text(321,110)[t]{\small $A_d^\ast ~y_{d_I}\delta^{KI}$}
\Text(270,130)[t]{$H^{(d)}$}
\Text(250,25)[t]{\small $y_{d_I}$}
\Text(330,27)[t]{\small $y_{u_J}V_{JK}$}
\DashArrowLine(290,100)(250,40){4}
\DashArrowLine(290,100)(330,40){4}
\DashArrowLine(290,100)(290,135){4}
\Text(290,5)[t]{$\mbox{\boldmath$d)$}$}
\end{picture}
\end{center}
\caption{Vertex corrections in the $SU(2)\times U(1)$ symmetry limit
  for vanishing electroweak gauge couplings.  Diagrams a) and b) give
  rise to corrections $(\Delta_u\mathbf{Y}_d)^{JI}$.  Corrections
  $(\Delta_d\mathbf{Y}_d)^{JI}$ are generated by similar diagrams with
  outgoing $H^{(u)}$ replaced by incoming $H^{(d)}$ and factors
  $\mu^\ast$ and $A_u^\ast$ in diagrams a) and b) replaced by $A_d$
  and $\mu$, respectively.  Diagrams c) and d) give rise to
  corrections $(\Delta_d\mathbf{Y}_u)^{JI}$.  Corrections
  $(\Delta_u\mathbf{Y}_u)^{JI}$ are generated by similar diagrams with
  outgoing $H^{(d)}$ replaced by incoming $H^{(u)}$ and factors
  $\mu^\ast$ and $A_d^\ast$ in diagrams c) and d) replaced by $A_u$
  and $\mu$, respectively.}
\label{fig:bcrs1}
\end{figure}

In this approach below the sparticle mass scale we get the effective
Lagrangian whose part relevant for the neutral Higgs boson couplings
to the down-type quarks has the form~\cite{CHPO,BAKO}
\begin{eqnarray}
{\cal L}_{\rm eff} 
=-\epsilon_{ij}H^{(d)}_i\overline{d_R}\cdot(
\mathbf{Y}_d+\Delta_d\mathbf{Y}_d)\cdot q_{jL}
-H^{(u)\ast}_i\overline{d_R}\cdot \Delta_u\mathbf{Y}_d\cdot q_{iL}
+{\rm H.c.}\label{eqn:effL_1}
\end{eqnarray}
where $\epsilon_{21}=-\epsilon_{12}=1$ and $\mathbf{Y}_{d,u}$ are
Yukawa coupling matrices.  As in the papers \cite{ISRE1,AMGIISST} we
make further assumption (which by itself is not related to the
symmetry limit) that the effects of the electroweak gauge couplings
$g_1$ and $g_2$ are negligible. We will examine numerically the
validity of this approximation in sec. 5. With this assumption the
only diagrams giving rise to the correction $\Delta_u\mathbf{Y}_d$ are
shown in Figs. \ref{fig:bcrs1}a and \ref{fig:bcrs1}b. The loop induced
term $-\epsilon_{ij}H^{(d)}_i\Delta_d\mathbf{Y}_d$ is always
subleading in the large $\tan\beta$ limit and can be neglected.  In
the basis in which
\begin{eqnarray}
\mathbf{Y}_d={\rm diag}(y_d), \phantom{aaa}
\mathbf{Y}_u={\rm diag}(y_u)\cdot V\label{eqn:yukbasis}
\end{eqnarray}
the correction $\Delta_u\mathbf{Y}_d$ can be easily seen to have the
structure~\cite{ISRE1}
\begin{eqnarray}
\left(\Delta_u\mathbf{Y}_d\right)^{JI}=-y_{d_J}
\left(\epsilon_0\delta^{JI}+\epsilon_Yy^2_tV^{3J\ast}V^{3I}\right)\equiv
-y_{d_J}\left(\tilde\epsilon_J\delta^{JI}+\epsilon_Yy^2_t\lambda_0^{JI}\right)
\label{eqn:DuYd_str}
\end{eqnarray}
where the contributions proportional to $y^2_u$ and $y_c^2$ have been 
neglected,
\be\label{epJ}
\tilde\epsilon_J\equiv\epsilon_0+\epsilon_Yy^2_tV^{3J\ast}V^{3J}
\approx \epsilon_0+\epsilon_Yy^2_t\delta^{J3}
\ee
and
\be
\lambda_0^{JI}=V^{3J\ast}V^{3I}\quad {\rm for}\quad J\neq I
\quad {\rm and}\quad \lambda_0^{JJ}=0\quad {\rm for}\quad J=I.
\ee
The approximation in (\ref{epJ}) is justified by the hierarchical
structure of the elements of the CKM matrix.  The correction
$\Delta_d\mathbf{Y}_d$ has the same structure (\ref{eqn:DuYd_str}) as
$\Delta_u\mathbf{Y}_d$ but is subleading for large $\tan\beta$.

As follows from figs. \ref{fig:bcrs1}a and \ref{fig:bcrs1}b, in
$\left(\Delta_u\mathbf{Y}_d\right)^{JI}$ the correction $\epsilon_0$
depends on the gluino mass and the masses of the $\tilde D^c_J$ and
$\tilde Q_J$ squarks, while $\epsilon_Y$ depends on the $\mu$
parameter and the masses of the $\tilde Q_J$ and $\tilde U^c_3$
squarks.  Note that in this limit the parameter $\mu$ plays the role
of the mass of the higgsinos. The dependence of $\epsilon_0(\tilde
D^c_J,\tilde Q_J)$ and $\epsilon_Y(\tilde Q_J,\tilde U^c_3)$ on the
generation indices can be neglected if the squark masses are not
significantly split. It will be taken into account in sec.
\ref{sec:su2xu1flav}

Decomposing the two Higgs doublets as
\begin{eqnarray}\label{DECOMP}
H^{(d)}_1={v_d\over\sqrt2}+{1\over\sqrt2}\left(c_\alpha H^0-s_\alpha h^0
+is_\beta A^0 -ic_\beta G^0 \right)\nonumber \\ 
H^{(u)\ast}_2={v_u\over\sqrt2}+{1\over\sqrt2}\left(s_\alpha H^0+c_\alpha h^0
-ic_\beta A^0 -is_\beta G^0 \right)
\end{eqnarray}
(where $c_\alpha\equiv\cos\alpha$, etc.) we find first
\begin{eqnarray}
{\cal L}_{\rm mass}^{(d)}=
{\overline d_R}\cdot\left({v_d\over\sqrt2}y_d 
+{v_d\over\sqrt2}\Delta_d\mathbf{Y}_d
-{v_u\over\sqrt2}\Delta_u\mathbf{Y}_d
\right)\cdot d_L +{\rm H.c.}
\label{eqn:dmasscorr}
\end{eqnarray}
Comparison with (\ref{eqn:effmassterm}) gives then
\begin{eqnarray}
\left(\Delta m_d\right)^{JI}=-{v_d\over\sqrt2}
\left(\Delta_d\mathbf{Y}_d -\tan\beta ~
\Delta_u\mathbf{Y}_d\right)^{JI}
\approx{v_d\over\sqrt2}\tan\beta ~
\left(\Delta_u\mathbf{Y}_d\right)^{JI}.
\label{eqn:dmdsymlim}
\end{eqnarray} 
In our conventions
\be
m_{d_J}=-{v_d\over\sqrt2} y_{d_J}, \qquad
m_{u_J}= {v_u\over\sqrt2} y_{u_J}
\ee
and $v_d^2/\cos^2\beta=v_u^2/\sin^2\beta=1/\sqrt2G_F\approx 
(246~{\rm GeV})^2$. Consequently
\begin{eqnarray}
&&\left(\Delta m_d\right)^{JI}=m_{d_J}\tan\beta\left(\tilde\epsilon_J
\delta^{JI}+\epsilon_Yy^2_t\lambda_0^{JI}\right)\nonumber\\
&&\left(\Delta m_d^\dagger\right)^{JI}
=m_{d_I}\tan\beta\left(\tilde\epsilon_I
\delta^{JI}+\epsilon_Yy^2_t\lambda_0^{JI}\right)\label{eqn:dmd_symlim}
\label{eqn:deltamasy}\\
&&\kappa_{d_J}\approx\epsilon_{d_J}\tan\beta 
=\tilde\epsilon_J\tan\beta\nonumber
\end{eqnarray}
with $\epsilon_{d_J}$ defined in (\ref{eqn:eps_def}) and
$\tilde\epsilon_J$ given in (\ref{epJ}). The last equation tells us
that in the approximation considered here
$\epsilon_{d_J}=\tilde\epsilon_J$. Thus using (\ref{eqn:corrmass}) we
have
\be\label{basic}
m_{d_J}=\frac{\overline m_{d_J}}{1+\tilde\epsilon_J\tan\beta},
\ee
with $\overline m_{d_J}$ denoting the running quark masses as
discussed in Section 2.  Note that the large $\tan\beta$ corrections
to $m_d$ and $m_s$ in contrast to $m_b$ do not involve the top Yukawa
coupling.

Similarly the corrections to the up-type quarks Yukawa coupling can be
summarized by the effective Lagrangian
\begin{eqnarray}
{\cal L}_{\rm eff} 
=-\epsilon_{ij}H^{(u)}_i\overline{u_R}\cdot
\left(\mathbf{Y}_u+\Delta_u\mathbf{Y}_u\right)\cdot q_{jL}
-H^{(d)\ast}_i\overline{u_R}\cdot \Delta_d\mathbf{Y}_u\cdot q_{iL}
+{\rm H.c.}\label{eqn:effL_2}
\end{eqnarray}
The correction $\Delta_d\mathbf{Y}_u$ is generated by the diagrams
shown in figs.  \ref{fig:bcrs1}c and \ref{fig:bcrs1}d.  In the basis
(\ref{eqn:yukbasis}) one has
\be
\left(\Delta_d\mathbf{Y}_u\right)^{JI}=y_{u_J}V_{JI}
\left(\epsilon^\prime_0+\epsilon^\prime_Yy^2_{d_I}\right)~.
\label{eqn:DdYu_str}
\ee
where the factor $\epsilon^\prime_0$ depends on the masses of gluino
and $\tilde U^c_J$ and $\tilde Q_I$ squarks, while $\epsilon^\prime_Y$
depends on the $\mu$ parameter and the masses of $\tilde Q_I$ and
$\tilde D^c_I$ squarks. For squark masses not significantly split the
generation dependence of $\epsilon^\prime_0(\tilde U^c_J,\tilde Q_I)$
and $\epsilon^\prime_Y(\tilde Q_I,\tilde D^c_I)$ on the generation
indices can be dropped.  The correction $\Delta_u\mathbf{Y}_u$ has the
same general structure.  The mass term for the up-type quarks takes
then the form
\begin{eqnarray}
{\cal L}_{\rm mass}^{(u)}=-
{\overline u_R}\cdot\left({v_u\over\sqrt2}y_uV 
+{v_u\over\sqrt2}\Delta_u\mathbf{Y}_u
+{v_d\over\sqrt2}\Delta_d\mathbf{Y}_u
\right)\cdot u_L +{\rm H.c.}
\label{eqn:umasscorr}
\end{eqnarray}
Note that the effects of the correction $\Delta_d\mathbf{Y}_u$ are
suppressed by $\tan\beta$ with respect to the effects of the
correction $\Delta_u\mathbf{Y}_u$ which itself is not $\tan\beta$
enhanced with respect to the tree level mass term.  The corrected mass
term (\ref{eqn:umasscorr}) is diagonalized by the rotations
\be\label{urotations}
u_L\rightarrow V^\dagger\cdot \mathbf{U}_L u_L,\qquad 
{\overline u}_R\rightarrow {\overline u}_R\mathbf{U}_R^\dagger 
\ee
that with $\mathbf{U}_{L,R}=\mathbf{1} +\mathbf{\Delta U}_{L,R}$
satisfy the conditions (\ref{barmu}) and (\ref{eqn:udiagcond}) with
\begin{eqnarray}
\left(\Delta m_u\right)^{JI}={v_u\over\sqrt2}
\left(\Delta_u\mathbf{Y}_u\cdot V^\dagger +\cot\beta ~
\Delta_d\mathbf{Y}_u\cdot V^\dagger\right)^{JI}
\label{eqn:dmusymlim}
\end{eqnarray}
Since $\Delta m_u$ is not $\tan\beta$ enhanced, $\mathbf{U}_{L,R}$ are
in general very close to unit matrices. Consequently
\be\label{basicu}
m_{u_J}\approx \overline m_{u_J}~.
\ee
\subsection{Neutral Higgs-fermion vertices}

Using the effective Lagrangian (\ref{eqn:effL_1}) and the
decomposition (\ref{DECOMP}) it is easy to identify in the
$SU(2)\times U(1)$ symmetry limit with zero $g_1$ and $g_2$ the vertex
corrections $\Delta F_{L(R)}^{dS}$ for $S^0=H^0$, $h^0$, $A^0$ and
$G^0$ defined in eq.  (\ref{eqn:Sverts}).  We find:
\begin{eqnarray}
&&\left(F_L^{dS}\right)^J={m_{d_J}\over v_d}x_d^S\nonumber
=\left(F_R^{dS}\right)^{J\ast}\\
&&\left(\Delta F_L^{dS}\right)^{JI}={m_{d_J}\over v_d}
\left(\tilde\epsilon_J\delta^{JI}+\epsilon_Yy^2_t\lambda_0^{JI}\right)x_u^S
=\left(\Delta F_R^{dS}\right)^{IJ\ast}\label{FdS}
\end{eqnarray}
where 
\be\label{xS}
x^S_d=(c_\alpha,-s_\alpha,is_\beta,-ic_\beta),\qquad
x^S_u=(s_\alpha,c_\alpha,-ic_\beta,-is_\beta)
\ee
for $S^0=H^0,h^0,A^0$ and $G^0$, respectively. In obtaining
(\ref{FdS}) we have set $\Delta_d\mathbf{Y}_d$ to zero as done in
refs. \cite{ISRE1} and \cite{AMGIISST}. We will proceed in the same
manner in deriving the formulae for the neutral Higgs couplings below.
However, at the end of this subsection we will list for completeness
the corrections to these formulae due to a non-vanishing
$\Delta_d\mathbf{Y}_d$.

It is now easy to check that the diagonal part (\ref{eqn:Sffdiag}) of
the $\bar dS^0d$ vertex is
\begin{eqnarray}\label{DIAGF}
\left[X^S\right]^{JJ}\equiv
\left(\eta_L^{dS}{{\overline m}_{d_J}\over1+\epsilon_{d_J}\tan\beta}
+\Delta F^{dS}_L\right)^{JJ}=
{{\overline m}_{d_J}\over v_d(1+\tilde\epsilon_J\tan\beta)}
\left(x_d^S+\tilde\epsilon_Jx_u^S\right)~.
\end{eqnarray}

For the flavour changing couplings $\left[X^S_{RL}\right]^{JI}$ with
$J>I$, neglecting $\Sigma^d_{VL}$ and $\Sigma^d_{VR}$ and ignoring the
contribution of $\Delta_d\mathbf{Y}_d$ one gets from (\ref{eqn:XRL})
\begin{eqnarray}
&&\left[X^S_{RL}\right]^{JI}
={m_{d_J}\over v_d}\epsilon_Yy^2_t\lambda^{JI}_0\left[
x^S_u-{\tan\beta\over1+\tilde\epsilon_J\tan\beta}x^S_d
-{\tilde\epsilon_J\tan\beta\over1+\tilde\epsilon_J\tan\beta}
x^S_u\right]\nonumber\\
&&\phantom{aaaaaa}
={{\overline m}_{d_J}\over v_d(1+\tilde\epsilon_J\tan\beta)^2}
\epsilon_Yy^2_t\lambda^{JI}_0\left(x_u^S-x^S_d\tan\beta\right)
\label{XRL}
\end{eqnarray}
in agreement with equation (9) of ref.~\cite{ISRE1}. Notice that effects 
of the genuine vertex correction $\Delta F_L^{dS}$ (proportional to 
$x_u^S$) are subleading compared to the effects of the rotation 
$\mathbf{\Delta D}_L\propto\Delta m_d$. In the case of split mass 
parameters $m^2_Q$, the factor $\epsilon_Y$ in (\ref{XRL}) depends on 
$m^2_Q$ of the $J$-th generation and $m_U^2$ of the third generation.

Similarly, neglecting $\Delta_d\mathbf{Y}_d$ we find from (\ref{eqn:XLR}) 
an explicit expression for $\left[X^S_{LR}\right]^{JI}$ with $J>I$
\begin{eqnarray}
\left[X^S_{LR}\right]^{JI}
={m_{d_I}\over v_d}\epsilon_Yy^2_t\lambda^{JI}_0\left[x^{S*}_u
-\tan\beta{m_{d_J}\over {\overline m}_{d_J}}
\epsilon_Yy^2_t\delta^{J3}x^{S*}_u
-\tan\beta{{\overline m}_{d_I}m^2_{d_J}\over m_{d_I}{\overline m}^2_{d_J}}
\left(x_d^{S*}+\tilde\epsilon_Jx^{S*}_u\right)\right]\nonumber\\
={{\overline m}_{d_I}\over v_d(1+\tilde\epsilon_J\tan\beta)^2}
\epsilon_Yy^2_t\lambda^{JI}_0\left(x_u^{S*}-x^{S*}_d\tan\beta\right)~.
\phantom{aaaaaaaaaa}
\label{XLR}
\end{eqnarray}
that has not been given in~\cite{ISRE1}.  In particular this formula
does not follow from the equation (9) of this paper, that for $J>I$ is
valid only for the $\left[X^S_{RL}\right]^{JI}$ couplings.  The
expression (\ref{XLR}) has been derived under the assumption that
$\epsilon_Y(\tilde Q_J,\tilde U^c_3)\approx\epsilon_Y(\tilde Q_I,
\tilde U^c_3)$. Note, that in this case $\left[X^S_{LR}\right]^{JI}$
can be obtained from $\left[X^S_{RL}\right]^{JI}$ by simply replacing
$\overline m_{d_J}$ by $\overline m_{d_I}$ and complex conjugation of
$x_d^{S}$ and $x_u^{S}$ without any other changes in the indices.  The
couplings $\left[X^S_{RL}\right]^{JI}$ and
$\left[X^S_{LR}\right]^{JI}$ for $J<I$ can be found by applying the
rule (\ref{eqn:rules}) to (\ref{XLR}) and (\ref{XRL}), respectively.

In the same approximation it is easy to derive the relation between
the CKM matrix $V_{JI}$ of the MSSM and the effective matrix
$V_{JI}^{\rm eff}$.  Using the hierarchy of quark masses and CKM
matrix entries (and neglecting $\mathbf{\Delta U}_L$) one obtains from
eqs.~(\ref{eqn:CKMeff}), (\ref{eqn:DLsol}), (\ref{eqn:dmd_symlim}) the
relations~\cite{BLPORA,BAKO,ISRE1}
\begin{eqnarray}
&&V_{JI}=V_{JI}^{\rm eff} ~
\left[{1+\tilde\epsilon_3\tan\beta\over1+\epsilon_0\tan\beta}\right]
\phantom{aaa}{\rm for}\phantom{aa} 
(JI)=(13), ~(23), ~(31) ~{\rm and} ~(32),\nonumber\\
&&V_{JI}=V_{JI}^{\rm eff}\phantom{aaa}{\rm otherwise.}
\label{eqn:CKMcorr}
\end{eqnarray}

We can now summarize the results for the neutral Higgs-quark vertices.
In the case of $B$-physics the pairs $(J,I)=(3,2),(3,1),(2,3)$ and
$(1,3)$ matter.  Combining (\ref{XRL}), (\ref{XLR}), the corresponding
formulae for $J<I$, and (\ref{eqn:CKMcorr}) and neglecting the
dependence of $\epsilon_0$ and $\epsilon_J$ on the generation indices
we find the final formulae for the neutral Higgs couplings relevant
for $B$-physics that exhibit all large $\tan\beta$ enhanced
corrections:
\be
\left[X^S_{RL}\right]^{JI}
=\frac{g_2}{2 M_W\cos\beta}
\frac{{\overline m}_{d_J} V^{3J*}_{\rm eff}V^{3I}_{\rm eff}}
{(1+\tilde\epsilon_3\tan\beta)(1+\epsilon_0\tan\beta)}
\epsilon_Yy^2_t \left(x_u^S-x^S_d\tan\beta\right)~,
\label{BXRLFIN}
\ee
\be
\left[X^S_{LR}\right]^{JI}
=\frac{g_2}{2 M_W\cos\beta}
\frac{{\overline m}_{d_I} V^{3J*}_{\rm eff}V^{3I}_{\rm eff}}
{(1+\tilde\epsilon_3\tan\beta)(1+\epsilon_0\tan\beta)}
\epsilon_Yy^2_t \left(x_u^{S*}-x^{S*}_d\tan\beta\right)~,
\label{BXLRFIN}
\ee
that are valid for both $J>I$ and $J<I$.

In the case of $K$-physics the pairs $(J,I)=(2,1)$ and $(1,2)$ matter
and we find
\be
\left[X^S_{RL}\right]^{JI}
=\frac{g_2}{2 M_W\cos\beta}
{\overline m}_{d_J} V^{3J*}_{\rm eff}V^{3I}_{\rm eff}
\frac{(1+\tilde\epsilon_3\tan\beta)^2}{(1+\epsilon_0\tan\beta)^4}
\epsilon_Yy^2_t \left(x_u^S-x^S_d\tan\beta\right)~.
\label{KXRLFIN}
\ee
\be
\left[X^S_{LR}\right]^{JI}=\frac{g_2}{2M_W\cos\beta}
{\overline m}_{d_I} V^{3J*}_{\rm eff}V^{3I}_{\rm eff}
\frac{(1+\tilde\epsilon_3\tan\beta)^2}{(1+\epsilon_0\tan\beta)^4}
\epsilon_Yy^2_t \left(x_u^{S*}-x^{S*}_d\tan\beta\right)~.
\label{KXLRFIN}
\ee
It is evident from (\ref{xS}) and (\ref{BXRLFIN})-(\ref{KXLRFIN}) that
the flavour violating couplings of $G^0$ vanish in this limit.

Finally, we would like to list the corrections to (\ref{FdS}),  
(\ref{DIAGF}), (\ref{XRL}) and (\ref{XLR}) that result from 
\begin{eqnarray}
\left(\Delta_d\mathbf{Y}_d\right)^{JI}=
-y_{d_J}\left(\tilde e_J\delta^{JI}+e_Yy^2_t\lambda_0^{JI}\right)
\label{eqn:DdYd_str}
\end{eqnarray}
with $e_0$, $e_Y$ and $\tilde e_J$ replacing $\epsilon_0$,
$\epsilon_Y$ and $\tilde\epsilon_J$ in $\Delta_u\mathbf{Y}_d$,
respectively. We find:
\begin{eqnarray}\label{FdScorr}
&&\delta\left(\Delta F_L^{dS}\right)^{JI}=
-{m_{d_J}\over v_d}\left(\tilde e_J\delta^{JI}+
e_Yy^2_t\lambda_0^{JI}\right)x_d^S
\end{eqnarray}
and
\begin{eqnarray}
\left(x_d^S+\tilde\epsilon_Jx_u^S\right) \to
\left(x_d^S-\tilde e_Jx_d^S+\tilde\epsilon_Jx_u^S\right)
\end{eqnarray}
in (\ref{DIAGF}). In $[X^S_{RL}]^{JI}$ and $[X^S_{LR}]^{JI}$ the
correction $\Delta_d\mathbf{Y}_d$ produces the following subleading
terms:
\begin{eqnarray}\label{XRLcorr}
\delta [X^S_{RL}]^{JI}=
{{\overline m}_{d_J}\over v_d(1+\tilde\epsilon_J\tan\beta)^2}
y^2_t\lambda^{JI}_0\left[-e_Y + (e_0\epsilon_Y
-e_Y\epsilon_0)\tan\beta\right]x^S_d
\end{eqnarray}
\begin{eqnarray}
\delta [X^S_{LR}]^{JI}= 
{{\overline m}_{d_I}\over v_d(1+\tilde\epsilon_J\tan\beta)^2}
y^2_t\lambda^{JI}_0\left[-e_Y + (e_0\epsilon_Y
-e_Y\epsilon_0)\tan\beta\right]x^{S*}_d
\end{eqnarray}
which are smaller than the $x^S_d\tan\beta$ terms in eqs. (\ref{XLR}),
(\ref{XRL}) but can be of the same order of magnitude as the $x^S_u$
ones.  The analysis of sec. \ref{sec:comp} shows that the above terms
are not numerically important.

\subsection{Charged Higgs-fermion vertices}

In the same approximation we can consider also the corrections to the
charged Higgs boson and charged Goldstone boson couplings.  In order
to find explicit expressions for the charged scalar ($H^\pm$) and
charged Goldstone boson ($G^\pm$) couplings defined in (\ref{LH}) and
(\ref{LG}), we combine the effective Lagrangians (\ref{eqn:effL_1})
and (\ref{eqn:effL_2}).  Performing the rotations
(\ref{eqn:grotations}) and (\ref{urotations}) we find
\begin{eqnarray}
&&{\cal L}^{\rm tot}_{\rm eff}=\overline{u_R} ~\mathbf{U}_R^\dagger
\left[H_1^{(u)}\left(y_u+\Delta_u\mathbf{Y}_u\cdot V^\dagger\right)
-H_2^{(d)*} \Delta_d\mathbf{Y}_u\cdot V^\dagger\right]
\mathbf{U}_L\cdot V^{\rm eff} ~d_L\nonumber\\
&&\phantom{aaaa}
+\overline{u_L} ~V^{\rm eff}\cdot \mathbf{D}_L^\dagger 
\left[-H_2^{(d)*}\left(y_d+(\Delta_d\mathbf{Y}_d)^\dagger\right)
-H_1^{(u)} (\Delta_u\mathbf{Y}_d)^\dagger\right]
\mathbf{D}_R ~d_R
\label{eqn:chcorr}
\end{eqnarray}
where we have used the definition (\ref{eqn:CKMeff}) of $V^{\rm eff}$.

Expressing $H_2^{(d)*}$ and $H_1^{(u)}$ in terms of $H^+$ and
$G^+$ as follows
\begin{eqnarray}
H^{(d)*}_2 = H^+\sin\beta - G^+\cos\beta, \qquad
H^{(u)}_1 = H^+\cos\beta + G^+\sin\beta
\label{PHYS}
\end{eqnarray}
we calculate, the couplings $\left[P^H_{RL}\right]^{JI}$,
$\left[P^H_{LR}\right]^{JI}$, $\left[P^G_{RL}\right]^{JI}$,
$\left[P^G_{LR}\right]^{JI}$ in (\ref{LH}) and (\ref{LG}) and the
parameters $\epsilon^{HL}_{JI}$, $\epsilon^{HR}_{JI}$,
$\epsilon^{GL}_{JI}$ and $\epsilon^{GR}_{JI}$ defined by eqs.
(\ref{PHIGGS}) and (\ref{PGOLD}).

For $G^+$ couplings, using (\ref{PHYS}) and comparing terms in square
brackets of eq.  (\ref{eqn:chcorr}) with eqs.  (\ref{eqn:dmasscorr}),
(\ref{eqn:dmdsymlim}), (\ref{eqn:umasscorr}), (\ref{eqn:dmusymlim}) we
find
\begin{eqnarray}
&&{\cal L}^{G^+}_{\rm eff}=
{\sqrt2\over v}G^+\overline{u_R}\mathbf{U}_R^\dagger
\left[m_u+\Delta m_u\right]
\mathbf{U}_L\cdot V^{\rm eff}d_L\nonumber\\
&&\phantom{aaaa}
-{\sqrt2\over v}G^+\overline{u_L}V^{\rm eff}\cdot 
\mathbf{D}_L^\dagger\left[m_d+(\Delta m_d)^\dagger\right]
\mathbf{D}_Rd_R.
\label{eqn:gcorr}
\end{eqnarray}
Subsequently using the relations (\ref{barmd}), (\ref{barmu}), we
easily find($\sqrt2/v=g_2/\sqrt2M_W$):
\be
{\cal L}^{G^+}_{\rm eff}={g_2\over\sqrt2M_W} G^+ 
\overline{u_R} ~\overline m_u V^{\rm eff} d_L-{g_2\over\sqrt2M_W} 
G^+\overline{u_L}V^{\rm eff}\overline m_d d_R ~.\label{eqn:SMGcoupl}
\ee
This compared with (\ref{PGOLD}) implies
\be\label{F2}
\left[P^G_{RL}\right]^{JI}= {g_2\over\sqrt2M_W}\overline m_{u_J} 
V^{\rm eff}_{JI}, \qquad \epsilon^{GL}_{JI}=0.  
\ee
\be\label{F3}
\left[P^G_{LR}\right]^{JI}=-{g_2\over\sqrt2M_W}V^{\rm eff}_{JI} ~
\overline m_{d_I},\qquad  \epsilon^{GR}_{JI}=0.
\ee

The vanishing of $\epsilon^{GL}_{JI}$ and $\epsilon^{GR}_{JI}$ is easy
to understand. Indeed, in the effective Lagrangian constructed before
the electroweak symmetry breaking and defined by eqs.
(\ref{eqn:effL_1}) and (\ref{eqn:effL_2}) the Higgs mechanism must
operate in the standard way (e.g. in the $R_\xi$ gauge the tree level
exchange of $G^+$ must cancel the gauge dependence of the amplitude
generated by the $W^+$ exchange) which requires that the $G^+$
couplings to quarks be precisely as in the SM with the mass parameters
corresponding to the tree-level quark masses in the effective theory
(that is ${\overline m}_{d(u)}$) and that the CKM matrix in the $G^+$
couplings is the same as the one in the $W^+$ couplings (that is
$V^{\rm eff}$).  The same argument shows that $\epsilon^{GL}_{JI}$ and
$\epsilon^{GR}_{JI}$ must vanish in the $SU(2)\times U(1)$ symmetry
limit also without the approximation of vanishing electroweak gauge
couplings.

For $H^+$ couplings, from (\ref{eqn:chcorr}) and (\ref{PHYS}) we find
\begin{eqnarray}
{\cal L}^{H^+}_{\rm eff}=
{\sqrt2\over v}H^+\overline{u_R} ~\mathbf{U}_R^\dagger
\left[\cot\beta\left(m_u+{v_u\over\sqrt2}\Delta_u\mathbf{Y}_u
\cdot V^\dagger\right)-\tan\beta~{v_d\over\sqrt2}\Delta_d\mathbf{Y}_u
\cdot V^\dagger\right]\mathbf{U}_L\cdot V^{\rm eff}d_L\nonumber\\
+{\sqrt2\over v}H^+\overline{u_L} ~V^{\rm eff}\cdot\mathbf{D}_L^\dagger 
\left[\tan\beta\left(m_d-{v_d\over\sqrt2}
(\Delta_d\mathbf{Y}_d)^\dagger\right)
-\cot\beta~{v_u\over\sqrt2}(\Delta_u\mathbf{Y}_d)^\dagger\right]
\mathbf{D}_R ~d_R\phantom{a}\nonumber\\
\label{eqn:chcor}
\end{eqnarray}
We observe that in the first line of eq.  (\ref{eqn:chcor})
$\Delta_d\mathbf{Y}_u$ cannot be neglected as it is multiplied by
$\tan\beta$.  On the other hand $\Delta_u\mathbf{Y}_u$ can be
neglected and $\mathbf{U}_{L,R}$ can be approximated by unit matrices.
Consequently we find for the $\bar u_RH^+d_L$ vertex
\begin{eqnarray}\label{PH}
\left[P^H_{RL}\right]^{JI}\approx{g_2\over\sqrt2}{m_{u_J}\over M_W}
\cos\beta\left[
V_{\rm eff}^{JI}-\tan\beta\left(\epsilon_0^\prime V_{\rm eff}^{JI}
+\epsilon_Y^\prime V^{JM}y^2_{d_M}V^{KM\ast}V_{\rm eff}^{KI}\right)\right]
\end{eqnarray}
Approximating next $m_{u_J}$ by $\overline m_{u_J}$ and expressing $V$
through $V^{\rm eff}$ by means of (\ref{eqn:CKMcorr}), we find
\begin{eqnarray}\label{F1}
\left[P^H_{RL}\right]^{JI}={g_2\over\sqrt2}{m_{\overline u_J}\over M_W}
\cos\beta V_{\rm eff}^{JI}\left[1-\tan\beta
\left(\epsilon_0^\prime+\epsilon_Y^\prime y^2_b \delta^{I3}\right)
-\Delta_{IJ}\right],
\phantom{a}
\end{eqnarray}
and consequently
\begin{eqnarray}
\epsilon^{HL}_{JI}=
\tan\beta\left(\epsilon_0^\prime+\epsilon_Y^\prime y^2_b \delta^{I3}\right)
+\Delta_{JI}~ \label{F1A}
\end{eqnarray}
where
\be\label{delta}
\Delta_{JI}=y^2_by^2_t \frac{\epsilon_Y\epsilon_Y^\prime \tan^2\beta}
{1+\epsilon_0\tan\beta}
\times\left\{ \begin{array}{ll}
+1 &{(J,I)=(1,3),(2,3)}\\
-1 & {(J,I)=(3,1),(3,2)} \\
 0 &{\rm otherwise}
\end{array} \right.
\ee 
This agrees with ref.  \cite{AMGIISST}, where the presence of
$\Delta_{JI}$ has been pointed out.  

In calculating the $\overline{u_L} ~H^+d_R$ vertex in
(\ref{eqn:chcor}) we observe that the effects of
$\Delta_d\mathbf{Y}_d$ are not $\tan\beta$ enhanced with respect to
the tree level coupling and can therefore be neglected.  The second
term involving $\Delta_u\mathbf{Y}_d$ is suppressed for large
$\tan\beta$ with respect to the first term and can also be neglected.
Consequently
\begin{eqnarray}
\left[P^H_{LR}\right]^{JI}\approx
{g_2\over\sqrt2M_W}\tan\beta 
\left[ V_{\rm eff}^{JI }m_{d_I} +V_{\rm eff}^{JK} 
\left(m_d\mathbf{\Delta D}_R-\mathbf{\Delta D}_Lm_d 
\right)^{KI}\right]\label{eqn:phlrsymlim}.
\end{eqnarray}

Using the formulae (\ref{eqn:DJI1}), (\ref{eqn:DJI2}),
(\ref{eqn:deltamasy}), (\ref{eqn:CKMcorr}) and retaining only terms
dominant for large $\tan\beta$ (and unsupressed by quark mass ratios
or small CKM matrix elements), we find
\be\label{PH1}
\left[P^H_{LR}\right]^{JI}= {g_2\over\sqrt2M_W}\tan\beta ~
V_{\rm eff}^{JI}{\overline m_{d_I}\over1+ \tilde\epsilon_J\tan\beta},
\qquad \epsilon^{HR}_{JI}=
\frac{\tilde\epsilon_J\tan\beta}{1+\tilde\epsilon_J\tan\beta}
\ee
for all $(J,I)$ with $\tilde\epsilon_J$ defined in (\ref{epJ}).

Inspecting (\ref{PH1}) we find that in the case of the vertices
involving $V^{\rm eff}_{ts}$ and $V^{\rm eff}_{td}$ the corrections
depend on the top Yukawa coupling $y_t^2$ while the corrections to
vertices $V^{\rm eff}_{cb}$ and $V^{\rm eff}_{ub}$ do not.

We would like to emphasize that whereas the rule (\ref{PH1}) for
$(J\not=3,I)$ and accidentally for $J=I=3$ is equivalent to expressing
in the tree level formulae $V_{JI}$ and $m_{d_I}$ through $V^{\rm
  eff}_{JI}$ and $\overline m_{d_I}$ respectively, for $J=3$ and
$I=1,2$ it is more involved.  Expressing in these cases only $V_{JI}$
and $m_{d_I}$ through $V^{\rm eff}_{JI}$ and $\overline m_{d_I}$,
would give $(1+\tilde\epsilon_3\tan\beta)/(1+\epsilon_0\tan\beta)^2$
as opposed to the correct result $1/(1+\tilde\epsilon_3\tan\beta)$.

\subsection{Chargino-Quark-Squark Vertices}
\label{subsec:cqvert1}

As argued in subsection 2.6, to account for rotations
(\ref{eqn:rotations}) performed on the quark fields in the higher
dimensional operators resulting {}from integrating out charginos,
their Wilson coefficients should be calculated by using the modified
chargino-quark-squark couplings.

Modifications of $c_L^i$ given in (\ref{RLiNew}) are trivial.  In the
$SU(2)\times U(1)$ symmetry limit the modifications (\ref{RRiNew}) of
$c_R^i$ can be easily given in closed form if we note that they are
exactly as in the expression (\ref{eqn:phlrsymlim}) for the couplings
$[P^H_{LR}]^{JI}$.  Thus, in $c^i_R$ in (\ref{RLi}) one should replace
as in (\ref{PH1})
\begin{eqnarray}\label{RRiN}
V^{JI} m_{d_I}\rightarrow  V_{\rm eff}^{JI } 
{\overline m_{d_I}\over1+ \tilde\epsilon_J\tan\beta}~.
\end{eqnarray}

\subsection{Introducing Flavour Dependence}
\label{sec:su2xu1flav}

Let us next generalize the main results of this section to the case of
flavour dependent $\epsilon_0$, $\epsilon_Y$, $\epsilon_0^\prime$ and
$\epsilon_Y^\prime$. In this case (\ref{eqn:DuYd_str}) and
(\ref{eqn:deltamasy}) are replaced by
\begin{eqnarray}
\left(\Delta_u\mathbf{Y}_d\right)^{JI}=-y_{d_J}
\left(\epsilon^{(J)}_0\delta^{JI}+
\epsilon^{(JI)}_Yy^2_tV^{3J\ast}V^{3I}\right)\equiv
-y_{d_J}\left(\tilde\epsilon_J\delta^{JI}+
\epsilon^{(JI)}_Yy^2_t\lambda_0^{JI}\right)
\label{eqn:fDuYd_str}
\end{eqnarray}
\begin{eqnarray}
&&\left(\Delta m_d\right)^{JI}=m_{d_J}\tan\beta\left(\tilde\epsilon_J
\delta^{JI}+\epsilon^{(JI)}_Yy^2_t\lambda_0^{JI}\right)
\label{eqn:fdmd_symlim}
\end{eqnarray}
where now
\be\label{fepJ}
\tilde\epsilon_J= \epsilon^{(J)}_0+\epsilon^{(33)}_Yy^2_t\delta^{J3}
\ee
and it is $\tilde\epsilon_J$ given in (\ref{fepJ}) that enters the
relation (\ref{basic}).

In agreement with the typical pattern of the soft SUSY breaking mass
parameters, in what follows we will assume that
\be\label{assume}
\epsilon^{(11)}_Y=\epsilon^{(12)}_Y=\epsilon^{(21)}_Y=\epsilon^{(22)}_Y,
\qquad \epsilon^{(31)}_Y=\epsilon^{(32)}_Y, \qquad
\epsilon^{(13)}_Y=\epsilon^{(23)}_Y~.
\ee
With this assumption, instead of the relation (\ref{eqn:CKMcorr}) for
$(JI)=(13)$, $(23)$, $(31)$ and $(32)$ we find:
\begin{eqnarray}
&&V_{JI}=V_{JI}^{\rm eff} ~
\left[{1+\tilde\epsilon_3\tan\beta\over1+\tilde\epsilon_0\tan\beta}\right]
\label{eqn:fCKMcorr}
\end{eqnarray}
with $\tilde\epsilon_J$ given in (\ref{fepJ}) and $\tilde\epsilon_0$ 
defined through
\be\label{tep0}
\tilde\epsilon_0=\epsilon^{(3)}_0+y^2_t(\epsilon^{(33)}_Y-\epsilon^{(31)}_Y)
=\tilde\epsilon_3-y_t^2\epsilon^{(31)}_Y
\ee
Note that $\tilde\epsilon_0$ differs from $\epsilon^{(J)}_0$ that 
enters (\ref{fepJ}).

Similarly we find for $J>I$
\begin{eqnarray}
&&\left[X^S_{RL}\right]^{JI}
={{\overline m}_{d_J}\over v_d(1+\tilde\epsilon_J\tan\beta)^2}
\epsilon^{(JI)}_Yy^2_t\lambda^{JI}_0\left(x_u^S-x^S_d\tan\beta\right)
\phantom{aaaa}\label{fXRL}
\end{eqnarray}
\begin{eqnarray}
\left[X^S_{LR}\right]^{JI}
={{\overline m}_{d_I}\over v_d(1+\tilde\epsilon_J\tan\beta)^2}
\epsilon^{(IJ)}_Yy^2_t\lambda^{JI}_0r_{JI}
\left(x_u^{S\ast}-x^{S\ast}_d\tan\beta\right)~.
\label{fXLR}
\end{eqnarray}
where
\be
r_{JI}={1+\left[\tilde\epsilon_J+(\tilde\epsilon_I-\tilde\epsilon_J)
(\epsilon^{(JI)}_Y/\epsilon^{(IJ)}_Y)\right]
\tan\beta\over(1+\tilde\epsilon_I\tan\beta)}~.
\ee
$\left[X^S_{RL}\right]^{JI}$ and $\left[X^S_{LR}\right]^{JI}$ for $J<I$ 
can be obtained by using the rules in (\ref{eqn:rules}). These improvements 
can be easily incorporated in the formulae (\ref{BXRLFIN})--(\ref{KXLRFIN}) 
taking into account that in expressing $\lambda_0^{JI}$ through 
$V^{JI}_{\rm eff}$, the formula (\ref{eqn:fCKMcorr}) instead of 
(\ref{eqn:CKMcorr}) should be used.

%

Next for the right-handed charged Higgs couplings we find
\be\label{PH3}
\phantom{aaaaaa}
\left[P^H_{LR}\right]^{JI}= {g_2\over\sqrt2M_W}\tan\beta ~
V_{\rm eff}^{JI}{\overline m_{d_I}\over1+\epsilon^{(I)}_0\tan\beta},
\phantom{aaa} J=1,2 \phantom{aaa} I=1,2
\ee

\be\label{PH4}
\left[P^H_{LR}\right]^{33}= {g_2\over\sqrt2M_W}\tan\beta ~
V_{\rm eff}^{33}{\overline m_{d_3}\over1+ \tilde\epsilon_3\tan\beta},
\phantom{aaaaaaaaaaaaaaa}
\ee

\be\label{PH5}
\left[P^H_{LR}\right]^{J3}= {g_2\over\sqrt2M_W}\tan\beta ~
V_{\rm eff}^{J3}{\overline m_{d_3}\over1+\tilde\epsilon_0\tan\beta},
\qquad
J=1,2\phantom{aaaaa}
\ee
with $\tilde\epsilon_0$ defined in (\ref{tep0}). Finally
\bea
\left[P^H_{LR}\right]^{3I}= {g_2\over\sqrt2M_W}\tan\beta ~
V_{\rm eff}^{JI } \overline m_{d_I}
\left[{1\over1+\epsilon^{(I)}_0\tan\beta}-
{\epsilon^{(3I)}_Y y_t^2\tan\beta\over
(1+\tilde\epsilon_0\tan\beta)
(1+ \tilde\epsilon_3\tan\beta)}\right.\phantom{aaa}\label{PH6}
\\
\left.+\frac{(\epsilon_Y^{(3I)}-\epsilon_Y^{(I3)}) y_t^2\tan\beta}
        {(1+\tilde\epsilon_0\tan\beta)(1+\epsilon^{(I)}_0\tan\beta)}
\right]\qquad I=1,2 \nonumber
\eea

Flavour dependence in the formulae (\ref{F1})-(\ref{delta}) for the
couplings $\left[P_{RL}^H\right]^{JI}$ will be introduced in sec.
\ref{sec:comp}.3.

\subsection{Summary}

In this section we have derived explicit expressions for the flavour
changing neutral scalar couplings $\left[X^S_{RL}\right]^{JI}$ and
$\left[X^S_{LR}\right]^{JI}$ and the charged scalar couplings
$\left[P^H_{RL}\right]^{JI}$, $\left[P^H_{LR}\right]^{JI}$,
$\left[P^G_{RL}\right]^{JI}$ and $\left[P^G_{LR}\right]^{JI}$ at large
$\tan\beta$ in the $SU(2)\times U(1)$ symmetry limit assuming the
dominance of $\alpha_s$ and the top and bottom Yukawa couplings. The
main results are given in (\ref{XRL}), (\ref{XLR}),
(\ref{BXRLFIN})-(\ref{KXLRFIN}), (\ref{F2}), (\ref{F3}), (\ref{F1})
and (\ref{PH1}). The important formula \cite{BAKO,ISRE1} that relates
$V_{JI}$ to $V^{\rm eff}_{JI}$ is given in (\ref{eqn:CKMcorr}). We
have also indicated how these formulae should be changed to account
for flavour violating effects in the case of split squark mass
parameters. This set of formulae allow to calculate scalar
contributions to any flavour violating process. In addition we have
presented the rules for incorporating large $\tan\beta$ effects in the
Wilson coefficients of the higher dimension operators generated in the
course of integrating out sparticles.


\section{Comparison with the Literature}
\label{sec:complit}
\setcounter{equation}{0}

The formulae presented in sections 2 and 3 generalize and in certain
cases correct those present in the literature.  Let us discuss this in
details.

The results in section 2 are new as they go beyond the limit of zero
electroweak gauge couplings and also beyond the $SU(2)\times U(1)$
symmetry limit.  We devote the next section to the detailed comparison
of the approximation of sec. 3 and the complete approach of sec. 2.
Our results of section 3 agree with the recent analysis in
\cite{AMGIISST} if one makes the following replacements in our
formulae
\be
\epsilon_0^\prime \to -\epsilon_0^\prime, \quad 
\epsilon_Y y_t^2 \to \epsilon_1, \quad 
\epsilon_Y^\prime y_b^2 \to -\epsilon_1^\prime 
\ee
with $\epsilon_0$ unchanged. We would like also to compare our results
with those of \cite{DEGAGI} in order to have a dictionary between the
two notations. This will prove helpful for discussing $\bar B\to
X_s\gamma$ decay.

The explicit expressions for $\left[P^{H(G)}_{RL}\right]^{JI}$ with
$J=3$, $I=1,2,3$ and for $\left[P^{H(G)}_{LR}\right]^{JI}$ with
$J=1,2,3$ and $I=3$ given in eq. (17) of ref.~\cite{DEGAGI} do not
take into account the modifications of the CKM factors summarized in
(\ref{eqn:CKMcorr}).  However, correcting the formula (17)
of~\cite{DEGAGI} has to be done with care. In particular, identifying
$V$ used in that paper with $V_{\rm eff}$ would generally give wrong
results for $\left[P^{H(G)}_{LR}\right]^{JI}$, while it is correct for
$\left[P^{H(G)}_{RL}\right]^{JI}$ if the term $\Delta_{JI}$ given in
eq. (\ref{delta}) is neglected.

In order to see this explicitly let us observe that the couplings
$\left[P^{H(G)}_{RL}\right]^{JI}$ involve left-handed down quarks that
enter also the $W^\pm$-quark couplings. As the corrections resulting
from rotating the $u_L$ and $u_R$ fields can be neglected, the CKM
matrix in $\left[P^{H(G)}_{RL}\right]^{JI}$ after the rotations
(\ref{eqn:rotations}) becomes automatically $V^{\rm eff}$ except for
$V$ in the last term in eq.  (\ref{PH}). Approximating $V$ in the
latter term with $V^{\rm eff}$ we find $\left[P^H_{RL}\right]^{JI}$ in
(\ref{F1}) with $\Delta_{JI}$ set to zero. This result, obtained in
the $SU(2)\times U(1)$ symmetry limit, agrees then with
$\left[P^H_{RL}\right]^{JI}$ in equation (17) of~\cite{DEGAGI}
provided the CKM matrix there is identified with $V^{\rm eff}$. We
find also the relation between $\epsilon^{HL}_{JI}$ introduced by us
and the parameters $\epsilon_J^\prime(I)$ in~\cite{DEGAGI}:
\be
\label{REL1}
\epsilon^{HL}_{JI}=\epsilon_J^\prime(I)\tan\beta\qquad\quad J=t,
\qquad I=d,s,b, \qquad (\Delta_{JI}\to 0) ~. 
\ee
However, as emphasized in ref. \cite{AMGIISST}, $\Delta_{JI}$ cannot
be generally neglected for $|\epsilon_Y\tan\beta|$ and
$|\epsilon_Y^\prime\tan\beta|$ larger than $0.5$ and it could be
important for $\epsilon_0^\prime\approx -\epsilon_Y^\prime$ when the
$\ord(\tan\beta)$ term in $\epsilon^{HL}_{JI}$ is small.

However, we will see in section \ref{sec:comp} that even with the
$\Delta_{JI}$ term included, $\epsilon^{HL}_{JI}$ obtained in the
approach of sec. \ref{sec:su2xu1} deviate rather significantly from
$\epsilon^{HL}_{JI}$ computed as in sec. 2. The replacement of
$\epsilon^\prime_0+\epsilon^\prime_Yy^2_b\delta^{I3}$ in eq.
(\ref{F1A}) obtained in the strict $SU(2)\times U(1)$ symmetry limit
and neglecting couplings other that $\alpha_s$ and Yukawas by
$\epsilon_J^\prime(I)$ given in eq. (16) of \cite{DEGAGI} does not
help much. To get a reliable approximation one has to include
additional terms that depend on the electroweak gauge couplings which
are numerically important but have been neglected in all previous
analyses. In contrast to $\left[P^H_{RL}\right]^{JI}$, the coupling
$\left[P^G_{RL}\right]^{JI}$ of the charged Goldstone boson as used in
ref. \cite{DEGAGI} agrees with our if $V$ of \cite{DEGAGI} is
interpreted as $V_{\rm eff}$. As we will see, $\epsilon^{GL}_{JI}$
computed in the approach of sec. 2 are indeed negligible.

In the case of $\left[P^{H(G)}_{LR}\right]^{JI}$ the right-handed down
quarks are rotated differently from the left-handed down quarks
present in the $W^\pm$-quark couplings and this mismatch results in
additional $\tan\beta$ enhanced corrections.  Interpreting $V$ in
$\left[P^{H(G)}_{LR}\right]^{JI}$ in equation (17) in~\cite{DEGAGI} as
$V^{\rm eff}$, would generally miss them.  However, it turns out that
for $J=1,2,3$ and $I=3$ considered in~\cite{DEGAGI}, the results
correct in the $SU(2)\times U(1)$ symmetry limit can be obtained by
identifying $V$ in $\left[P^{H(G)}_{LR}\right]^{JI}$ of this paper not
with $V_{\rm eff}$ as in the case of $\left[P^{H(G)}_{RL}\right]^{JI}$
but with the original CKM matrix $V$.  Indeed, expressing the $V$
elements in $\left[P^{H(G)}_{LR}\right]^{JI}$ of~\cite{DEGAGI} in
terms of $V^{\rm eff}$ by using the rules in (\ref{eqn:CKMcorr}), the
formulae for these couplings given in~\cite{DEGAGI} can be transformed
into ours provided the following replacements are simultaneously made:
\be\label{REL2}
\frac{V^{JI}}{V_{\rm eff}^{JI}}\frac{1}{1+\epsilon_I\tan\beta}
\rightarrow 1-\epsilon^{HR}_{JI}\qquad\quad J=u,c,t, \qquad I=b
\ee
in the case of $\left[P^{H}_{LR}\right]^{JI}$ and
\be\label{REL3}
\frac{V^{JI}}{V_{\rm eff}^{JI}}
\frac{1+\epsilon^\prime_I(J)\tan\beta}{1+\epsilon_I\tan\beta}
\rightarrow 1+\epsilon^{GR}_{JI} \qquad J=u,c,t \quad I=b
\ee
in the case of $\left[P^{H(G)}_{LR}\right]^{JI}$ (note that $\epsilon_I$ 
of~\cite{DEGAGI} corresponds to $\tilde\epsilon_I$ in our paper).
As in the $SU(2)\times U(1)$ symmetry limit
\be
\epsilon_b^\prime(t)=\epsilon_b=\tilde\epsilon_3, \qquad 
\epsilon_b^\prime(c)=\epsilon_b^\prime(u)=\epsilon_0, 
\ee
with $\tilde\epsilon_3$ and $\epsilon_0$ defined in (\ref{epJ}), the
formulae given above reduce indeed to the corresponding formulae for the 
parameters $\epsilon_{JI}^{H(G)R}$ given in (\ref{F3}) and (\ref{PH1}). 
It should be emphasized that the factor $V^{JI}/V^{JI}_{\rm eff}$ in 
(\ref{REL3}) that we introduced here is essential for vanishing of 
$\epsilon^{GR}_{JI}$ in the $SU(2)\times U(1)$ symmetry limit as required 
by general arguments. We will see, however, in the next section that for 
split squark masses the formula (\ref{PH1}) for $\epsilon^{HR}_{JI}$ has 
to be modified along the lines of section 3.6 in order to approximate well 
the results of the complete calculation. We shall also explain in more 
detail the status of the recipe (\ref{REL3}).


\section{The Size of the Large \boldmath{$\tan\beta$} Corrections: 
Full Calculation and the \boldmath{$SU(2)\times U(1)$} Limit}
\label{sec:comp}
\setcounter{equation}{0}

This section is devoted to the numerical evaluation and the assessment
of the importance of the corrections discussed in sections 2 and 3.
We will also compare results of the complete diagrammatic calculation
of these corrections with the approximation based on the $SU(2)\times
U(1)$ symmetry limit discussed in section 3. This will allow to test
the quality of the latter approach and to find its limitations.  We
will also present analytic expressions for various couplings that
approximate the full approach much better than the formulae of section
3.

\subsection{Full Calculation}

The one-loop contributions of sparticles to quark self-energies
$\Sigma_{mL(R)}^q(0)$, $\Sigma_{VL(R)}^q(0)$ and various vertex
formfactors $\Delta F_{L(R)}$ at vanishing external momenta, that are
the main ingredients of the full method of section 2, can be easily
calculated by using the Feynman rules of the MSSM collected in
\cite{ROS} and the standard one-loop functions. We recall the relevant
expressions in the Appendix A.1.

In order to find the original mass parameters $m_{d_J}$ and the
original CKM matrix $V^{JI}$ that enter these rules we proceed as
follows.  We begin by computing $\epsilon_{d_J}$ defined in
(\ref{eqn:eps_def}). Since to a good approximation
$(\Sigma^d_{mL})^{JJ}\propto m_{d_J}$ the quark mass cancels out in
the ratio in (\ref{eqn:eps_def}). Therefore, in the actual calculation
of $\epsilon_{d_J}$ the MSSM mass parameters $m_{d_J}$ can be replaced
by ${\overline m}_{d_J}$ that are known experimentally. Note also,
that to compute diagonal self energies one can use experimentally
known $V^{\rm eff}$ instead of the yet unknown $V$. Having obtained
$\epsilon_{d_J}$ in this manner we can calculate
$m_{d_J}\approx{\overline m}_{d_J}/(1+\epsilon_{d_J}\tan\beta)$.

In order to find $V$ we proceed iteratively: starting with $m_{d_J}$
and $V=V^{\rm eff}$ we compute $\Delta^\prime m_d$ by means of
(\ref{eqn:dmd}) and (\ref{eqn:dmdprime}) and subsequently using
(\ref{eqn:DLsol}) and (\ref{eqn:DRsol}) we find the rotations
$\mathbf{D}_{L,R}=\mathbf{1+\Delta D}_{L,R}$.  This allows to find $V$
in the first approximation by inverting the relation
(\ref{eqn:CKMeff}) in which $\mathbf{\Delta U}_L$ can be set to zero.
This is next used to compute $\Delta^\prime m_d$ anew and so on. The
procedure converges quickly yielding $V$ and the rotations
$\mathbf{D}_{L,R}$.  Having these, we can compute all the necessary
formfactors $\Delta F_{L,R}$ and the couplings $X_{RL}$, $X_{LR}$,
$P_{RL}$ and $P_{LR}$.

The simple and transparent expressions derived in secs. 3.2-3.5 are
very convenient, as they make the effects of large $\tan\beta$
explicit and allow for easy qualitative discussion of the dominant
effects. However, as the comparison with the complete calculation will
reveal, they are not very accurate numerically and it is desirable to
improve the accuracy of the approach of section 3 retaining, however,
its simplicity. It turns out that for the CKM matrix and the couplings
$[X_{RL}]$, $[X_{LR}]$, and $[P^H_{LR}]$ this can be achieved by
simply using the expressions (\ref{eqn:fCKMcorr})-(\ref{PH6}) with
$\tilde\epsilon_J$ and $\epsilon_Y^{(JI)}$ calculated directly from
the full one loop correction $\Delta m_d$ to the down-quark mass
matrix given in (\ref{eqn:dmd}):
\begin{eqnarray}
y^2_t\epsilon_Y^{(JI)} = 
{(\Delta m_d)^{JI}\over m_{d_J}\lambda_0^{JI}\tan\beta}
\phantom{aa}{\rm for}\phantom{aa}J\neq I, 
\phantom{aaa}
\tilde\epsilon_J ={(\Delta m_d)^{JJ}\over m_{d_J}\tan\beta}~.
\label{eqn:epsfromdmd}
\end{eqnarray}
The complete formula for $(\Delta m_d)^{JI}$ is given in (\ref{eqn:dmd})
with the relevant expressions for quark self-energies given in the Appendix 
A.2. Note that with $\epsilon_Y^{(JI)}$ and $\tilde\epsilon_J$ all parameters 
entering (\ref{eqn:fCKMcorr})-(\ref{PH6}), in particular $\tilde\epsilon_0$ 
in (\ref{tep0}), can be calculated. In this manner also the effects of the 
electroweak couplings $g_1$ and $g_2$ present in $(\Delta m_d)^{JI}$ are 
automatically taken into account. Simple but accurate expressions for the 
couplings $[P^H_{RL}]$ can also be given. They will be discussed separately 
below.

\subsection{The neutral Higgs boson sector}

In order to implement the formulae of section 3 we need the quantities
$\epsilon_0$, $\epsilon_Y$ defined by eq. (\ref{eqn:DuYd_str}) and
$\epsilon^\prime_0$, $\epsilon^\prime_Y$ of eq. (\ref{eqn:DdYu_str}).
Evaluating diagrams shown in fig. \ref{fig:bcrs1} we find
\begin{eqnarray}
\epsilon_0=-{2\alpha_s\over3\pi}{\mu\over m_{\tilde g}}
H_2\left(x^{Q/g},x^{D/g}\right), \phantom{aaaa}
\epsilon_Y={1\over16\pi^2}{A_t\over\mu}~
H_2\left(x^{Q/\mu}_l,x^{U/\mu}_l\right)\label{eqn:e}\\
\epsilon^\prime_0=-{2\alpha_s\over3\pi}{\mu\over m_{\tilde g}}
H_2\left(x^{Q/g},x^{U/g}\right), \phantom{aaaa}
\epsilon^\prime_Y={1\over16\pi^2}
{A_b\over\mu}~ H_2\left(x^{Q/\mu}_l,x^{D/\mu}_l\right)
\label{eqn:eprime}
\end{eqnarray}
where $x^{Q/g}\equiv m^2_Q/m^2_{\tilde g}$, $x^{D/g}\equiv
m^2_D/m^2_{\tilde g}$, $x^{Q/\mu}_l\equiv m^2_Q/|\mu|^2$ etc., and
$m^2_Q$, $m^2_D$, $m^2_U$, $A_t$, and $A_b$ are the parameters of the
soft supersymmetry breaking in the MSSM Lagrangian.\footnote{Our
  convention for $A_u$ and $A_d$ parameters is fixed by the form of
  the left-right squark mixing terms in the squark mass squared
  matrices which read $-m_u(A_u+\mu\cot\beta)$ and
  $-m_d(A_d+\mu\tan\beta)$ for the up and down squarks, respectively.}
The function $H_2(x,y)$ is defined as
\begin{eqnarray}
H_2(x,y) = {x\ln x\over(1-x)(x-y)}+{y\ln y\over(1-y)(y-x)}~.
\label{eqn:hfun}
\end{eqnarray}
As discussed in sec. 3, for squark mass parameters $m^2_Q$, $m^2_D$,
$m^2_U$ having different values for different generations (e.g. if the
the third generation squarks are split from the others as happens in
the minimal SUGRA scenario) $\epsilon_0$, $\epsilon_Y$,
$\epsilon_0^\prime$ and $\epsilon_Y^\prime$ depend on the generation
indices. We have given the relevant expressions in subsection
\ref{sec:su2xu1flav}. In our comparison of the approximate and full
methods, we parametrize the squark spectrum with four independent mass
parameters: $m^2\equiv m^2_{Q_{1,2}}=m^2_{D_{1,2}}=m^2_{U_{1,2}}$ for
the squarks of the first two generations and $m^2_{Q_3}$, $m^2_{D_3}$,
$m^2_{U_3}$ for the third one. Consequently, in eqs.
(\ref{BXRLFIN})-(\ref{KXLRFIN}) $\tilde\epsilon_3$ always depends on
the third generation squark masses but as in sec. \ref{sec:su2xu1flav}
we distinguish $\epsilon_0$ arising from expressing $V_{JI}$ in terms
of $V_{JI}^{\rm eff}$, which also depends on the third generation squark 
masses, and $\epsilon_0$ that result from expressing $m_{d_{1,2}}$ in 
terms of $\overline{m}_{d_{1,2}}$ which depends on $m^2$.


The formulae (\ref{eqn:e}) for $\epsilon_0$ and $\epsilon_Y$ apply in
the strict limit of $SU(2)\times U(1)$ symmetry supplemented by the
approximation $g_1=g_2=0$. Under these conditions only the higgsino of
mass $|\mu|$ couples to quarks through Yukawa couplings. In
application to the relation between $m_{d_J}$ and $\overline{m}_{d_J}$
and to the neutral Higgs boson couplings it is easy to take into
account the contribution of the true charginos. Simplifying appropriately
the full expression for $(\Sigma_{mL}^d)^{JJ}$ (or the full expression
for the one loop vertex diagram in which $H^0$ or $h^0$ couples to the
up-type squarks and the charginos connect the external $d$-quarks) and
comparing with the $\epsilon_Y$ contribution to $\Delta m_d$ (to
$\Delta F^{dS}_L$) one gets
\begin{eqnarray}
\epsilon_Y={1\over16\pi^2}\sum_{l=1}^2Z_-^{2l}{A_t\over m_{C_l}}
Z_+^{2l} ~H_2\left(x^{Q/C}_l,x^{U/C}_l\right)\label{eqn:eY}
\end{eqnarray}
where $x^{Q/C}_l\equiv m^2_Q/m^2_{C_l}$ etc., and the matrices $Z_+$
and $Z_-$ are defined in ref.~\cite{ROS}. Similarly one could take
into account also the left-right mixing of squarks. The formula
(\ref{eqn:eY}) goes beyond the $SU(2)\times U(1)$ symmetry limit in
that it takes into account the mixing of higgsinos and gauginos which
depends on vacuum expectation values and on the gauge couplings $g_1$
and $g_2$ and can be suitably extended to include also the mixing of
left and right-handed squarks proportional to VEVs. However, it still
neglects $g_1$ and $g_2$ in the couplings of charginos to quarks and
the contribution of the neutralinos.

\begin{figure}[htbp]
\begin{center}
\epsfig{file=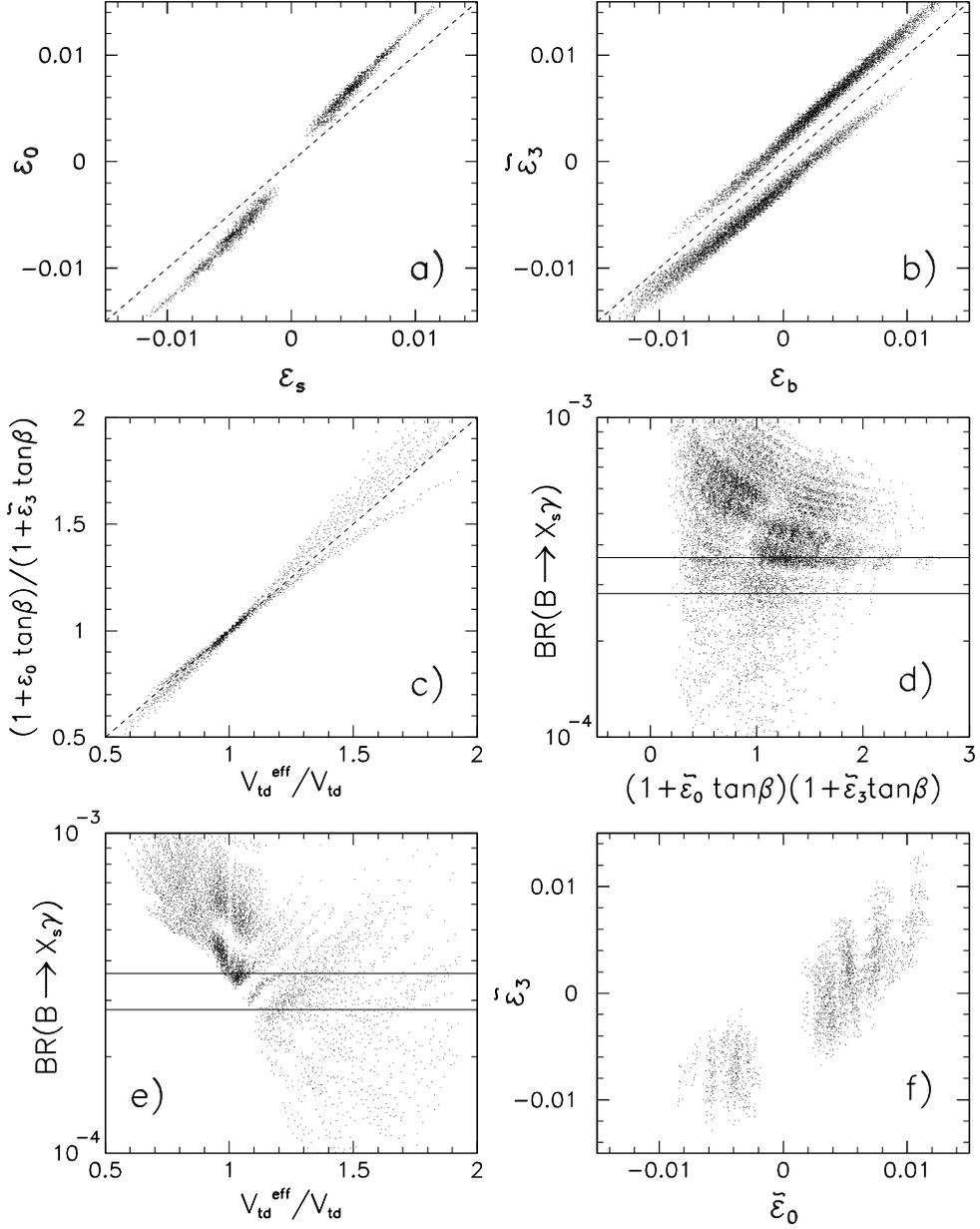,width=\linewidth}
\end{center}
\caption{Comparison of the results obtained in the complete 
  calculation with the ones obtained in the approximation based on the
  $SU(2)\times U(1)$ symmetry limit for a sample of the sparticle mass
  spectra for $\tan\beta=50$ and $M_A=200$ GeV. See the text for
  further explanations.}
\label{fig:eps1}
\end{figure}

The magnitude of $\epsilon_0$ and $\epsilon_Y$ can be easily estimated
from eq. (\ref{eqn:e}) remembering that $H_2(1,1)=-1/2$ and that
$H_2(x,x)$ varies between $-0.18$ for $x=4$ and $-1.7$ for $x=0.1$.
Figures \ref{fig:eps1}a,b show the comparison of
$\epsilon_{d_2}\approx\epsilon_{d_1}$ and of $\epsilon_{d_3}$ computed
from the full one-loop self energies (eq. (\ref{eqn:eps_def})) with
$\epsilon_0$ and $\tilde\epsilon_3$, respectively, for a sample of
points in the MSSM parameter space corresponding to sparticles heavier
than 500 GeV. As seen in figs.  \ref{fig:eps1}a and \ref{fig:eps1}b,
typically $|\epsilon_{d_3}|$ and $|\epsilon_{d_2}|$ ($|\epsilon_0|$
and $|\tilde\epsilon_3|$) are of order $\sim5\times10^{-3}$ and reach
$\sim1\times10^{-2}$ only for very special combinations of squark and
gluino masses, $|\mu|$ and/or $|A_t|$. 

The two distinct bands seen in panels a) and b) of the figure
\ref{fig:eps1} correspond to $\mu<0$ ($\epsilon_0<\epsilon_{d_2}<0$
and $\tilde\epsilon_3<\epsilon_{d_3}$) and $\mu>0$
($\epsilon_0>\epsilon_{d_2}>0$ and $\tilde\epsilon_3>\epsilon_{d_3}$).
The discrepancy between $\epsilon_0$ ($\tilde\epsilon_3$) and
$\epsilon_{d_2}$ ($\epsilon_{d_3}$) (quantified by the deviation of
the bands from the diagonal) is partly due to the absence in
$\epsilon_0$ given by (\ref{eqn:e}) of the neutralino contribution to
the d-quark self energies but mostly to the fact that in the full
approach the chargino couplings depend also on the gauge coupling
constants so that charginos do contribute also to $\epsilon_0$ and
hence also to $\epsilon_{d_1}$ and $\epsilon_{d_2}$. Neglecting the
contribution of $\Delta_d\mathbf{Y}_d$ in the formula
(\ref{eqn:dmdsymlim}) corresponds in the full approach to neglecting
the contribution of $A_b$ (of $\mu$) to the mixing of the left and
right sbottoms (stops) and is justified for $|A_b|\ll|\mu|\tan\beta$
and $|\mu|\ll A_t\tan\beta$. More detailed comparison shows that the
differences between $\epsilon_0$ ($\tilde\epsilon_3$) and
$\epsilon_{d_2}$ ($\epsilon_{d_3}$) are of order 15\%-20\% when the
epsilons assume phenomenologically relevant values.

The gap in values of $\epsilon_{d_2}$ ($\epsilon_0$) seen in panel a)
of fig. \ref{fig:eps1} reflects the existence of a lower bound on $\mu
m_{\tilde g}$ in the scan which follows from the imposed condition
$m_{C_1}, m_{\tilde g}>500$ GeV.  Note also, that while $\epsilon_0$
and $\epsilon_Y$ as given by (\ref{eqn:e}) or (\ref{eqn:eY}) are UV
finite, $\epsilon_{d_I}$ have to be renormalized in the $\overline{\rm
  MS}$ scheme. We have checked however, that the variation of
$\epsilon_{d_I}$ with the renormalization scale $Q$ is very small for
500 GeV$<Q<m_{\tilde g}$.

Figure \ref{fig:eps1}c shows the comparison of\footnote{We have
  checked that the full approach based on eq. (\ref{eqn:CKMeff}) also
  gives $V^{\rm eff}_{td}/V_{td}\approx V^{\rm eff}_{ts}/V_{ts}\approx
  V^{\rm eff}_{ub}/V_{ub}\approx V^{\rm eff}_{cb}/V_{cb}$.}  $V^{\rm
  eff}_{td}/V_{td}$ obtained from the complete expression
(\ref{eqn:CKMeff}) and in the approximation (\ref{eqn:CKMcorr}) for a
sample of sparticle parameters and $\tan\beta=50$. It is to be noted
that the approximation (\ref{eqn:CKMcorr}) with $\epsilon_0$ and
$\epsilon_Y$ computed as in eqs. (\ref{eqn:e}) and (\ref{eqn:eY}),
works only qualitatively.

Figures \ref{fig:eps1}a - \ref{fig:eps1}c show that in general the effects 
of the electroweak couplings are non-negligible. They could be included in 
the framework of the $SU(2)\times U(1)$ symmetry limit by computing 
diagrams with gaugino exchange but since the full expression for the 
supersymmetric contribution to the down quark self energy are simple enough, 
it is more practical to use the formulation of section \ref{sec:su2xu1flav} 
with $\epsilon_Y^{(JI)}$ and $\tilde\epsilon_J$ computed as in 
(\ref{eqn:epsfromdmd}). The formula (\ref{eqn:fCKMcorr}) improved in this 
way gives perfect approximation to the ratios $V^{\rm eff}_{td}/V_{td}$ 
etc. obtained in the approach of sec. 2.  Obviously, $\tilde\epsilon_J$ 
given by (\ref{eqn:epsfromdmd}) coincide with the factors $\epsilon_{d_J}$ 
of sec.~2.

The panel d) of figure \ref{fig:eps1} shows for $M_A=200$ GeV  
and $\tan\beta=50$ the correlation\footnote{For details of the 
  $BR(\bar B\rightarrow X_s\gamma)$ calculation see sec. 6.} 
of $BR(\bar B\rightarrow X_s\gamma)$ with the factor 
$(1+\tilde\epsilon_0\tan\beta)(1+\tilde\epsilon_3\tan\beta)$ 
that enters the denominator of the flavour violating neutral Higgs boson 
couplings (\ref{BXRLFIN}) and (\ref{BXLRFIN}) improved as in sec.
\ref{sec:su2xu1flav}. We have used here the eq. (\ref{eqn:epsfromdmd})
to determine $\tilde\epsilon_0$ and $\tilde\epsilon_3$. We observe
that the experimentally acceptable value of $BR(\bar B\rightarrow
X_s\gamma)$ can be realized both for
$(1+\tilde\epsilon_0\tan\beta)(1+\tilde\epsilon_3\tan\beta)$ greater
than one and smaller than one corresponding respectively to the
suppression and the enhancement of the actual scalar FCNC compared to
the naive one loop estimates of ref.~\cite{CHSL,BOEWKRUR,BUCHROSL1}.

In contrast, it is interesting to observe in the panel e) of figure
\ref{fig:eps1} that the correlation between $V^{\rm eff}_{td}/V_{td}$
and $\bar B\rightarrow X_s\gamma$ has quite different implications.
Here the requirement of acceptable rate of the $\bar B\rightarrow
X_s\gamma$ decay eliminates (for $M_A=200$ GeV and $\tan\beta=50$) all
points with $V^{\rm eff}_{td}/V_{td}<1$.

Both facts can be qualitatively understood by investigating signs of
$\tilde\epsilon_0$ and $\epsilon^{(31)}_Y$ and remembering that, as
follows from eqs. (\ref{eqn:e}),
$\tilde\epsilon_0\approx\epsilon_0\propto\mu$ and
$\epsilon_Y^{(31)}\approx\epsilon_Y\propto-\mu A_t$ (the function
$H_2$ is negative). Let us consider the case $\tilde\epsilon_0<0$,
$\tilde\epsilon_3>0$ which according to eq. (\ref{eqn:fCKMcorr}) leads
to $V^{\rm eff}_{td}/V_{td}<1$. This requires $\mu<0$ and $A_t>0$.
However, the relative sign of the chargino-stop and $H^+$
contributions to the amplitude of the $\bar B\rightarrow X_s\gamma$
decay is opposite to the sign of $\mu A_t$. Therefore, $\mu A_t>0$ is
a necessary condition for the cancellation to occur \cite{CAGANIWA2}.
Hence, for $\tilde\epsilon_0<0$, $\tilde\epsilon_3>0$ no cancellation
between the chargino-stop and $H^+$ contributions is possible which
explains why $V^{\rm eff}_{td}/V_{td}<1$ is incompatible with the
$BR(\bar B\rightarrow X_s\gamma)$ constraint.

On the other hand, for
$(1+\tilde\epsilon_0\tan\beta)(1+\tilde\epsilon_3\tan\beta)<1$
(enhancement of the flavour changing scalar couplings) that holds
always for $\mu<0$ such a cancellation is possible for $A_t<0$. This
is further illustrated in figure \ref{fig:eps1}f which shows the
correlation of $\tilde\epsilon_0$ and $\tilde\epsilon_3$ for those
sparticle parameters which for $\tan\beta=50$ and $M_A=200$ GeV give
$2.81\times10^{-4}<BR(\bar B\rightarrow X_s\gamma)<3.65\times10^{-4}$.
It should be stressed that the frequently used argument~\cite{ISRE1}
that obtaining acceptable $\bar B\rightarrow X_s\gamma$ rate requires
$\mu>0$, which implies the suppression of the scalar FCNC, applies
only in particular SUSY breaking scenarios. For example, in the
minimal SUGRA, due to the fixed point structure of the relevant RGEs,
not too large values of $A_0$ (common value of $A_t$, $A_b$ etc. at
the high scale) always lead to a fixed sign (positive in our
convention) of $A_t$. However, fixed sign of $A_t$ needs not be a
common feature of all SUSY breaking scenarios and can be reversed even
in the minimal SUGRA if the initial value $|A_0|$ is sufficiently
large (see e.g. ref.~\cite{CHELOLPO}). It is therefore justified to
study the phenomenological consequences of both possibilities:
$(1+\tilde\epsilon_0\tan\beta)(1+\tilde\epsilon_3\tan\beta)$ greater 
than one and smaller than one.

Finally, let us stress that in the case of the enhancement of the
scalar flavour changing couplings the value of
$(1+\epsilon_0\tan\beta)(1+\tilde\epsilon_3\tan\beta)$ with
$\epsilon_0$ and $\tilde\epsilon_3$ calculated from eqs.
(\ref{eqn:e}), (\ref{eqn:eY}) is smaller than
$(1+\tilde\epsilon_0\tan\beta)(1+\tilde\epsilon_3\tan\beta)$ with
$\tilde\epsilon_0$ and $\tilde\epsilon_3$ calculated using
(\ref{eqn:epsfromdmd}) by 10\%$-$35\% which means that the naive
approximation gives too big an enhancement. Similarly, in the case of
the suppression of the scalar flavour changing couplings the value of
$(1+\epsilon_0\tan\beta)(1+\tilde\epsilon_3\tan\beta)$ obtained by
means of (\ref{eqn:e}) and (\ref{eqn:eY}) is larger by 10\%$-$25\%
than the corresponding factor calculated using (\ref{eqn:epsfromdmd}).
Thus the approximation of sec 3. overestimates the effects of resummed
large $\tan\beta$ corrections in the neutral Higgs boson couplings.
 
\begin{figure}[htbp]
\begin{center}
\epsfig{file=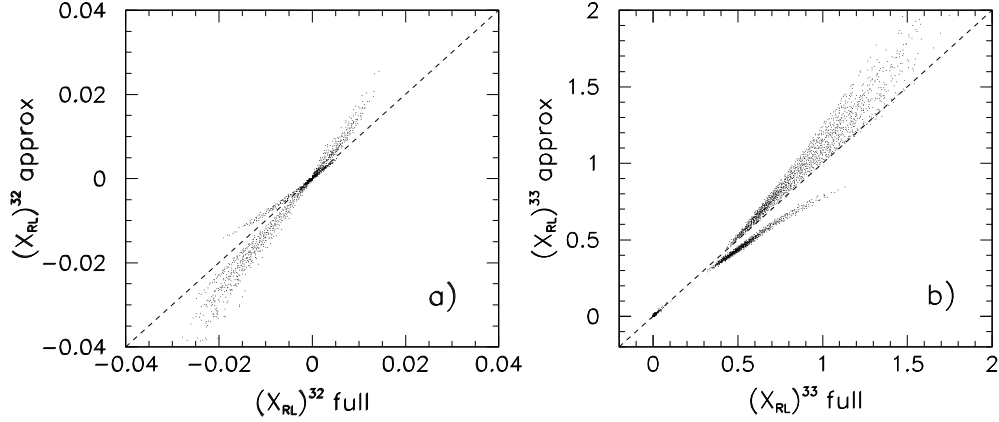,width=\linewidth}
\end{center}
\caption{\protect As in figure \ref{fig:eps1} but for the 
couplings $[X_{RL}]^{JI}$ of $H^0$ to the down quarks.}
\label{fig:eps2}
\end{figure}

This is reflected in figure \ref{fig:eps2}a where we show the
comparison of the couplings $\left[X_{RL}^H\right]^{32}$ of $H^0$ to
the down quarks obtained from the complete calculation and from the
approximate formulae (\ref{BXRLFIN}) with $\epsilon_0$ and
$\epsilon_Y$ computed by means of eqs. (\ref{eqn:e}), (\ref{eqn:eY}).
In this figure positive (negative) values of
$\left[X_{RL}^H\right]^{32}$ correspond to negative (positive) value
of the product $\mu A_t$. Clearly visible shorter band corresponds to
positive $\mu$ (suppression of the flavour changing scalar couplings
by the factor $(1+\epsilon_0\tan\beta)(1+\tilde\epsilon_3\tan\beta)$)
and the longer one to $\mu<0$ (enhancement). For points for which the
coupling $\left[X_{RL}^H\right]^{32}$ has a non-negligible value the
difference between the full and approximate calculation varies between
10\%$-$25\% for $\mu>0$ (suppression) and 10\%$-$40\% for $\mu<0$
(enhancement). Other flavour changing couplings:
$\left[X_{RL}^H\right]^{31}$, $\left[X_{RL}^H\right]^{23}$ etc.
exhibit similar behaviour. For completness, in figure \ref{fig:eps2}b
we show the comparison of the diagonal couplings
$\left[X_{RL}^H\right]^{33}$ of $H^0$ obtained from the complete
calculation and from the approximate formulae (\ref{DIAGF}).

Similarly as in the case of the ratio $V^{\rm eff}_{td}/V_{td}$, the
quality of the approximation of section 3 seen in figure \ref{fig:eps2} 
can be greatly improved by using the formulae (\ref{DIAGF}) and 
(\ref{BXRLFIN}), (\ref{BXLRFIN}) improved according to the rules of
sec. \ref{sec:su2xu1flav} with $\tilde\epsilon_0$ and
$\tilde\epsilon_J$ and $\epsilon^{(JI)}_Y$ computed as in eqs.
(\ref{eqn:epsfromdmd}). Since the $\tan\beta$ enhanced parts of the
flavour changing couplings are proportional to $A_t$, such an
approximation is not good only if the parameter $A_t$ is small but
then also the couplings
$\left[X_{RL}^S\right]^{JI}$, $\left[X_{LR}^S\right]^{JI}$ for $J\neq
I$ become phenomenologically uninteresting.

For (almost) degenerate squarks of the first two generations the
approximation of subsection 3.6 for couplings
$\left[X_{RL}^S\right]^{JI}$ and $\left[X_{LR}^S\right]^{IJ}$ where
$I=3$ and/or $J=3$ with $\tilde\epsilon_0$ and $\tilde\epsilon_J$ and
$\epsilon^{(JI)}_Y$ given by eqs. (\ref{eqn:epsfromdmd}) could break
down also for unrealistically large differences $m^2-m^2_{Q_3}$. This
is because, as we have explained in section 3.1, we are working in the
scenario in which the left-left block of the up-type squark mass
squared matrix is $Vm^2_QV^\dagger$ and is therefore not diagonal in
the generation space as we allow for $m^2_{Q_3}\neq m^2\equiv
m^2_{Q_1}=m^2_{Q_2}$. However, due to the special structure of the CKM
matrix this flavour violation remains negligible as long as
$|(m^2_{Q_3}-m^2)V_{tb}V_{cb}^\ast|$ remains small compared to
$m^2_{Q_3}$. If this condition is not satisfied then, the couplings of
stops to charginos and down quarks depart from the simple structure
(\ref{RLi}) and in consequence $(\Delta^\prime m_d)^{JI}$ is no longer
proportional to $\lambda^{JI}_0$. Since in our scans
$|(m^2_{Q_3}-m^2)V_{tb}V_{cb}^\ast|\ll m^2_{Q_3}$ the effects of this
additional flavour violation never manifest themselves. The main
effects of flavour violation arise then from flavour dependence of
$\epsilon_0$ and $\epsilon_Y$ and when the approximate formulae
(\ref{BXRLFIN})-(\ref{KXLRFIN}) are modified according to the 
prescriptions of section \ref{sec:su2xu1flav} with epsilon parameters 
calculated by means of (\ref{eqn:epsfromdmd}) they always agree very well 
with the results of the full calculation.

\subsection{Charged Higgs Couplings}

In this subsection we will asses the magnitude of the factors
$\epsilon_{JI}^{HL}$, $\epsilon_{JI}^{HR}$ and $\epsilon_{JI}^{GL}$,
$\epsilon_{JI}^{GR}$ introduced in subsection 2.5 to parametrize the
corrections to the charged Higgs and Goldstone boson couplings 
and test the quality of the approximations developed in subsection
3.3.

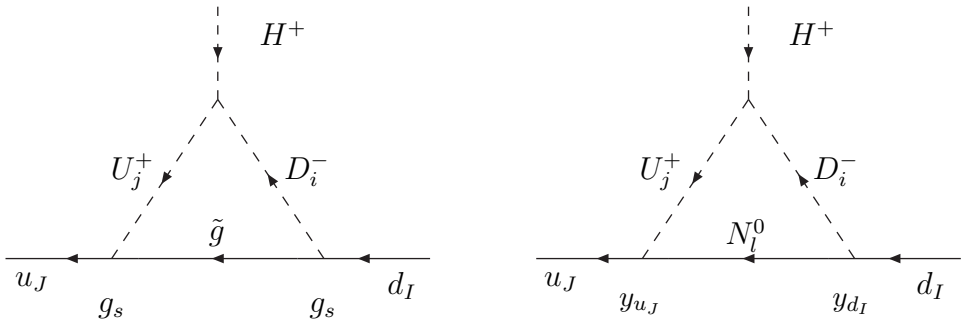
\begin{figure}[htbp] 
\begin{center}
\begin{picture}(400,160)(0,0)
\ArrowLine(120,40)(60,40)
\ArrowLine(60,40)(10,40)
\ArrowLine(170,40)(120,40)
\Text(20,35)[t]{$u_J$}
\Text(160,35)[t]{$d_I$}
\Text(50,25)[t]{$g_s$}
\Text(130,25)[t]{$g_s$}
\Text(90,55)[t]{$\tilde{g}$}
\Text(58,80)[t]{$U^+_j$}
\Text(124,80)[t]{$D^-_i$}
\Text(115,130)[t]{$H^+$}
\DashArrowLine(90,100)(50,40){4}
\DashArrowLine(130,40)(90,100){4}
\DashArrowLine(90,135)(90,100){4}
\ArrowLine(320,40)(260,40)
\ArrowLine(260,40)(210,40)
\ArrowLine(370,40)(320,40)
\Text(220,35)[t]{$u_J$}
\Text(360,35)[t]{$d_I$}
\Text(290,55)[t]{$N^0_l$}
\Text(258,80)[t]{$U^+_j$}
\Text(324,80)[t]{$D^-_i$}
\Text(315,130)[t]{$H^+$}
\Text(250,27)[t]{\small $y_{u_J}$}
\Text(330,27)[t]{\small $y_{d_I}$}
\DashArrowLine(290,100)(250,40){4}
\DashArrowLine(330,40)(290,100){4}
\DashArrowLine(290,135)(290,100){4}
\end{picture}
\end{center}
\caption{Dominant corrections to the  
charged Higgs boson couplings to quarks.}
\label{fig:chargedvert}
\end{figure}

The factors $\epsilon_0^\prime$, $\epsilon_Y^\prime$, given in eq.
(\ref{eqn:eprime}) entering the expression (\ref{F1A}) for
$\epsilon^{HL}_{JI}$ obtained in the $SU(2)\times U(1)$ symmetry limit
and for vanishing electroweak gauge couplings can be easily extended
beyond this approximation. Improvement analogous to the replacement of
$\epsilon_Y$ of eq. (\ref{eqn:e}) by $\epsilon_Y$ of eq.
(\ref{eqn:eY}) in the neutral Higgs boson sector has been given in
ref.~\cite{DEGAGI}. Simplifying appropriately the full MSSM expressions
for the diagrams shown in fig. \ref{fig:chargedvert} allows to replace
$\mu$ by the contribution of the four neutralino mass eigenstates and
to take into account the mixing of the left and right squarks of the
third generation. As a result
$\epsilon^\prime_0+\epsilon^\prime_Yy^2_b$ in the formulae
(\ref{eqn:DdYu_str}) and (\ref{F1A}) for $(\Delta_d\mathbf{Y}_u)^{JI}$
and $\epsilon^{HL}_{JI}$, respectively are replaced by the flavour
dependent quantity $\epsilon^\prime_J(I)$ of~\cite{DEGAGI}:
\begin{eqnarray}
&&\epsilon^\prime_J(I)=-{2\alpha_s\over3\pi}{\mu\over \tilde m_g}
\left[c^2_{\tilde u}c^2_{\tilde d}H_2\left(x^{u/g}_2,x^{d/g}_1\right)
     +s^2_{\tilde u}c^2_{\tilde d}H_2\left(x^{u/g}_1,x^{d/g}_1\right)
\right.\nonumber\\
&&\left.\phantom{aaaaaaaaaaaaa}
     +c^2_{\tilde u}s^2_{\tilde d}H_2\left(x^{u/g}_2,x^{d/g}_2\right)
     +s^2_{\tilde u}s^2_{\tilde d}H_2\left(x^{u/g}_1,x^{d/g}_2\right)
\right]\label{eqn:epb_t_gamb}\label{eqn:eprimb_gamb}\\
&&\phantom{aaaaa}
  -{y^2_b\over16\pi^2}\sum_{l=1}^4Z^{3l}{A_b\over m_{N_l}}Z^{4l}
\left[c^2_{\tilde u}c^2_{\tilde b}H_2\left(x^{u/N}_{1l},x^{b/N}_{2l}\right)
     +s^2_{\tilde u}c^2_{\tilde b}H_2\left(x^{u/N}_{2l},x^{b/N}_{2l}\right)
\right.\nonumber\\
&&\phantom{aaaaaaaaaaaaa}\left.
     +c^2_{\tilde u}s^2_{\tilde b}H_2\left(x^{u/N}_{1l},x^{b/N}_{1l}\right)
     +s^2_{\tilde u}s^2_{\tilde b}H_2\left(x^{u/N}_{2l},x^{b/N}_{1l}\right)
\right]\delta^{I3}
\nonumber
\end{eqnarray}
where $x^{u/g}_k\equiv M^2_{\tilde u_k}/m^2_{\tilde g}$
($x^{d/g}_k\equiv M^2_{\tilde d_k}/m^2_{\tilde g}$), $k=1,2$ with
$\tilde u_k$ ($\tilde d_k$) being the up (down) squark of the $J$-th
($I$-th) generation and $c_{\tilde u}$ and $c_{\tilde d}$ etc.
denoting cosines and sines of the appropriate mixing angles. (For the
first two generations $c_{\tilde u, \tilde d}\approx1$, $s_{\tilde u,
  \tilde d}\approx0$.) Similarly, $x^{u/N}_{kl}\equiv M^2_{\tilde
  u_k}/m^2_{N_l}$, $k=1,2$, $l=1,4$ etc.  In the formula
(\ref{eqn:epb_t_gamb}) $y_b$ is given by
\begin{eqnarray}
y_b=-{\sqrt2\over v_d}{\overline{m}_b\over(1+\tilde\epsilon_3\tan\beta)}~.
\label{eqn:yb}
\end{eqnarray}
Using the identity $Z_N^{3j}m_{N_l}Z_N^{4j}=-\mu^\ast$ it is easy to
see that in the $SU(2)\times U(1)$ limit
$\epsilon^\prime_J(I)\rightarrow\epsilon^\prime_0+\epsilon^\prime_Yy^2_b$.

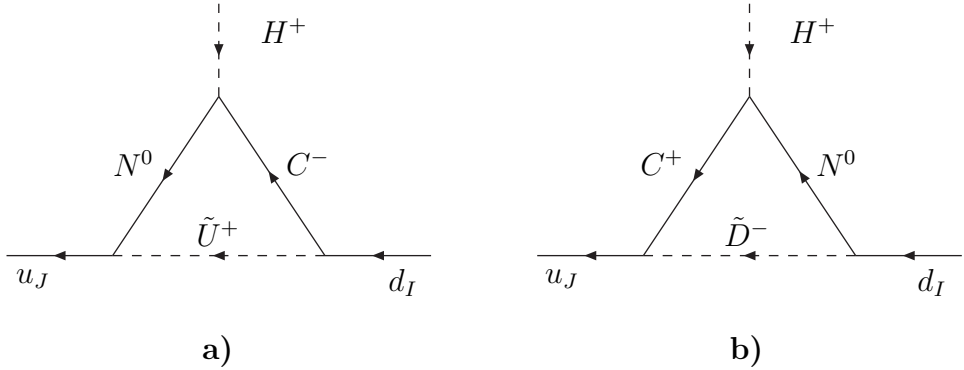
\begin{figure}[htbp] 
\begin{center}
\begin{picture}(400,160)(0,0)
\DashArrowLine(130,40)(50,40){4}
\ArrowLine(50,40)(10,40)
\ArrowLine(170,40)(130,40)
\Text(20,35)[t]{$u_J$}
\Text(160,35)[t]{$d_I$}
\Text(90,55)[t]{$\tilde{U}^+$}
\Text(58,80)[t]{$N^0$}
\Text(124,80)[t]{$C^-$}
\Text(115,130)[t]{$H^+$}
\ArrowLine(90,100)(50,40)
\ArrowLine(130,40)(90,100)
\DashArrowLine(90,135)(90,100){4}
\Text(90,10)[t]{\bf a)}
\DashArrowLine(330,40)(250,40){4}
\ArrowLine(250,40)(210,40)
\ArrowLine(370,40)(330,40)
\Text(220,35)[t]{$u_J$}
\Text(360,35)[t]{$d_I$}
\Text(290,55)[t]{$\tilde{D}^-$}
\Text(258,80)[t]{$C^+$}
\Text(324,80)[t]{$N^0$}
\Text(315,130)[t]{$H^+$}
\ArrowLine(290,100)(250,40)
\ArrowLine(330,40)(290,100)
\DashArrowLine(290,135)(290,100){4}
\Text(290,10)[t]{\bf b)}
\end{picture}
\end{center}
\caption{Additional important corrections to the charged Higgs boson
couplings to quarks.}
\label{fig:chargedvert_extra}
\end{figure}

\begin{figure}[htbp] 
\begin{center}
\begin{picture}(400,160)(0,0)
\DashArrowLine(50,40)(130,40){4}
\ArrowLine(10,40)(50,40)
\ArrowLine(170,40)(130,40)
\Text(20,35)[t]{$u^c_J$}
\Text(160,35)[t]{$q_I$}
\Text(90,58)[t]{$\tilde{U}^c$}
\Text(70,87)[t]{$\chi$}
\Text(52,63)[t]{$\chi$}
\Text(120,89)[t]{$\tilde H^{(d)}$}
\Text(137,65)[t]{$\tilde H^{(u)}$}
\Text(115,130)[t]{$H^{(d)}$}
\ArrowLine(70,70)(90,100)
\ArrowLine(70,70)(50,40)
\Line(67,70)(73,70)
\Line(72,67)(68,73)
\ArrowLine(110,70)(130,40)
\ArrowLine(110,70)(90,100)
\Line(108,67)(112,73)
\Line(113,70)(107,70)
\DashArrowLine(90,100)(90,135){4}
\Text(90,10)[t]{\bf a)}
\DashArrowLine(330,40)(250,40){4}
\ArrowLine(210,40)(250,40)
\ArrowLine(370,40)(330,40)
\Text(220,35)[t]{$u^c_J$}
\Text(360,35)[t]{$q_I$}
\Text(290,58)[t]{$\tilde Q$}
\Text(267,89)[t]{$\tilde H^{(d)}$}
\Text(248,63)[t]{$\tilde H^{(u)}$}
\Text(321,87)[t]{$\chi,\psi$}
\Text(339,63)[t]{$\chi,\psi$}
\Text(315,130)[t]{$H^{(d)}$}
\ArrowLine(270,70)(250,40)
\ArrowLine(270,70)(290,100)
\Line(267,70)(273,70)
\Line(272,67)(268,73)
\ArrowLine(310,70)(330,40)
\ArrowLine(310,70)(290,100)
\DashArrowLine(290,100)(290,135){4}
\Line(308,67)(312,73)
\Line(313,70)(307,70)
\Text(290,10)[t]{\bf b)}
\end{picture}
\end{center}
\caption{Additional contribution to $\Delta_d\mathbf{Y}_u$ for nonzero
  electroweak gauge couplings. $\chi$ and $\psi$ denote $U(1)$ and
  $SU(2)$ gauginos, respectively.}
\label{fig:chargedvert_extra_lim}
\end{figure}
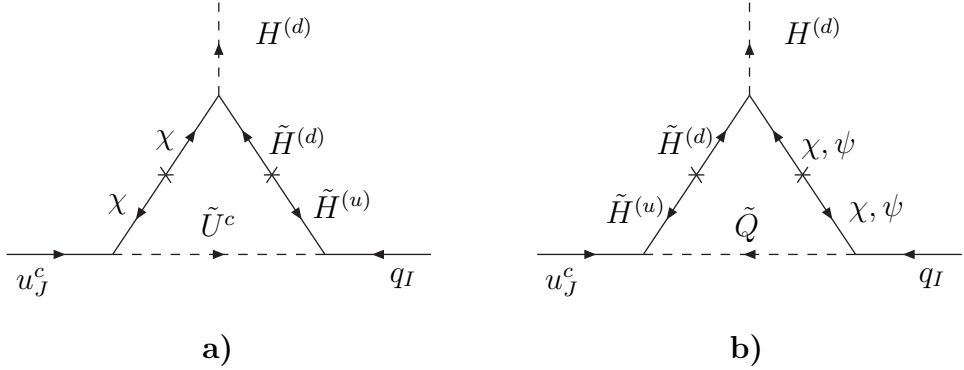

Similarly as in the case of the replacement of $\epsilon_Y$ of eq.
(\ref{eqn:e}) by that of eq. (\ref{eqn:eY}), the use of
$\epsilon^\prime_J(I)$ does not improve the approximation to
$\epsilon^{HL}_{JI}$ obtained in the full approach of sec. 2.5. This
is because while going beyond the $SU(2)\times U(1)$ symmetry limit in
taking into account vacuum expectation value dependent sparticle
mixing, it still neglects contributions which are present in the
$SU(2)\times U(1)$ symmetry limit for nonzero gauge couplings $g_1$
and $g_2$. The latter effects can be incorporated taking into account
also the $SU(2)\times U(1)$ symmetry breaking effects by computing
diagrams shown in fig.  \ref{fig:chargedvert_extra} (in the
$SU(2)\times U(1)$ symmetry limit they correspond to diagrams shown in
\ref{fig:chargedvert_extra_lim} contributing to $\Delta_d\mathbf{Y}_u$
for nonzero $g_1$ and $g_2$).  Taking into account their dominant
parts (which are finite) amounts to adding to $\epsilon^\prime_J(I)$
the following terms:
\begin{eqnarray}
&&\delta_a\epsilon^\prime_J(I)={1\over16\pi^2}
{g^2_2\over\sqrt2c_W}\sum_{j,l}Z_+^{1l}Z_N^{1j}
a^{lj}{m_{N_j}\over m_{C_l}}H_2\left(x^{N/C}_{jl},x^{Q/C}_{Jl}\right)
\nonumber\\
&&\phantom{aaaaaa}+{1\over16\pi^2}
{2\over3}{g^2_2s_W\over c^2_W}\sum_{j,l}Z_+^{2l}Z_N^{4j}
a^{lj}{m_{N_j}\over m_{C_l}}H_2\left(x^{N/C}_{jl},x^{U/C}_{Jl}\right)
\nonumber\\
&&\delta_b\epsilon^\prime_J(I)=-{1\over16\pi^2}{g^2_2\over2c^2_W}
\sum_{j,l}Z_+^{2l}\left({1\over3}s_WZ_N^{1j}-c_WZ_N^{2j}\right)
a^{lj}{m_{N_j}\over m_{C_l}}H_2\left(x^{N/C}_{jl},x^{Q/C}_{Il}\right)
\nonumber
\end{eqnarray}
where $x^{N/C}_{jl}=m^2_{N_j}/m^2_{C_l}$,
$x^{Q/C}_{Jl}=m^2_{Q_J}/m^2_{C_l}$, etc. and
$a^{lj}=Z_-^{2l}(s_WZ_N^{1j}+c_WZ_N^{2j})-\sqrt2Z_-^{1l}Z_N^{3j}$.

\begin{figure}[htbp]
\begin{center}
\epsfig{file=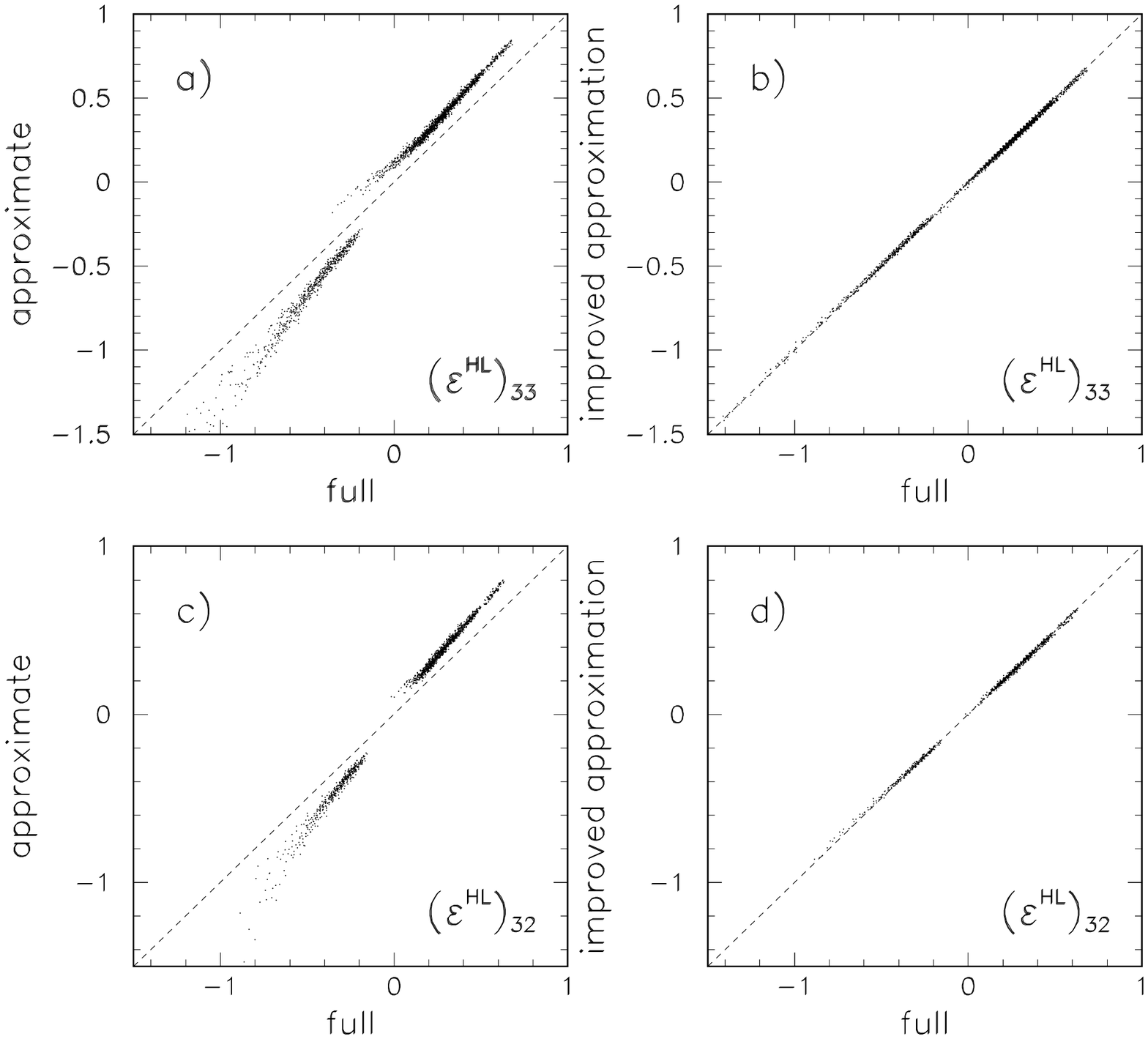,width=0.62\linewidth}
\end{center}
\caption{\protect As in figure \ref{fig:eps1} but for the corrections 
  $\epsilon^{HL}_{33}$ and $\epsilon^{HL}_{32}$ to the couplings of
  $H^+$ to the left-handed down quarks. Only points yielding
  acceptable $BR(\bar B\rightarrow X_s\gamma)$ are shown.}
\label{fig:epshl}
\end{figure}

Only the sum of all these corrections including the delta term
(\ref{delta}) reproduces the correction $\epsilon^{HL}_{JI}$ computed
as in sec 2.5:
\begin{eqnarray}
\epsilon^{HL}_{JI}\approx\tan\beta\left[\epsilon^\prime_J(I)+
\delta_a\epsilon^\prime_J(I)+\delta_b\epsilon^\prime_J(I)\right]
+\Delta_{JI}
\label{eqn:approximation}
\end{eqnarray}
While in most cases the term proportional to $\alpha_s$ in
$\epsilon^\prime_J(I)$ is dominant, the term
$\delta_a\epsilon^\prime_J(I)+\delta_b\epsilon^\prime_J(I)$ is usually
more important than the term $\Delta_{JI}$ and even than $y_b^2$
dependent part of the $\epsilon^\prime_J(I)$ term. Let us also stress
that the quality of the approximation of $\epsilon^{HL}_{JI}$ by the
rhs depends crucially on how well $\tilde\epsilon_3$ approximates the 
factor $\epsilon_{d_3}$ (\ref{eqn:eps_def}) needed to obtain $y_b$ and
how well epsilons reproduce the ratio $V^{JI}/V^{JI}_{\rm eff}$. Since 
the approximations of secs. 3.2-3.3 are usually not better than 15\%-20\%, 
the contributions to the lhs and rhs of eq. (\ref{eqn:approximation}) not
proportional to $\alpha_s$ usually differ by about 20\%-50\%.
Reliable approximation can be obtained only by using in (\ref{delta})
$\tilde\epsilon_0$ and $\epsilon^{(32)}_Y$ computed as in eq. 
(\ref{eqn:epsfromdmd}) instead of $\epsilon_0$ and $\epsilon_Y$. The
approximation (\ref{eqn:approximation}) works then to an accuracy
5\%-10\%. This is demonstrated in figure \ref{fig:epshl} where in
panels a and b (panels c and d) we compare $\epsilon^{HL}_{33}$
($\epsilon^{HL}_{32}$) calculated as in section 2.5 with the
approximation based on the formula (\ref{F1A}) and with the
approximation based on eq. (\ref{eqn:approximation}), respectively. In
panel a (and c) we have used in the $\Delta_{JI}$ term (\ref{delta})
$\epsilon_0$ and $\epsilon_Y$ given by eqs. (\ref{eqn:e}), (\ref{eqn:eY}) 
whereas in panel b (and d) $\tilde\epsilon_0$ and $\epsilon^{(32)}_Y$ from 
the formula (\ref{eqn:epsfromdmd}).
The difference between the results of the full and approximate
calculations is quantified by the deviation of the points from the
diagonal line.  The dramatic improvement of the approximation is
clearly seen in panels b and d of figure \ref{fig:epshl}.

\begin{figure}[htbp]
\begin{center}
\epsfig{file=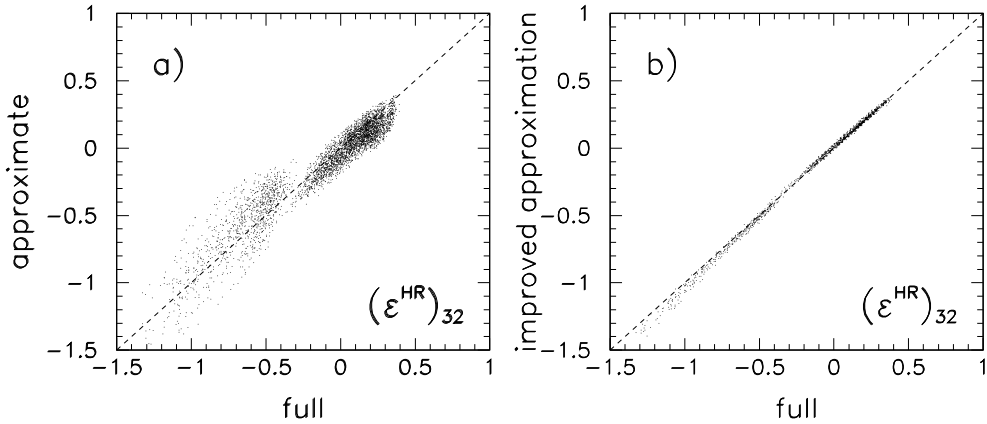,width=\linewidth}
\end{center}
\caption{\protect As in figure \ref{fig:eps1} but for the corrections 
  $\epsilon^{HR}_{32}$ to the couplings of $H^+$ to the right-handed
  down quarks. Only points yielding acceptable $BR(\bar B\rightarrow
  X_s\gamma)$ are shown.}
\label{fig:epshr}
\end{figure}

Next we consider the correction $\epsilon^{HR}_{JI}$. As in the
approximation based on the $SU(2)\times U(1)$ symmetry limit, also in
the complete calculation of sec. 2.5 it is entirely determined by the
rotations in the down quark sector (the genuine vertex corrections and
the contribution of the vector self energies are negligible) and can
be obtained from the formula (\ref{eqn:phlrsymlim}) with exact
matrices $\mathbf{D}_L$ and $\mathbf{D}_R$. The quality of the
approximation (\ref{PH1}) depends therefore crucially on how well the
matrices $\mathbf{D}_L$ and $\mathbf{D}_R$ are approximated. For split
squarks the full flavour dependence of the correction $(\Delta
m_d)^{JI}$ is due also to the flavour dependence of $\epsilon_0$ and
$\epsilon_Y$ and, consequently, the matrices $\mathbf{D}_L$ and
$\mathbf{D}_R$ obtained in the strict $SU(2)\times U(1)$ symmetry
limit with universal $\epsilon_0$ and $\epsilon_Y$ deviate from the
true ones. To obtain a reliable approximation for $\epsilon^{HR}_{JI}$
in the more realistic situation, in which squarks from the third
generation are split from the others, one has to use flavour dependent
expressions (\ref{PH3})-(\ref{PH6}) and the improved formulae in
(\ref{eqn:epsfromdmd}).  The results of this improvement are shown in
figure \ref{fig:epshr} where we compare $\epsilon^{HR}_{32}$ obtained
in the full approach based on eqs. (\ref{PLR}), (\ref{PLRform}) with
the approximation (\ref{PH1}) using universal $\epsilon_0$ and
$\epsilon_Y$ obtained from eqs. (\ref{eqn:e}) and (\ref{eqn:eY})
(panel a) and with the $\epsilon^{HR}_{32}$ obtained from eqs.
(\ref{PH6}) and (\ref{eqn:epsfromdmd}) (panel b).

Finally we consider the corrections of the charged Goldstone bosons to
quarks defined in eq. (\ref{PGOLD}). As we argued, in the $SU(2)\times
U(1)$ symmetry limit $\epsilon_{JI}^{GL}=\epsilon_{JI}^{GR}=0$ also
for nonzero electroweak gauge couplings $g_1$ and $g_2$.  To
understand our results, it is important to stress at this point that
integrating out sparticles and constructing the effective Lagrangian
in the $SU(2)\times U(1)$ symmetry limit without making any assumption
about dominance of some particular couplings (like $\alpha_s$, or
$y_t$) requires also the inclusion of the vector parts of the quark
self energies and of all vertex diagrams, in particular the ones shown
in fig.  \ref{fig:chargedvert_extra_lim}.  In other words, the whole
procedure of sec. 2 has to be used with the exception that the Higgs
boson vacuum expectation values are equal zero.  The argument for
vanishing $\epsilon_{JI}^{GL}$ and $\epsilon_{JI}^{GR}$ applies then
up to $1/M^2_{\rm SUSY}$ corrections.

To go beyond the $SU(2)\times U(1)$ symmetry limit we write the
factors $\epsilon_{JI}^{GL}$ and $\epsilon_{JI}^{GR}$ as
\begin{eqnarray}
&&\epsilon_{GL}^{JI}={\sqrt2M_W\over g_2}
{\left[\mathbf{U}_R^\dagger\cdot\left(\Delta\hat F_L^G\cdot V^\dagger
-{g_2\over\sqrt2M_W}\Delta m_u\right)\cdot\mathbf{U}_L\right]^{JK}
V_{\rm eff}^{KI}\over\overline{m}_{u_J}V_{\rm eff}^{JI}}\label{eqn:epsgl}\\
&&\epsilon_{GR}^{JI}=-{\sqrt2M_W\over g_2}
{V_{\rm eff}^{JK}\left[\mathbf{D}_L^\dagger\cdot
\left(V^\dagger\cdot \Delta\hat F_R^G +{g_2\over\sqrt2M_W}
(\Delta m_d)^\dagger\right)\cdot \mathbf{D}_R\right]^{KI}
\over V_{\rm eff}^{JI} \overline{m}_{d_I}}\label{eqn:epsgr}
\end{eqnarray}
where $\Delta\hat F^G_{L,R}$ are defined in eqs. (\ref{PRLform}),
(\ref{PLRform}). (Similar expressions can be written down for
corrections $\epsilon_{HL}^{JI}$ and $\epsilon_{HR}^{JI}$.) As can be
seen by using the formula (\ref{eqn:chcorr}), in the $SU(2)\times
U(1)$ symmetry limit $\Delta\hat F^G_L={g_2\over\sqrt2M_W}\Delta
m_uV$, $\Delta\hat F^G_R=-{g_2\over\sqrt2M_W}V(\Delta m_d)^\dagger$
and the corrections indeed vanish.  (It is also easy to see that
$\Sigma_{VL}^u$ and $\Sigma_{VR}^d$ cancel out in (\ref{eqn:epsgl})
and in (\ref{eqn:epsgr}), respectively.)
Because both, $V^\dagger\Delta\hat F^G_R$ and $(g_2/\sqrt2M_W)(\Delta
m_d)^\dagger$ contain contributions $\propto\tan\beta$, any mismatch
between them could in principle give rise to $\tan\beta$ enhanced
terms in $\epsilon_{JI}^{GR}$ and this motivated the authors of ref.
\cite{DEGAGI} to keep the correction $\epsilon_{JI}^{GR}$ in their
$BR(\bar B\to X_s\gamma)$ calculation (no such enhancement was
expected in $\epsilon_{JI}^{GL}$). Exact numerical evaluation of the
formulae (\ref{eqn:epsgl}), (\ref{eqn:epsgr}) reveals, however, that
the cancellation between $\Delta\hat F^G_L\cdot V^\dagger$ and
$-(g_2/\sqrt2M_W)\Delta m_d$ in (\ref{eqn:epsgl}) and between
$V^\dagger\cdot\Delta\hat F^G_R$ and $(g_2/\sqrt2M_W)(\Delta
m_d)^\dagger$ in (\ref{eqn:epsgr}) is surprisingly accurate when these
quantities are calculated as in sec. 2 which means that the
$SU(2)\times U(1)$ symmetry breaking corrections to $\Delta\hat
F^G_{L,R}$ and $\Delta m_{u,d}$ are tiny compared to the corrections
introduced by switching on the electroweak gauge couplings. We find
that $|\epsilon_{JI}^{GL}|$ and $|\epsilon_{JI}^{GR}|$ are of the same
order of magnitude and are $\sim{\cal O}(10^{-4})$ in most of the
parameter space (for sparticles heavier than 500 GeV and
$\tan\beta=50$). These are small corrections compared to 1 and do not
influence appreciably the phenomenology we are going to discuss.  Note
also, that approximating $\Delta m_d$ or $\Delta\hat F^G_{L,R}$ by
dropping some ``small'' terms or using simplified couplings in the
vertices as compared to the ones given in ref. \cite{ROS} spoils this
delicate cancellation and generates
$|\epsilon_{JI}^{GL,R}|\sim10^{-(1-2)}$ !

Finally let us explain the status of the rules (\ref{REL2}) and
(\ref{REL3}).  As long as one considers only ${\cal O}(\alpha_s)$
corrections to the effective Lagrangian couplings the rules
(\ref{REL2}) and (\ref{REL3}) are the true equalities. This has to be
so, because when the charged Goldstone boson vertices corrected as in
\cite{DEGAGI} are used in the one loop expressions for the Wilson
coefficients $C_{7,8}$ of the (chromo)magnetic operators they
reproduce correctly \cite{CAGANIWA2} appropriate $\tan\beta$ enhanced
terms in the two loop contributions to $C_{7,8}$ calculated in
\cite{CIDEGAGI2}. The ideology of the effective Lagrangian serves then
only to correctly resumm these terms. Indeed, in this case $V^{\rm
  eff}=V$, $\mathbf{D}_{L,R}=\mathbf{I}$ and inserting $\Delta\hat
F^G_R$ in the form
\be
(\Delta\hat F^G_R)^{JI} = 
-{g\over\sqrt2M_W}V^{JI}m_{d_I}\epsilon^\prime_I(J)\tan\beta
\ee
as in ref. \cite{DEGAGI} in (\ref{eqn:epsgr}) one recovers the rule
(\ref{REL3}).\footnote{However, calculating $\epsilon^\prime_I(J)$
  from the triangle diagram of fig. 1a of \cite{DEGAGI} with
  simplified $G^+$ couplings to up and down squarks would result in
  $\epsilon^{GR}_{JI}\sim10^{-2}$, much bigger than ${\cal
    O}(10^{-4})$ obtained with the full $G^+$ couplings given in
  \cite{ROS}. In particular, we have found that for non-negligible
  sbottom mixing (if, say, $|A_b|\sim|A_t|$) neglecting in this vertex 
  terms $\propto y^2_t$ and $\propto g_2^2$ has dramatic effect on the
  cancellation in (\ref{eqn:epsgr}). On the other hand, neglecting
  simultaneously the contributions of the vector self energies to
  $\Delta\hat F^G_R$ and to $\Delta m_d$ does not affect it.}
However, when the corrections depending on the top Yukawa coupling to
the vertices of the effective Lagrangian are taken into account, the
true formula (\ref{eqn:epsgr}) for $\epsilon^{GR}_{JI}$ is more
involved and cannot be obtained from the rule (\ref{REL3}) just by
including the terms $\propto y^2_t$ to $\epsilon^\prime_I(J)$ as in
\cite{DEGAGI}. Formulae (\ref{REL2}) and (\ref{REL3}) are then only
the substitution rules allowing to translate the formulae of
\cite{DEGAGI} into ours.


\section{\boldmath{$\Delta M_{d,s}$}, \boldmath{$B^0_{d,s}\rightarrow 
    \mu^+\mu^-$} and \boldmath{$\bar B\rightarrow X_s \gamma$}}
\setcounter{equation}{0}

\begin{figure}[htbp]
\begin{center}
\begin{picture}(160,80)(0,0)
\ArrowLine(50,40)(20,70)
\ArrowLine(20,10)(50,40)
\Vertex(50,40){5}
\ArrowLine(140,10)(110,40)
\ArrowLine(110,40)(140,70)
\DashLine(50,40)(110,40){3}
\Vertex(110,40){2}
\Text(80,30)[]{\small $h^0$,$H^0$,$A^0$}
\Text(35,70)[]{\small $b_R$}
\Text(40,10)[]{\small $s_L,d_L$}
\Text(125,70)[]{\small $l^-$}
\Text(125,10)[]{\small $l^+$}
\Text(15,40)[]{\small $\tan^2\beta$}
\Text(135,40)[]{\small $\tan\beta$}
\end{picture}
\end{center}
\caption{Diagrams giving dominant contribution to 
$B^0_{s,d}\rightarrow l^+l^-$ amplitudes at large $\tan\beta$.}
\label{fig:bmumu}
\end{figure}
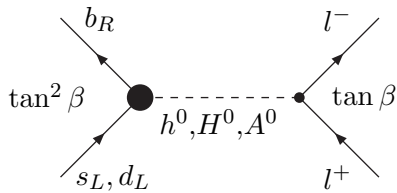

\begin{figure}[htbp]
\begin{center}
\begin{picture}(340,80)(0,0)
\ArrowLine(50,20)(10,10)
\ArrowLine(90,10)(50,20)
\Vertex(50,20){5}
\ArrowLine(10,70)(50,60)
\ArrowLine(50,60)(90,70)
\Vertex(50,60){5}
\DashLine(50,20)(50,60){3}
\Text(78,40)[]{\small $h^0$,$H^0$,$A^0$}
\Text(25,0)[]{\small $b_R$}
\Text(75,0)[]{\small $s_L$}
\Text(75,75)[]{\small $b_R$}
\Text(25,75)[]{\small $s_L$}
\ArrowLine(170,20)(130,10)
\ArrowLine(210,10)(170,20)
\Vertex(170,20){5}
\ArrowLine(130,70)(170,60)
\ArrowLine(170,60)(210,70)
\Vertex(170,60){5}
\DashLine(170,20)(170,60){3}
\Text(198,40)[]{$h^0$,$H^0$,$A^0$}
\Text(145,0)[]{$b_L$}
\Text(195,0)[]{$s_R$}
\Text(195,75)[]{$b_L$}
\Text(145,75)[]{$s_R$}
\ArrowLine(290,20)(250,10)
\ArrowLine(330,10)(290,20)
\Vertex(290,20){5}
\ArrowLine(250,70)(290,60)
\ArrowLine(290,60)(330,70)
\Vertex(290,60){5}
\DashLine(290,20)(290,60){3}
\Text(318,40)[]{$h^0$,$H^0$,$A^0$}
\Text(265,0)[]{$b_R$}
\Text(315,0)[]{$s_L$}
\Text(315,75)[]{$b_L$}
\Text(265,75)[]{$s_R$}
\end{picture}
\end{center}
\caption{Double penguin diagrams contributing to $\Delta M_s$.}
\label{fig:2pg}
\end{figure}
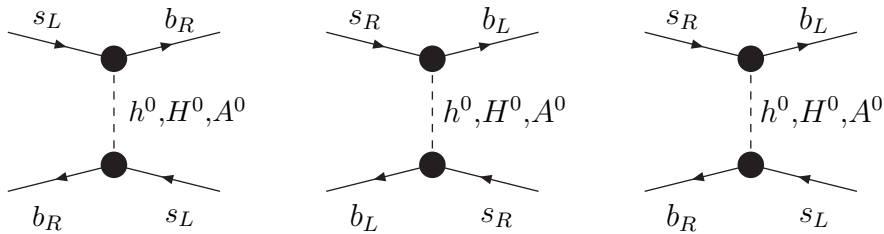

The $B^0_{s,d}\to\mu^+\mu^-$ decays and the $B^0_s$-$\bar B^0_s$
mixing attracted recently a renewed attention due to the observation
that for large values of $\tan\beta$ their amplitudes can receive very
large contributions from diagrams depicted in figures \ref{fig:bmumu}
and \ref{fig:2pg} in which the black blobs represent the flavour
changing couplings $\left[X^S_{RL}\right]^{JI}$ and
$\left[X^S_{LR}\right]^{JI}$ discussed in the preceding sections.
These contributions have been found to {\it increase} by orders of
magnitude the branching ratios for the rare decays
$B^0_{s,d}\to\mu^+\mu^-$ \cite{BAKO,CHSL,HULIYAZH,BOEWKRUR} and to
{\it decrease} substantially the $B^0_s$-$\bar B^0_s$ mass difference
$\Delta M_s$~\cite{BUCHROSL1}.

As demonstrated in~\cite{BAKO,CHSL,HULIYAZH,BOEWKRUR}, for
$\tan\beta\sim50$ and non-negligible values of the parameter $A_t$ the
$B^0_{s,d}\to\mu^+\mu^-$ amplitudes are totally dominated by the
diagram of fig. \ref{fig:bmumu}. In the absence of any other
constraints on the MSSM parameter space, the corresponding branching
ratios, which behave as $|A_t\tan^3\beta/M_A^2|^2$, can be enhanced by
up to three orders of magnitude relative to the SM predictions and can
even exceed the present experimental bounds:
\begin{eqnarray}
BR(B^0_d\rightarrow\mu^+\mu^-)<2.1\times10^{-7}\phantom{aa}{\rm BaBar}~ 
\cite{ALEXAN},
\label{eqn:BaBarlimit}\\
BR(B^0_s\rightarrow\mu^+\mu^-)<2.0\times10^{-6}\phantom{aa}{\rm CDF}~
\cite{CDF}~.
\phantom{a}
\label{eqn:CDFlimit}
\end{eqnarray}

The $B^0_{s,d}$-$\bar B^0_{s,d}$ mass difference in turn receives several 
contributions:
\be
\Delta M_s=|(\Delta M_s)^{\rm SM}+(\Delta M_s)^{H^\pm}+
(\Delta M_s)^{\chi^\pm}+(\Delta M_s)^{\rm DP}|\equiv 
(\Delta M_s)^{\rm SM}|1+f_s|
\label{DMS0}
\ee
(by definition $\Delta M_s$ is a positive definite quantity). For
large $\tan\beta$ and non-negligible $A_t$ the contribution $(\Delta
M_s)^{\rm DP}$ of the double scalar penguin (DP) shown in fig.
\ref{fig:2pg} is the dominant correction to the SM contribution
$(\Delta M_s)^{\rm SM}$. Both, $(\Delta M_s)^{\rm DP}$ and the
contribution $(\Delta M_s)^{H^\pm}$ of the box-diagrams with top and
charged Higgs bosons $H^\pm$ have the signs opposite to $(\Delta
M_s)^{\rm SM}$. While generally smaller than $(\Delta M_s)^{\rm SM}$,
their sum leads for large $\tan\beta$ to a significant decrease of the
predicted $\Delta M_s$ (i.e. to $f_s<0$) independently of the choice
of supersymmetric parameters.  We will asses the relative magnitudes
of the double penguin, charged Higgs and chargino box diagrams in
secs. 6.1.3, 6.1.4.

As is evident from the comparison of figures \ref{fig:bmumu} and
\ref{fig:2pg} there must exist a strong correlation between the
enhancement of $BR(B^0_{s(d)}\to\mu^+\mu^-)$ and the suppression of
$\Delta M_s$. In particular for $0<(1+f_s)<1$ the experimental lower
bound $\Delta M_s>15/$ps puts an upper bound on the possible
enhancement of $BR(B^0_{s(d)}\to\mu^+\mu^-)$. Of interest is also the
case $(1+f_s)<0$ corresponding to a very large negative $(\Delta
M_s)^{\rm DP}$ (that can be realized for very special values of
supersymmetric parameters) which has quite different implications than
the case $0<(1+f_s)<1$.  The main result concerning this correlation has
been presented in \cite{BUCHROSL2}. Our purpose now is to explore it
in more detail investigating in particular its dependence on the MSSM
parameters and elucidating the impact of the $\tan\beta$ enhanced
corrections to $\Delta M_s$ on the standard Unitarity Triangle (UT)
analysis which is necessary to determine the CKM matrix element
$V_{td}$ needed to predict accurately $BR(B^0_d\to\mu^+\mu^-)$.

Large $\tan\beta$ effects manifest themselves also in the transition
$\bar B\to X_s\gamma$, which being already relatively well measured,
plays an important role in constraining the allowed region in the
space of supersymmetric parameters ~\cite{DEGAGI,CIDEGAGI2,CAGANIWA2}.
Exploring the correlation between the increase of
$BR(B^0_{s(d)}\to\mu^+\mu^-)$ and suppression of $\Delta M_s$ it is
therefore important to incorporate this constraint in the
analysis.\footnote{Large $\tan\beta$ effects have been also
  investigated in $(g-2)_\mu$. However, they depend on the slepton
  sector parameters which can be correlated with the squark sector
  ones only within a particular scenario like e.g.  minimal
  supergravity~\cite{DEDRNI} or gauge mediation.}

In this section we collect the formulae for
$BR(B^0_{s(d)}\to\mu^+\mu^-)$ and $(\Delta M_s)^{\rm DP}$ in the large
$\tan\beta$ limit and discuss the importance of various contributions.
We recall also the UT analysis and include a subsection devoted to
the calculation of $BR(\bar B\to X_s\gamma)$.  As in
\cite{BUCHROSL1,BUCHROSL2} we will concentrate on the scenario with
heavy sparticles and the mass scale of the Higgs sector close to the
electroweak scale but will comment also on possible effects of lighter
sparticles. The numerical analysis of the correlation of the
enhancement of $BR(B^0_{s(d)}\to\mu^+\mu^-)$ and the suppression of
$\Delta M_s$ will be the subject of the next section.

\subsection{Formulae for calculating \boldmath{$\Delta M_s$}} 

\subsubsection{The Effective Hamiltonian}

The effective Hamiltonian for $B_s^0$-$\bar B^0_s$ mixing has the form
as follows
\begin{eqnarray}
{\cal H}^{\Delta B=2}_{\rm eff}=
{G_F^2M^2_W\over16\pi^2}(V_{\rm eff}^{tb\ast}V_{\rm eff}^{ts})^2
\sum_i C_i(\mu)Q_i~.
\label{eqn:heff_dF2}
\end{eqnarray}
Here $Q_i$ are the $\Delta B=2$ operators and $G_F$ is the Fermi
constant.  It should be stressed that with the improvements in the
calculation of Higgs boson vertices the factorization of the CKM
elements in (\ref{eqn:heff_dF2}) is only approximate. Still, it is an
excellent approximation to the full calculation and allows for a
transparent discussion of the unitarity triangle (UT). The set of
dimension six $\Delta B=2$ operators consists of
\begin{eqnarray}\label{OPS}
&&Q^{\rm VLL}=\left(\overline{b_L}\gamma_\mu s_L\right)
\left(\overline{b_L}\gamma^\mu s_L\right)\nonumber\\
&&Q_1^{\rm LR}=\left(\overline{ b_L}\gamma_\mu s_L\right)
\left(\overline{ b_R}\gamma_\mu s_R\right)\nonumber\\
&&Q_2^{\rm LR}=\left(\overline{ b_R} s_L\right)
\left(\overline{ b_L}s_R\right) \label{eqn:ops_dF2}\\
&&Q_1^{\rm SLL}=\left(\overline{ b_R} s_L\right)
\left(\overline{ b_R}s_L\right) \nonumber\\
&&Q_2^{\rm SLL}=\left(\overline{ b_R}\sigma_{\mu\nu} s_L\right)
\left(\overline{ b_R}\sigma^{\mu\nu}s_L\right) \nonumber
\end{eqnarray}
where $\sigma_{\mu\nu}={1\over2}[\gamma_\mu,\gamma_\nu]$ and the
colour indices are contracted within the brackets. The additional
operators $Q^{\rm VRR}$, $Q_1^{\rm SRR}$ and $Q_2^{\rm SRR}$ are
obtained from $Q^{\rm VLL}$, $Q_1^{\rm SLL}$ and $Q_2^{\rm SLL}$ by
replacing $L$ with $R$. Similar Hamiltonians, with appropriate changes
of the quark fields and the CKM matrix elements describe also
$B_d^0$-$\bar B^0_d$ and $K^0$-$\bar K^0$ transitions. In the SM only
the Wilson coefficients of $Q^{\rm VLL}$ are relevant and the details
of their contributions are well known (see e.g. \cite{BUER}).

The basic formula for the $B_q^0$-$\bar B_q^0$ mass difference reads
\be\label{DMSTOT}
\Delta M_q=\frac{G_F^2 M_W^2}{6\pi^2} M_{B_q}\eta_B F^2_{B_q}
\hat B_{B_q} |V^{\rm eff}_{tq}|^2 |F^q_{tt}|
\ee
where $q=s$ or $d$. We recall also that \cite{BUCHROSL1}
\begin{eqnarray}
F^s_{tt}&=&\left[S_0(x_t)+ \frac{1}{4 r}C_{\rm new}^{\rm VLL}(\mu_S)\right]
\nonumber\\
&+&{1\over4r}C^{\rm VRR}_1(\mu_S)+ \bar P_1^{\rm LR} C^{\rm LR}_1(\mu_S) 
+\bar P_2^{\rm LR} C^{\rm LR}_2(\mu_S) \label{hds2}\\
&+& \bar P_1^{\rm SLL}\left[C^{\rm SLL}_1(\mu_S)+C^{\rm SRR}_1(\mu_S)\right]
+\bar P_2^{\rm SLL}\left[C^{\rm SLL}_2(\mu_S)+C^{\rm SRR}_2(\mu_S)\right]
\nonumber
\end{eqnarray}
with $r=0.985$~\cite{BUJAWE} describing $\ord(\alpha_s)$ QCD
corrections to $S_0(x_t)$ in the SM and $\eta_B=0.55$. The factor
$f_s$ introduced in eq. (\ref{DMS0}) is given by $1+f_s\equiv
F^s_{tt}/S_0(x_t)$. Next
\begin{equation}\label{pab}
\bar P^a_i={P_i^a\over4\eta_B\hat B_{B_s}} 
\end{equation}
where $\hat B_{B_{s}}$ is the non-perturbative parameter related to
the matrix elements of $Q^{\rm VLL}$.  The coefficients $P_i^a$
include NLO QCD renormalization group factors
\cite{CET0,BUMIUR,BUJAUR} that sum up large logarithms between $\mu_S$
and $\mu_b=\ord(m_b)$ scales. Explicit formulae for $P_i^a$ in terms
of these QCD factors and non-perturbative parameters $B_i$ can be found
in~\cite{BUJAUR,BUCHROSL1}.

In~\cite{BUCHROSL1}, except for $B^{VLL}$, we have set the non-perturbative 
parameters $B_i$ in the B-system to unity as the results of lattice 
calculations were not available at that time. Meanwhile all these parameters 
have been calculated in \cite{BECIREVIC}. Translating these results into our 
operator basis by means of the formulae given in~\cite{BUJAUR} we find
\be
B^{VLL}=0.87,   \phantom{aa}
B^{LR}_1=1.75,  \phantom{aa}
B^{LR}_2=1.16,   \phantom{aa}
B^{SLL}_1=0.80, \phantom{aa}
B^{SLL}_2=0.71 
\ee
for $\mu_b=4.6$~GeV. This gives
\be
P_1^{LR}=-2.03, \quad P_2^{LR}=2.56,\quad P_1^{SLL}=-1.06,\quad
P_2^{SLL}=-2.05
\label{eqn:Piss}
\ee
and using (\ref{pab}) with $\hat B_{B_s}=1.3$ we find
\be
\bar P_1^{LR}=-0.71,   \phantom{aa}
\bar P_2^{LR}= 0.90,    \phantom{aa}
\bar P_1^{SLL}=-0.37,  \phantom{aa}
\bar P_2^{SLL}=-0.72.
\label{eqn:Pis}
\ee 
We observe substantial suppression of $4\bar P_1^{SLL}$ relative to
the vacuum insertion estimates used in \cite{BUCHROSL1,ISRE1} that
resulted in $-0.53$. The formula for $\varepsilon_K$ in terms of
$F^{\varepsilon}_{tt}$ as well as the expressions analogous to
(\ref{hds2}) for $F^{\varepsilon}_{tt}$ can be found in ref.
\cite{BUCHROSL1}.

In our numerical analysis of section \ref{sec:num} we have included
complete expressions for all the one loop box diagrams contributing to
Wilson coefficients $C_i(\mu_S)$ relevant for $B^0_{s,d}$-$\bar
B^0_{s,d}$ mass difference and $\varepsilon_K$ as well as the complete
expressions for the contribution of the double penguin diagrams shown
in fig. \ref{fig:2pg}. In the box diagrams involving charged Higgs and
Goldstone bosons we have used vertices with corrections
$\epsilon^{HL(R)}_{JI}$ and $\epsilon^{GL(R)}_{JI}$ calculated as in
sec. 2. Similarly, in the diagrams shown in fig. \ref{fig:2pg} we have
used full flavour changing vertices $\left[X^S_{RL}\right]^{JI}$,
$\left[X^S_{LR}\right]^{JI}$ as given in eqs. (\ref{eqn:XRL}),
(\ref{eqn:XLR}), (\ref{eqn:XLRbis}) and (\ref{eqn:XRLbis}). The
necessary formulae are collected in the Appendix A. In the
numerical calculation for the scale $\mu_S$ we have taken
$\mu_S=M_{H^+}$.

Let us now concentrate on the most important of new contributions to
the Wilson coefficients $C_i$ relevant for $\Delta M_s$.

\subsubsection{Double Penguin Diagrams}

As shown in~\cite{HAPOTO,BAKO,BUCHROSL1} in supersymmetry with large
$\tan\beta$ contribution of the double-penguin diagrams shown
in fig.~\ref{fig:2pg} to the Wilson coefficients of the operators
$Q_2^{\rm LR}$, $Q_1^{\rm SLL}$ and $Q_1^{\rm SRR}$ can be
substantial.  Their Wilson coefficients $C_i$ relevant for the 
$B^0_s$-$\bar B^0_s$ mixing are then determined from the
matching conditions:
\begin{eqnarray}\label{match}
{G_F^2M^2_W\over16\pi^2}
\left(V_{\rm eff}^{tb\ast}V_{\rm eff}^{ts}\right)^2 C_2^{\rm LR}=
i\sum_{S^0=h^0,H^0,A^0}{i\over-M^2_{S^0}}
(-i)^2\left[X^S_{RL}\right]^{bs}\left[X^S_{LR}\right]^{bs}
\end{eqnarray}
with $X^S_{RL}X^S_{LR}$ replaced by $X^S_{RL}X^S_{RL}$ and
$X^S_{LR}X^S_{LR}$ for $C_1^{\rm SLL}$ and $C_1^{\rm SRR}$,
respectively.  In these cases a combinatorial factor $1/2$ has to be
included on the r.h.s of (\ref{match}).  The contribution of the
neutral Goldstone boson $G^0$ can be neglected as $X^G_{LR}$ and
$X^G_{RL}$ vanish in the $SU(2)\times U(1)$ symmetry limit.  (See eqs.
(\ref{XRL}) and (\ref{XLR}).)

In what follows it will be convenient to introduce the ``reduced''
couplings $\left[\overline X^S_{RL}\right]^{JI}$ defined by
\be\label{RED}
\left[X^S_{RL}\right]^{bs}\equiv G_F^{3/2} 2^{7/4}
V_{\rm eff}^{tb\ast}V_{\rm eff}^{ts} \left[\overline X^S_{RL}\right]^{bs}
\ee
with the same definition for $\left[\overline X^S_{LR}\right]^{bs}$.

In the approximation of sec. \ref{sec:su2xu1}, using 
(\ref{BXRLFIN}) and (\ref{BXLRFIN}) we find for $J=3=b$, $I=2=s$
\be\label{XRLbar}
\left[\overline X^S_{RL}\right]^{bs}=
\frac{\overline m_b \overline m_t^2 
\epsilon_Y \tan\beta}{(1+\tilde\epsilon_3\tan\beta)
(1+\epsilon_0\tan\beta)} \left( x^S_u-x^S_d\tan\beta\right)
\ee
and
\be\label{XLRbar}
\left[\overline X^S_{LR}\right]^{bs}=
\frac{\overline m_s \overline m_t^2 
\epsilon_Y\tan\beta}{(1+\tilde\epsilon_3\tan\beta)
(1+\epsilon_0\tan\beta)}
\left(x^{S\ast}_u-x^{S\ast}_d\tan\beta\right)
\ee
where we used the approximations $1/\cos\beta\approx\tan\beta$,
$y^2_t\approx2\sqrt2G_F\overline m^2_t$ valid for $\tan\beta\gg1$.
$x^S_u$ and $x^S_d$ are defined in (\ref{xS}). Following the numerical 
analysis performed in sec. \ref{sec:comp}.2 reliable approximation to 
the results of the complete calculation requires, however, using the 
formulae of section 3.6 with the epsilon parameters extracted as in eq.
(\ref{eqn:epsfromdmd}). Still, the simple formulae (\ref{eqn:e}) used
in (\ref{XRLbar}), (\ref{XLRbar}) describe well the qualitative
behaviour of the double penguin contribution.

From (\ref{match}) and (\ref{RED}) we find then the double penguin
contributions to the Wilson coefficients in question:
\be\label{C2LR} 
C^{LR}_2=- \frac{G_F}{\sqrt{2}}\frac{(16\pi^2)^2}{M^2_W\pi^2}
\sum_{S^0=h^0,H^0,A^0}{1\over M^2_{S^0}} 
\left[\overline X^S_{RL}\right]^{bs}\left[\overline X^S_{LR}\right]^{bs}
\ee
and
\be\label{C1SLL} 
C^{SLL}_1=-\frac{1}{2} \frac{G_F}{\sqrt{2}}\frac{(16\pi^2)^2}{M^2_W\pi^2}
\sum_{S^0=h^0,H^0,A^0}{1\over M^2_{S^0}} 
\left[\overline X^S_{RL}\right]^{bs}\left[\overline X^S_{RL}\right]^{bs}
\ee
with $C^{SRR}_1$ obtained from $C^{SLL}_1$ by interchanging $L$ and $R$.

Using formulae (\ref{XRLbar}) and (\ref{XLRbar}) we find 
\be\label{XRLbarL}
\left[\overline X^S_{RL}\right]^{bs}=
\frac{\overline m_b \overline m_t^2 
\epsilon_Y \tan^2\beta}{(1+\tilde\epsilon_3\tan\beta)
(1+\epsilon_0\tan\beta)} 
\left[\sin(\alpha-\beta),\cos(\alpha-\beta),-i\right]
\ee
\be\label{XLRbarL}
\left[\overline X^S_{LR}\right]^{bs}=
\frac{\overline m_s \overline m_t^2 
\epsilon_Y \tan^2\beta}{(1+\tilde\epsilon_3\tan\beta)
(1+\epsilon_0\tan\beta)} 
\left[\sin(\alpha-\beta),\cos(\alpha-\beta),i\right]
\ee
for $S^0=\left[H^0,h^0,A^0\right]$. Consequently the dominant, ${\cal
  O}(\tan^4\beta)$, contributions to $C_2^{\rm LR}$, $C_1^{\rm SLL}$
and $C_1^{\rm LRR}$, that come solely from the double penguin diagrams
read
\begin{eqnarray}
C_2^{\rm LR}\approx
-{G_F{\overline m}_b{\overline m}_s{\overline m}_t^4\over\sqrt2\pi^2M^2_W}
{\tan^4\beta ~
\epsilon_Y^2 (16\pi^2)^2
\over(1+\tilde\epsilon_3\tan\beta)^2(1+\epsilon_0\tan\beta)^2}~
{\cal F}^+,
\label{C2LRA}
\end{eqnarray}
\begin{eqnarray}
C_1^{\rm SLL}\approx
-{G_F{\overline m}^2_b\overline{m}_t^4\over2\sqrt2\pi^2M^2_W}
{\tan^4\beta ~\epsilon_Y^2 (16\pi^2)^2
\over(1+\tilde\epsilon_3\tan\beta)^2(1+\epsilon_0\tan\beta)^2}~
{\cal F}^-,
\label{eqn:C1SLLA}
\end{eqnarray}
\begin{eqnarray}
C_1^{\rm SRR}=\frac{\overline m^2_s}{\overline m^2_b}C_1^{\rm SLL}.
\label{eqn:C1SRRA}
\end{eqnarray}
(Note that $\epsilon_Y$ as given by eq. (\ref{eqn:e}) is equal
$X_{Ct}/16\pi^2$ of ref.~\cite{BUCHROSL1} and
$G_F/\sqrt2=\pi\alpha_{\rm EM}/2s^2_WM^2_W$.) Here
\begin{eqnarray}
{\cal F}^\pm\equiv{\sin^2(\alpha-\beta)\over M^2_{H^0}}+
{\cos^2(\alpha-\beta)\over M^2_{h^0}}\pm{1\over M^2_{A^0}}
\approx{1\over M^2_{H^0}} \pm {1\over M^2_{A^0}}
\end{eqnarray}
where we have used the fact that in the MSSM for $\tan\beta\gg1$ and
$M_{A^0}>M_Z$ one has $\sin^2(\alpha-\beta)\approx1$ and
$\cos^2(\alpha-\beta)\approx0$. Because $M_{A^0}\approx M_{H^0}$, it
follows, that for large $\tan\beta$ the $H^0$ and $A^0$ contributions
to $C_1^{\rm SLL}$ and $C_1^{\rm SRR}$ cancel each other \cite{BAKO}
and the contribution of $h^0$ can be neglected. It is therefore the
coefficient $C_2^{\rm LR}$ that receives the largest contribution from
double penguin diagrams~\cite{BUCHROSL1}. Their contribution to the
Wilson coefficients relevant for the $B^0_d$-$\bar B^0_d$ mixing (for
$\varepsilon_K$) are given by the same formulae with $m_s$ replaced by
$m_d$ ($m_s$ replaced by $m_d$ and $m_b$ replaced by $m_s$). Our
result for $C_2^{\rm LR}$ agrees with the corrected version of eq.
(42) of~\cite{ISRE1} where the presence of additional $\tan\beta$
factors involving $\tilde\epsilon_3$ and $\epsilon_0$, not included in
\cite{BUCHROSL1} has been pointed out.

An important feature of the double penguin contribution to $C^{\rm
  LR}_2$, (and also to $C_1^{\rm SLL}$ and $C_1^{\rm SRR}$) is its
fixed negative sign that is the same as the sign of the contribution
of the charged Higgs boson box diagrams at large $\tan\beta$
\cite{BUCHROSL1}. Because the strong correction factor $\bar P_2^{\rm
  LR}$ in (\ref{hds2}) is positive the double penguin contribution
interferes destructively with the SM ones and leads to $1+f_s<1$.
Another interesting feature is its strong dependence on the left-right
mixing of the top squarks as $\epsilon_Y^2\propto A_t^2$.

Using the general formulae given in refs. \cite{BUCHROSL1,BUJAUR} and
collected in the preceding subsection we find the contribution of the
double penguin diagrams to $\Delta M_s$:
\be\label{DMDP1} 
(\Delta M_s)^{DP}=\frac{G_F^2 M_W^2}{24\pi^2} M_{B_s} F^2_{B_s}
|V^{\rm eff}_{ts}|^2 P_2^{LR} C_2^{LR}(\mu_S).
\ee
For $\mu_S=\ord(m_t)$ the factor $P_2^{LR}$ summarizing
renormalization group effects between $\mu_b\le \mu\le \mu_S$ and
including the relevant hadronic matrix elements is $P_2^{LR}\approx
2.5$.

Identifying $C_2^{\rm LR}$ in (\ref{C2LRA}) with $C_2^{LR}(\mu_S)$ we
find for large $\tan\beta$
\bea\label{DMDP2}
(\Delta M_s)^{DP}=-12.0/ps \times
  \Bigg[\frac{\tan\beta}{50}\Bigg]^4
  \Bigg[\frac{P_2^{LR}}{2.50}\Bigg]
  \Bigg[\frac{F_{B_s}}{230\ {\rm MeV}}\Bigg]^2 
  \Bigg[\frac{|V_{ts}|}{0.040}\Bigg]^2\phantom{aaaaaa}\nonumber\\
\times
\Bigg[\frac{\overline m_b(\mu_S)}{3.0 {\rm GeV}}\Bigg]
\Bigg[\frac{\overline m_s(\mu_S)}{0.06 {\rm GeV}}\Bigg]
\Bigg[\frac{\overline m_t^4(\mu_S)}{M_W^2 M^2_A}\Bigg] 
{(16\pi^2)^2\epsilon_Y^2\over
(1+\tilde\epsilon_3\tan\beta)^2(1+\epsilon_0\tan\beta)^2}  
\eea

\subsubsection{Charged Higgs Box Diagram Contributions}

The formulae for the charged Higgs boson $(H^\pm,G^\pm)$ box diagram
contributions to the Wilson coefficients of the effective Hamiltonians
describing $B^0_{s,d}$-$\bar B^0_{s,d}$ mixing have been presented in
section 4.1 of~\cite{BUCHROSL1} and have been confirmed
in~\cite{ISRE1}.  They do not, however take into account the
corrections to the $H^+$ couplings discussed in sections 2.5 and 3.4.
The corrections $\epsilon^{HL}_{JI}$ and $\epsilon^{HR}_{JI}$ can be
easily incorporated in those expressions of eqs. (4.4) and (4.5) of
\cite{BUCHROSL1} that contain a single $D_0$ or a single
$D_2$-function. To this end it suffices to make the following
replacements
\begin{eqnarray}
a^{JIk=1}_{L(R)}\rightarrow \left[P^H_{RL(LR)}\right]^{JI}/V_{\rm eff}^{JI}
\phantom{aaaaaa}
a^{JIk=2}_{L(R)}\rightarrow \left[P^G_{RL(LR)}\right]^{JI}/V_{\rm eff}^{JI}
\label{eqn:Hboxrule}
\end{eqnarray}
The two expressions in eqs. (4.4) and (4.5) of \cite{BUCHROSL1} which
contain three $D_2$-functions have been obtained by using the
unitarity of the CKM matrix. Therefore, in this case the replacement
(\ref{eqn:Hboxrule}) has to be supplemented by multiplication of the
second and third $D_2$-function in the square bracket in (4.4) of
\cite{BUCHROSL1} by $r$ and $r^2$, respectively and by $r^2$ and $r^4$
in (4.5) of \cite{BUCHROSL1} where
$r=(1+\tilde\epsilon_3\tan\beta)/(1+\tilde\epsilon_1\tan\beta)$ as 
follows from the formula (\ref{PH1}). For completeness we give the relevant 
expressions in the Appendix A.4.

In this section we would like to asses the magnitude of the charged
Higgs boson box diagram contribution to $\Delta M_s$ and $\Delta M_d$
with the corrections $\epsilon^{HL(R)}_{JI}$ included (as follows from
the numerical analysis of sec. 5, the impact of the corrections
$\epsilon^{GL(R)}_{JI}$ is negligible) and compare it with the
magnitude of the contribution of the double penguin diagrams.

As found in \cite{BUCHROSL1} the largest contribution to $\Delta M_s$
of the charged Higgs box diagrams, arises from the coefficient
$C_2^{LR}$.  This contribution is proportional to $\overline
m_s\overline m_b \tan^2\beta$ and its sign is opposite to the SM
contribution. The analogous contributions to $\Delta M_d$ and
$\varepsilon_K$ are much smaller being proportional to $\overline
m_d\overline m_b \tan^2\beta$ and $\overline m_d\overline m_s
\tan^2\beta$, respectively.

\begin{figure}[htbp]
\begin{center}
\epsfig{file=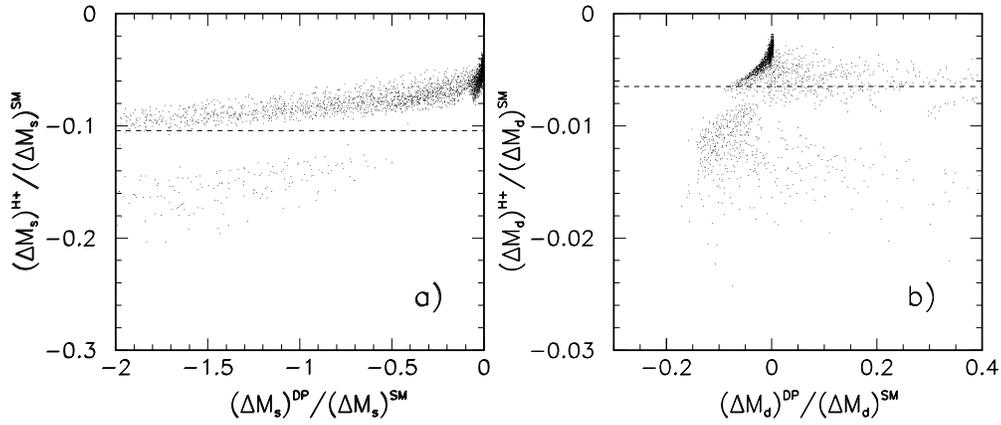,width=\linewidth}
\end{center}
\caption{Comparison of the double penguin and charged Higgs boson box 
  diagram contributions to the ratios $\Delta M_s/(\Delta M_s)^{\rm
    SM}$ (panel a) and $\Delta M_d/(\Delta M_d)^{\rm SM}$ (panel b)
  for $M_A=200$ GeV and $\tan\beta=50$. Only points giving acceptable
  $BR(\bar B\to X_s\gamma)$ are shown. Dashed lines show the ratios $(\Delta
  M_s)^{H^+}/(\Delta M_s)^{\rm SM}$ and $(\Delta M_d)^{H^+}/(\Delta
  M_d)^{\rm SM}$ in the absence of radiative corrections to the
  charged Higgs boson vertices.}
\label{fig:boxvsdp}
\end{figure}

An order of magnitude estimate of the charged Higgs boson box diagram 
contribution to $\Delta M_s$ for $M_{H^+}\approx \overline m_t$ can be  
obtained by taking into account only the contribution to $C_2^{LR}$ 
in (\ref{hds2}) of the box diagram with two $H^\pm$:
\be\label{DMSLR}
(\Delta M_s)^{LR}_{H^\pm}\approx-{G_F^2M^2_W\over 18\pi^2}M_{B_s}F^2_{B_s}
P_2^{\rm LR} {\overline m_b\overline m_s\over M^2_W}
{\tan^2\beta\over(1+\tilde\epsilon_3\tan\beta)^2}
\ee
where $P_2^{\rm LR}\approx2.56$ and the factor 
$(1+\tilde\epsilon_3\tan\beta)^2$ in the denominator comes from the 
rule (\ref{PH1}). One should remember, however, that the box diagram with 
the $W^\pm H^\mp$ exchange also gives the contribution $\propto\tan^2\beta$
and with the same sign as the one with $H^\pm H^\mp$ exchange 
\cite{BUCHROSL1}.

In figure \ref{fig:boxvsdp} we compare for our scan over the MSSM
parameter space the relative magnitude of the (complete) charged Higgs
boson contributions to $\Delta M_s$ and $\Delta M_d$ with the
contributions of the double penguin diagrams. We observe that in the
case of $\Delta M_s$ the contributions of the box-diagrams is much
smaller than that of the double penguin diagrams for most of the
points in the parameter space.  The latter contribution is always
negative and in a large portion of the parameter space exceeds also
the SM one. The upper (lower) branch of points seen in panel a of
figure \ref{fig:boxvsdp}a corresponds to $\mu>0$ ($\mu<0$).  In fact,
for points corresponding to unnaturally large $A_t$ combined with
negative values of $\mu$ leading to enhancement of the effective
couplings, the ratio $(\Delta M_s)^{\rm DP}/(\Delta M_s)^{\rm SM}$ can
reach values even as big as $-10$. The same points lead also to
$(\Delta M_d)^{\rm DP}/(\Delta M_d)^{\rm SM}\simlt-0.2$. They are
excluded as they lead to unacceptably big $BR(B_s^0\to\mu^+\mu^-)$ -
see sec.  \ref{sec:correl}. In contrast, the double penguin
contribution to $\Delta M_d$ can be also positive. This is because in
this case the suppression by $\overline m_b\overline m_d$ of the
Wilson coefficient $C_2^{\rm LR}$ is stronger than the suppression of
$C_1^{\rm SLL}$ by the factor ${\cal F^-}$ in eq. (\ref{eqn:C1SLLA}).
Since $\bar P^{\rm SLL}_1$ in (\ref{hds2}) is negative this leads to
positive contribution to $(\Delta M_d)^{\rm DP}/(\Delta M_d)^{\rm SM}$. 
As is seen from fig.~\ref{fig:boxvsdp} for some special values
of the MSSM parameters $(\Delta M_d)^{\rm DP}/(\Delta M_d)^{\rm SM}$
can reach $\sim\pm0.2$ but is much smaller in most of the parameter
space. The contribution of the box diagrams involving $H^\pm$ to
$\Delta M_d$ for large $\tan\beta$ is always negligible
compared to the SM and double penguin contributions.

\subsubsection{Chargino Box Diagram Contributions}

In the scenario considered in~\cite{BUCHROSL1} and in the present
paper supersymmetric particles are heavier than the Higgs bosons
and the chargino box contribution $(\Delta M_s)^{\chi^\pm}$ is small
as it is suppressed by $1/M^2_{\rm SUSY}$. We have checked that in our
scan, with all sparticles heavier than 500 GeV, the ratio $(\Delta
M_s)^{\chi^\pm}/(\Delta M_s)^{\rm SM}$ can reach only $0.02$ for
acceptable values of the $\bar B\to X_s\gamma$ rate. However, we find
some differences with the formulae derived in \cite{ISRE1} and we want
to devote this subsection to clarify this issue.\footnote{P.H.Ch.
  would like to thank Gino Isidori for the discussion which allowed to
  identify the source of the discrepancy.}

The contributions of box diagrams with chargino exchanges to $\Delta
M_{s,d}$ at large $\tan\beta$ have been calculated in
\cite{BUCHROSL1,ISRE1} but only in \cite{ISRE1} they have been
presented in details. These authors have found that the most important
contribution of these diagrams is present in the coefficient
$C_1^{SLL}$ that behaves as $\overline m_t^4\overline m_b^2
\tan^2\beta$ and contributes to $\Delta M_{s,d}$ with the same sign as
the SM contribution. As stressed in \cite{ISRE1}, if the sparticles
are light this contribution could be relevant also for $\Delta M_d$
due to the absence of the suppression factor $\overline m_d$. These
statements have to be revised, however.

First of all, let us notice that, as follows from the exact formulae for the 
chargino contributions collected in the Appendix A.4, the Wilson coefficient 
$C_2^{\rm SLL}$ in (\ref{eqn:heff_dF2}) has exactly the same dependence on 
the chargino couplings as does $C_1^{\rm SLL}$. In fact, the term in the 
chargino box amplitude which has that combination of the chargino couplings 
gives rise to a single scalar left-left operator which, however, has colour 
indices contracted differently than the operators in the basis (\ref{OPS}) 
used also in \cite{ISRE1}. Making the Fierz transformation of this operator 
gives rise to the ${\cal O}^{\rm SLL}_1$ and ${\cal O}^{\rm SLL}_2$ 
operators with the Wilson coefficients given in the Appendix A.4. The authors 
of ref. \cite{ISRE1} incorrectly identified the scalar left-left operator 
with wrong contraction of the colour indices with ${\cal O}^{\rm SLL}_1$. 
This has the following consequences. Firstly, our  $C_1^{\rm SLL}$ has an 
additional factor $-1/2$ which results from the Fierz transformation.
This reverses the sign of the contribution of $C_1^{\rm SLL}$ and decreases 
it by one half as compared to ref. \cite{ISRE1}. Secondly, 
$C_2^{\rm SLL}=-(1/4)C_1^{\rm SLL}$ also contributes to the mass difference 
and further reduces the effects of $C_1^{\rm SLL}$. More precisely, using 
(\ref{hds2}) and (\ref{pab}) the $\tan^2\beta$ contribution of charginos 
to (\ref{DMSTOT}) enters in the combination
\bea
P_1^{\rm SLL}C_1^{\rm SLL}(\mu_S)+P_2^{\rm SLL}C_2^{\rm SLL}(\mu_S)
\approx{1\over2}P_1^{\rm SLL}C_1^{\rm SLL}(\mu_S)\nonumber
\eea
where we have used the fact that $P_2^{\rm SLL}\approx2P_1^{\rm SLL}$
(see eq. (\ref{eqn:Piss})). This reduces the $\tan^2\beta$
contribution of charginos by roughly another $1/2$.

\begin{figure}[htbp]
\begin{center}
\epsfig{file=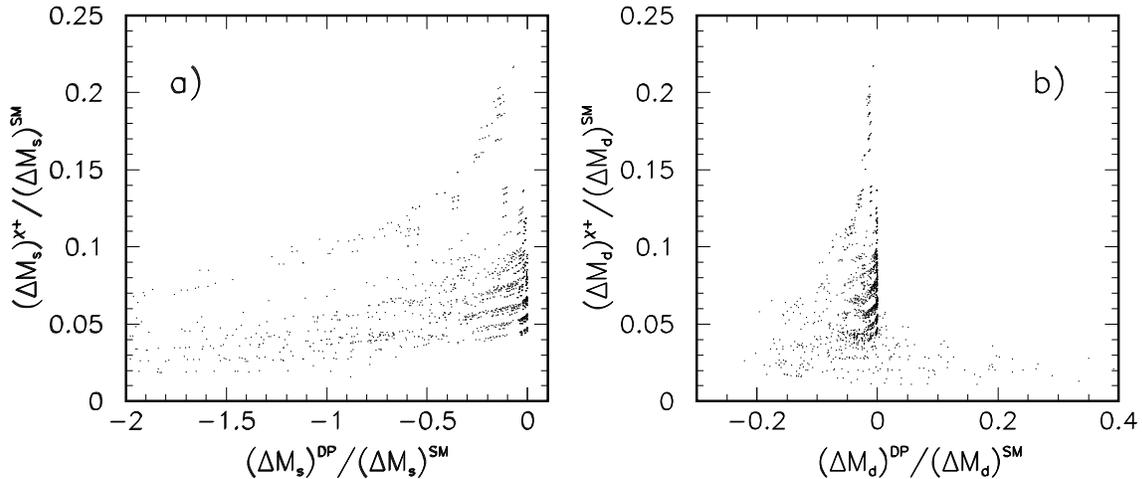,width=\linewidth}
\end{center}
\caption{Comparison of the double penguin and chargino box diagram 
  contributions to the ratios $\Delta M_s/(\Delta M_s)^{\rm SM}$
  (panel a) and $\Delta M_d/(\Delta M_d)^{\rm SM}$ (panel b) for 
  $M_A=200$ GeV and $\tan\beta=50$. Only points giving acceptable
  $BR(\bar B\to X_s\gamma)$ are shown. In this scan we allowed for 
  charginos and stop as light as 100 GeV.}
\label{fig:chvsdp}
\end{figure}

Thus, our dominant at large $\tan\beta$ part of the chargino
contribution to $\Delta M_{s,d}$ is effectively smaller by a factor of
approximately 4 compared to the one of ref. \cite{ISRE1} and has
opposite sign compared to the SM contribution. Neglecting the gauge
coupling constants as well as the Yukawa couplings other than top and
bottom one gets for the dominant chargino box diagram contribution
\begin{eqnarray}
C^{\rm SLL}_1(\mu_S)=-2
{\overline m^2_b\overline m^4_tA^2_t\over M^2_W M^4_{\tilde t_1}}
{\tan^2\beta\over(1+\tilde\epsilon_3\tan\beta)^2}
\sum_{i,j=1}^2Z_-^{2i}m_{C_i}Z_+^{2i}Z_-^{2j}m_{C_j}Z_+^{2j}\nonumber\\
\times\left[D_0(1,1)-2D_0(1,2) + D_0(2,2)\right]\phantom{aaaaaa}
\end{eqnarray}
where $\tilde t_1$ is the heavier stop and $D_0(l,k)$ stands for
$D_0(m_{C_i},m_{C_j},M_{\tilde t_k},M_{\tilde t_l})$. The epsilon
factor in the denominator appears here in agreement with the rule
(\ref{RRiN}).  With all these corrections we find that even for
$\tan\beta\sim50$ the contribution of charginos to $\Delta M_{s,d}$ is
in general dominated by the $C^{\rm VLL}$ Wilson coefficient which has
the same sign as the SM. Only for very big values of $A_t$ can the chargino
contribution to the Wilson coefficients of the scalar operators 
compete with the contribution of the standard VLL one.

In figure \ref{fig:chvsdp} we show the relative magnitude of the
chargino box and double penguin contributions to $\Delta M_s$ and
$\Delta M_d$ for a scan which included charginos and stop as light as
100 GeV. The biggest contribution of charginos allowed by $\bar B\to
X_s\gamma$ constraint is $\sim0.2$ of the SM one. If the $\bar B\to
X_s\gamma$ constraint is relaxed, the ratios 
$(\Delta M_s)^{\chi^\pm}/(\Delta
M_s)^{\rm SM}$ and $(\Delta M_d)^{\chi^\pm}/(\Delta M_d)^{\rm SM}$ can reach
$\sim0.3$.  The anti-correlation of the chargino and double penguin
contribution is clearly seen in figure \ref{fig:chvsdp} (it is even
better visible if the points with unacceptable $BR(\bar B\to
X_s\gamma)$ are retained).  The reason for it is simple: the biggest
double penguin contribution arises for small values of
$(1+\tilde\epsilon_3\tan\beta)$ when also the contribution the scalar
left-left operators generated by boxes is enhanced and cancels the
contribution of the ${\cal O}^{\rm VLL}$ generated by the chargino box
diagrams. The biggest total chargino box contribution arises when
$(1+\tilde\epsilon_3\tan\beta)$ is as big as possible to suppress
their negative contribution through the scalar left-left operators.
For $\mu<0$ this requires $A_t>0$ which is incompatible with $BR(\bar
B\to X_s\gamma)$ constraint. For $\mu>0$ the biggest values of
$(1+\tilde\epsilon_3\tan\beta)$ are obtained for $A_t<0$ which again
leads to unacceptable $\bar B\to X_s\gamma$ rate. (Cf. the signs of
$\epsilon_Y$ and $\epsilon_0$ in eqs. (\ref{eqn:e}).)

\subsection{The \boldmath{$B^0_{s,d}\to \mu^+\mu^-$} Decay}

The effective Hamiltonians describing these decays has the form
\be\label{eff:ham:bqll}
H_{\rm eff} =-\frac{2G_F}{\sqrt2}\frac{\alpha}{2\pi\sin^2\theta_W}
V_{tb}^{\rm eff*}V_{tq}^{\rm eff}
[c_A \Oi_A + c_A^\prime \Oi_A^\prime + c_S \Oi_S + 
c_S^\prime \Oi_S^\prime + c_P \Oi_P + c_P^\prime \Oi_P^\prime], 
\ee
where $q=s$ or $d$ and

\be\label{ops:bsll:av}
{\Oi}_A =(\overline{b_L} \gamma^\mu q_L) (\bar{l}\gamma_\mu \gamma_5 l),
\quad
{\Oi}'_A = (\overline{b_R} \gamma^\mu q_R) (\bar{l} 
\gamma_\mu\gamma_5 l),
\ee
\be\label{ops:bsll:s}
{\Oi}_S = m_b(\overline{b_R} q_L) (\bar{l}l),\quad
{\Oi}^\prime_S = m_q (\overline{b_L} q_R) (\bar{l}l),
\ee
\be\label{ops:bsll:p}
{\Oi}_P = m_b (\overline{b_R} q_L) (\bar{l}\gamma_5 l),\quad
{\Oi}^\prime_P =m_q (\overline{b_L} q_R) (\bar{l} \gamma_5 l).
\ee

For large $\tan\beta$ the contributions from neutral Higgs scalars
dominate.  Using the Lagrangian (\ref{LNEUTRAL}) for the $q=s$ case
and neglecting $m_s$ with respect to $m_b$ we find
\be\label{eff:Higgs}
{\cal H}_{\rm eff} = 
{G_F\over\sqrt2}\frac{4\pi\alpha}{M_W^2\sin^2\theta_W}
m_\mu V_{tb}^{\rm eff*}V_{tq}^{\rm eff}
\sum_{S=h^0,H^0,A^0}{1\over M^2_S} 
\left( \bar b_R\left[\overline X^S_{RL}\right]^{bs} s_L\right)
\left(\bar l [Z^S] l \right)
\ee
with the reduced coupling $\left[\overline X^S_{RL}\right]^{bs}$
defined in (\ref{RED}) and
\be
[Z^{H^0}]=-\frac{\cos\alpha}{\cos\beta}
 ,\qquad [Z^{h^0}]=\frac{\sin\alpha}{\cos\beta}
 ,\qquad [Z^{A^0}]=i\tan\beta \gamma_5~.
\ee

Comparing (\ref{eff:Higgs}) with (\ref{eff:ham:bqll}) and using
(\ref{XRLbar}) we find
\begin{eqnarray}\label{CS}
c_S\approx{m_\mu \overline m_t^2\over4M^2_W}
\frac{16\pi^2\epsilon_Y ~\tan^3\beta}
{(1+\tilde\epsilon_3\tan\beta)(1+\epsilon_0\tan\beta)}
\left[{\sin(\alpha-\beta)\cos\alpha\over M^2_{H^0}}
-{\cos(\alpha-\beta)\sin\alpha\over M^2_{h^0}} \right]~.
\end{eqnarray}

\begin{eqnarray}\label{CP}
c_P\approx
-\frac{m_\mu \overline m_t^2}{4 M^2_W}
\frac{16\pi^2\epsilon_Y ~\tan^3\beta}
{(1+\tilde\epsilon_3\tan\beta)(1+\epsilon_0\tan\beta)}
\left[{1\over M^2_{A^0}}\right]~.
\end{eqnarray}

The contributions of the operators ${\Oi}^\prime_S$ and
${\Oi}^\prime_P$ are strongly suppressed by $m_s$ with respect to
${\Oi}_S$ and ${\Oi}_P$.  Consequently following~\cite{BOBUKRUR} we
can write to an excellent approximation
\bea\label{BRbmumu}
  BR(B_s\to\mu^+\mu^-)&=&2.32\times 10^{-6}
\Bigg[\frac{\tau_{B_s}}{1.5\ ps}\Bigg]
  \Bigg[\frac{F_{B_s}}{230\ {\rm MeV}}\Bigg]^2 
  \Bigg[\frac{|V_{ts}|}{0.040}\Bigg]^2\nonumber\\
&\times&\left[|\tilde c_S|^2+|\tilde c_P+0.04(c_A-c_A^\prime)|^2\right] 
\eea
where $\tilde c_S$ and $\tilde c_P$ are the dimensionless Wilson 
coefficients
\be
\tilde c_S=M_{B_s} c_S, \qquad \tilde c_P=M_{B_s} c_P.
\ee
The coefficients $c_A$ and $c_A^\prime$ receive contributions from
$Z^0$-penguin diagrams and box diagrams and are weighted by the muon
mass.  For large $\tan\beta$ they can be safely neglected with respect
to $\tilde c_S$ and $\tilde c_P$ that grow like $\tan^3\beta$ and are
not chirally suppressed by the muon mass.

In the large $\tan\beta$ limit the contribution of $h^0$ to $c_S$ can
be neglected and setting $M^2_{H^0}\approx M^2_{A^0}$ we find from
(\ref{CS}) and (\ref{CP})
\be\label{CSCP}
c_S=c_P
\ee
with $c_P$ given in (\ref{CP}).
Consequently we find the branching ratio 
\bea\label{num:BRbmumu}
  BR(B_s\to\mu^+\mu^-)&=&3.5\times 10^{-5}
\Bigg[\frac{\tan\beta}{50}\Bigg]^6
\Bigg[\frac{\tau_{B_s}}{1.5\ ps}\Bigg]
  \Bigg[\frac{F_{B_s}}{230\ {\rm MeV}}\Bigg]^2 
  \Bigg[\frac{|V_{ts}|}{0.040}\Bigg]^2\nonumber\\
&\times&\frac{\overline m_t^4}{M^4_A} 
\frac{(16\pi^2\epsilon_Y)^2 }
{(1+\tilde\epsilon_3\tan\beta)^2(1+\epsilon_0\tan\beta)^2}.   
\eea
This result agrees with \cite{ISRE1}. Moreover one has
\be\label{Bdmumu}
\frac{BR(B^0_d\to\mu^+\mu^-)}{BR(B^0_s\to\mu^+\mu^-)} 
=\left[\frac{\tau_{B_d}}{\tau_{B_s}}\right]
  \left[\frac{F_{B_d}} {F_{B_s}}\right]^2 
  \left[\frac{|V_{td}^{\rm eff}|}{|V_{ts}^{\rm eff}|}\right]^2
\left[\frac{M_{B_d}} {M_{B_s}}\right]^5~
\ee
that is, the ratio of the branching fractions can depend on the SUSY
parameters only weakly through $|V_{td}^{\rm eff}/V_{ts}^{\rm eff}|$
which should be consistently determined from the unitarity triangle
analysis.

The presence of additional $\tan\beta$ dependence in the denominators
of eqs.~(\ref{DMDP2}) and~(\ref{num:BRbmumu}), not included
in~\cite{BUCHROSL1} and~\cite{BAKO,CHSL,BOEWKRUR,HULIYAZH}, has been
pointed out in~\cite{ISRE1}.  While we confirm these additional
factors, we would like to emphasize that depending on the sign of the
supersymmetric parameter $\mu$ they can suppress $\Delta M_s^{DP}$ and
$BR(B^0_s\to\mu^+\mu^-)$ relative to the estimates in the papers in
question, as stressed in~\cite{ISRE1}, but can also provide additional
enhancements.

\subsection{Correlation between \boldmath{$B^0_{s,d}\to\mu^+\mu^-$} and
\boldmath{$\Delta M_s^{DP}$}}
\label{sec:correl}

We are now in the position to give an explicit formula for a
correlation between the neutral Higgs contributions to
$B^0_s\to\mu^+\mu^-$ and $\Delta M_s^{DP}$ that we have pointed out
in~\cite{BUCHROSL1,BUCHROSL2}.  Indeed from (\ref{DMDP2}) and
(\ref{num:BRbmumu}) we find
\be\label{CORR}
{BR(B^0_s\to\mu^+\mu^-)}=\kappa ~10^{-6} 
\left[\frac{\tan\beta}{50}\right]^2
\left[\frac{200 {\rm GeV}}{M_{A^0}}\right]^2 
\left[\frac{|\Delta M_s^{DP}|}{2.12/ps}\right]
\ee
where
\be
\kappa=\left[\frac{2.50}{P_2^{LR}}\right]
\left[\frac{3.0 {\rm GeV}}{\overline m_b(\mu_S)}\right]
\left[\frac{0.06 {\rm GeV}}{\overline m_s(\mu_S)}\right]
\left[\frac{\tau_{B_s}}{1.5\ ps}\right]\approx 1~.
\ee
This relation depends sensitively on $M_{A^0}$ and $\tan\beta$ but it does 
not depend on $\epsilon_0$ and $\tilde\epsilon_3$. From~(\ref{Bdmumu}) a 
similar correlation between $BR(B^0_d\to\mu^+\mu^-)$ and $\Delta M_s^{DP}$ 
follows. If the flavour dependence in the epsilon parameters is taken into 
account as in section~\ref{sec:su2xu1flav} and formulae (\ref{fXRL}) and 
(\ref{fXLR}) are used, the r.h.s of (\ref{CORR}) is multiplied by the factor
$\epsilon^{(32)}/(\epsilon^{(23)}r_{32})$ which, however, does not depart 
significantly from unity. In our numerical analysis of section 7 we use 
the formulae of the full approach anyway.

In order to understand these results better, let us now assume that
$\Delta M_s$ has been measured and that appropriate supersymmetric
parameters can be found for which the MSSM considered here agrees with
$(\Delta M_s)^{\rm exp}$.  If $0<(1+f_s)<1$ this implies $(\Delta
M_s)^{\rm exp}<(\Delta M_s)^{\rm SM}$.  Then combining~(\ref{DMS0})
and~(\ref{CORR}) we find \cite{BUCHROSL2}
\bea
\label{CORRMAIN}
BR(B^0_s\to\mu^+\mu^-)&=&8.5\cdot 10^{-6} \kappa
\left[\frac{\tan\beta}{50}\right]^2 \left[\frac{200 {\rm
      GeV}}{M_{A^0}}\right]^2 \left[\frac{(\Delta M_s)^{\rm
      SM}}{18.0/ps}\right]
\nonumber\\
&\times& \left[1\mp \frac{(\Delta M_s)^{\rm exp}}{(\Delta M_s)^{\rm
      SM}} -\frac{|(\Delta M_s)^{H^\pm}|}{(\Delta M_s)^{\rm SM}}+
  \frac{(\Delta M_s)^{\chi^\pm}}{(\Delta M_s)^{\rm SM}}\right].
\eea 
with ``$\mp$'' corresponding to $0<(1+f_s)<1$ and $(1+f_s)<0$,
respectively.  Using~(\ref{Bdmumu}) analogous expression for
$BR(B^0_d\to\mu^+\mu^-)$ can be found. In writing~(\ref{CORRMAIN}) we
have taken into account that $(\Delta M_s)^{\rm DP}$ is always
negative and that for large $\tan\beta$ $(\Delta M_s)^{H^\pm}$ is
negative and $(\Delta M_s)^{\chi^\pm}$ mostly positive.
Formula~(\ref{CORRMAIN}) is valid provided the expression in square
brackets is positive and larger than $10^{-3}$.  Otherwise, other
contributions, in particular those coming from $Z^0$-penguins have to
be taken into account.  In our numerical analysis we take them into
account anyway.

Formula~(\ref{CORRMAIN}) demonstrates very clearly that if $(\Delta
M_s)^{\rm exp}$ will turn out to be close or larger than the SM value,
the order of magnitude enhancements of $BR(B^0_{s,d}\to\mu^+\mu^-)$ in
the scenario of the MSSM considered here with $0<(1+f_s)<1$ will be
excluded.  On the other hand large enhancements of
$BR(B^0_{s,d}\to\mu^+\mu^-)$ are in principle still possible if the
double-penguin contribution is so large that $(1+f_s)<0$ and the "$+$"
sign in~(\ref{CORRMAIN}) applies. For $\tan\beta<50$ obtaining
$(1+f_s)<0$ and the right magnitude of $\Delta M_s$ requires $\mu<0$
so that the couplings~(\ref{BXRLFIN}) are enhanced by the
$\epsilon$-factors in the denominator.  $\mu<0$ is excluded in
particular scenarios like minimal SUGRA, in which the sign of $A_t$ is
fixed and $\mu<0$ does not allow for satisfying the $\bar B\rightarrow
X_s \gamma$ constraint~\cite{CAGANIWA2}, but cannot be excluded in
general.

We will analyze (\ref{CORRMAIN}) numerically in detail in section 7.3.

\subsection{Unitarity Triangle}

The rate of the $B^0_d\to\mu^+\mu^-$ decay depends on the element
$V^{\rm eff}_{td}$ of the low energy CKM matrix. The value of $V^{\rm
  eff}_{td}$ cannot be extracted by using exclusively tree level
dominated processes and requires an analysis of the combination of
data on $|V_{ub}^{\rm eff}/V_{cb}^{\rm eff}|$, $\sin2\beta_{\rm ut}$
from the CP asymmetry $a_{\psi K_S}$, $\varepsilon_K$, $\Delta M_d$
and $\Delta M_s$ (the so-called UT analysis). While in the scenario of
the MSSM we are considering the extraction of the first two quantities
from the data is not affected by the new particles, $\varepsilon_K$,
$\Delta M_d$ and $\Delta M_s$ receive additional contributions which
can modify the extracted value of $V^{\rm eff}_{td}$.  The unitarity
triangle in the MSSM with large $\tan\beta$ has been discussed in
refs. \cite{BUCHROSL1,CHRO}. Here we recall this discussion briefly
and outline the procedure we use in sec. \ref{sec:num}.3 in order to
consistently determine $V^{\rm eff}_{td}$ for a given set of the MSSM
parameters.

Let us first recall that the contribution of the new particles enter 
$\varepsilon_K$, $\Delta M_d$ and $\Delta M_s$ through the factors
\be
F_{tt}^\varepsilon=S_0(x_t)[1+f_\varepsilon], \qquad
F_{tt}^d=S_0(x_t)[1+f_d], \qquad
F_{tt}^s=S_0(x_t)[1+f_s]
\ee
respectively where $S_0(x_t)$ is the universal contribution of the SM
particles. $F_{tt}^s$ is explicitly given in eq. (\ref{hds2}). The
hierarchy of the new contributions to the parameters $f_i$ is then as
follows:
\begin{itemize}
\item The contributions to $f_\varepsilon$ at large $\tan\beta$
  generated by the diagrams similar to the ones shown in figure
  \ref{fig:2pg} are fully negligible being strongly suppressed by
  $\overline m_s\overline m_d$, $\overline m_s^2$ and higher order
  terms in $\overline m_{d,s}$. Also the charged Higgs and chargino 
  box diagram contributions are
  known to be negligible for $\tan\beta\gg1$.  Consequently
  $f_\varepsilon$ can be set to zero and the constraint on the UT
  stemming from $\varepsilon_K$ is the same as in the SM.
\item The double penguin and charged Higgs boson contributions to
  $\Delta M_d$ are proportional to $\overline m_d\overline
  m_b\tan^4\beta$ and $\overline m_d\overline m_b\tan^2\beta$,
  respectively \cite{BUCHROSL1,ISRE1}. Detailed calculation shows that
  $|f_d|$ is at least one order of magnitude smaller than $|f_s|$.
  The latter, in the most likely scenario with $0<1+f_s<1$, is in turn
  bounded by $(\Delta M_s)_{\rm exp}$ to be smaller than $0.5$
  \cite{BUCHROSL1}.  Consequently $|f_d|\simlt0.05$ and can be safely
  neglected in view of the uncertainties in $F_{B_d}\sqrt{\hat B_{B_d}}$. 
  In the unlikely scenario in which $1+f_s<0$ and
  $|f_s|>1.5$, $|f_d|$ could reach $0.2$ and having negative sign
  would suppress $\Delta M_d$ relative to the SM prediction.  For
  fixed $F_{B_d}\sqrt{\hat B_{B_d}}$ and $(\Delta M_d)_{\rm exp}$,
  this would imply a larger $|V_{td}^{\rm eff}|$.
\item The new contributions to $f_s$ can be substantial
  \cite{BUCHROSL1,BUCHROSL2}.  They will be investigated numerically
  in detail in section 7 using the approach developed in sec. 2 which
  takes into account the resumation of leading higher order terms
  pointed out in \cite{ISRE1}. In our scenario $f_s$ is always
  negative and suppresses $\Delta M_s$ with respect to the
  expectations based on the SM.
\end{itemize}

In order to determine the impact of $f_s\neq0$ on the determination of
$|V_{td}^{\rm eff}|$ let us recall that 
\be\label{dmd}
\Delta M_d\propto F^2_{B_d}\hat B_{B_d}|V_{td}^{\rm eff}|^2 S_0(x_t)
\ee
\be\label{dms}
\Delta M_s\propto F^2_{B_s}\hat B_{B_s}|V_{ts}^{\rm eff}|^2 S_0(x_t)|1+f_s|
\ee
and 
\be\label{Rt}
|V_{td}^{\rm eff}|\propto\xi\sqrt{{\Delta M_d\over\Delta M_s}}\sqrt{|1+f_s|}, 
\phantom{aaa}{\rm where}\phantom{aaa}
\xi=\frac{F_{B_s}\sqrt{\hat B_{B_s}}}{F_{B_d}\sqrt{\hat B_{B_d}}}
\ee
Exact expressions can be found in \cite{BUCHROSL1,BUER}.

If the value of $F^2_{B_d}\hat B_{B_d}$ was known precisely
$|V_{td}^{\rm eff}|$ could be determined directly from eq. (\ref{dmd})
and the value of $f_s$ would not have any impact on it. In our
scenario $|V_{td}^{\rm eff}|$ would then assume the same value as in
the SM. Since $|V_{ts}^{\rm eff}|$ is almost fixed by the unitarity of
CKM matrix the eq.  (\ref{dms}) would then only provide a constraint
on the MSSM parameters space, the more stringent, the smaller was the
uncertainty in $F^2_{B_s}\hat B_{B_s}$ \cite{BUCHROSL1}. In other
words, a precise value of $F^2_{B_s}\hat B_{B_s}$ combined with a
precise measurement of $\Delta M_s$ could directly probe by means of
the formula (\ref{dms}) the value of $|1+f_s|$.

At present, however, $F^2_{B_d}\hat B_{B_d}$ (and also 
$F^2_{B_s}\hat B_{B_s}$) is poorly known and a more precise value of 
$|V_{td}^{\rm eff}|$ is obtained by using the relation (\ref{Rt}) in 
which $\xi$, calculated by QCD sum rules or lattice methods that are 
insensitive to new short distance contributions, is known with much better 
accuracy than are $F^2_{B_d}\hat B_{B_d}$ and $F^2_{B_s}\hat B_{B_s}$
separately. This means that $F^2_{B_d}\hat B_{B_d}$ and 
$F^2_{B_s}\hat B_{B_s}$ are positively correlated and cannot simultaneously 
assume values from the opposite extremes of their respective allowed ranges. 
Therefore, as seen in (\ref{Rt}) $0<1+f_s<1$ leads to a smaller value of 
$|V_{td}^{\rm eff}|$ and smaller angle $\gamma$ than does $f_s=0$. For 
example, if the value of $\Delta M_s$ is close to its present experimental 
bound, $|1+f_s|\approx0.6$ gives $|V_{td}^{\rm eff}|\approx7\times10^{-3}$.
This can be also viewed as follows: for a given experimental value of 
$\Delta M_s$, the factor $F^2_{B_s}\hat B_{B_s}$ must be larger if 
$0<|1+f_s|<1$ than if $f_s=0$. With $\xi$ fixed, this means that also 
$F^2_{B_d}\hat B_{B_d}$ must be larger implying a smaller value of 
$|V_{td}^{\rm eff}|$ and of the angle $\gamma$. Similarly, $1+f_s<0$ leads
to bigger $|V_{td}^{\rm eff}|$ and bigger $\gamma$ provided $|1+f_s|>1$
For example if $1+f_s\approx-1.3$ one has 
$|V_{td}^{\rm eff}|\approx1\times10^{-2}$ \cite{CHRO}.

In our global scans in sec. \ref{sec:num} we proceed as in
\cite{BUCHROSL2,CHRO}: For a given set of the MSSM parameters we scan
over the Wolfenstein parameters $\lambda$, $A$, $\bar\rho$ and
$\bar\eta$ which parametrize $V^{\rm eff}$ (see e.g.\cite{BUER}) as
well as over the non-perturbative parameters $F^2_{B_d}\hat B_{B_d}$
and $F^2_{B_s}\hat B_{B_s}$ in their respective ranges specified in
ref. \cite{BUPAST} and compute the quantities of interest only for
those $\lambda$, $A$, $\bar\rho$ and $\bar\eta$ for which
$\varepsilon_K$, $\Delta M_d$, $a_{\psi K_S}=\sin2\beta_{\rm ut}$,
$|V_{ub}^{\rm eff}/V_{cb}^{\rm eff}|$ and $\xi$ assume acceptable
values. This allows to take into account in $BR(B^0_d\to\mu^+\mu^-)$
also the small effects related to the dependence of $|V_{td}^{\rm eff}|$ 
on the supersymmetric parameters.

\subsection{\boldmath{$\bar B\rightarrow X_s\gamma$}}
\label{sec:bsg}

Finally we recall certain aspects of the $BR(\bar B\to X_s\gamma)$
calculation referring frequently to the formulae present in the
literature.  We would like to present for completeness simple recipes
allowing to use the corrections $\epsilon^{HL}$ etc. introduced in
sec.~2 in the existing formulae for the relevant Wilson coefficients.
We also show how to include the recently found correction depending on
the neutral Higgs boson masses \cite{AMGIISST} and on the couplings
$[X_{LR}]^{JI}$ in these formulae.

As far as the SM part of the computation is concerned, we closely
follow the approach of ref.~\cite{GAMI}. The two loop contribution of
$H^+$ is taken from ref.~\cite{CIDEGAGI1} (see also~\cite{CIROST}). As
in ref.  \cite{DEGAGI}, for our scenario with heavy sparticles we
include only one loop contribution of the supersymmetric particles as
given in refs.~\cite{BAGI,CIDEGAGI2}.

\subsubsection{Charged Goldstone and $H^+$ Contributions}

In order to see how to include our corrections in the calculation of
the $\bar B\to X_s\gamma$ rate let us recall that at large $\tan\beta$
the one-loop contributions of $G^+(H^+)$ to the Wilson coefficients
$C_{7(8)}^{(0)}$ are proportional to the product of the $\bar t_L
G^+(H^+)b_R$ vertex and the Hermitian conjugate of the $\bar t_R
G^+(H^+)s_L$ vertex. Neglecting the contributions originating in the
chiral flip on the external $b$ quark line and not including yet the
large $\tan\beta$ enhanced corrections one has
\be 
C^{(G^\pm)}_{7(8)}=-F_{7(8)}^{(2)}(x_t), \qquad
C^{(H^\pm)}_{7(8)}=F_{7(8)}^{(2)}(y_t)~.
\ee
Here $x_t=\overline m^2_t/M_W^2$, $y_t=\overline m^2_t/M_{H^+}^2$.
Combining the $G^+$ and $W^+$ contributions results in replacing
$-F_{7(8)}^{(2)}(x_t)$ by $F_{7(8)}^{(1)}(x_t)$. The explicit
expressions for $F_{7(8)}^{(1)}(x_t)$ and $F_{7(8)}^{(2)}(x_t)$ 
can be found in~\cite{DEGAGI}.

Let us consider the $G^+$ contribution first.  As we have demonstrated
in sec. 3, in the $SU(2)\times U(1)$ symmetry limit 
$\epsilon^{GL}_{JI}=\epsilon^{GR}_{JI}=0$ and the couplings of $G^+$
have the same form (\ref{eqn:SMGcoupl}) as in the SM. This means that
in the $SU(2)\times U(1)$ symmetry limit this contribution is
correctly accounted for in the SM contribution to $C_{7(8)}^{(0)}$.
Small departures of $\epsilon^{GR}_{JI}$ and $\epsilon^{GL}_{JI}$ from
zero which arise beyond this approximation can be also included but
our numerical studies of sec. 5 show that they are too small to have
any impact on the $\bar B\rightarrow X_s\gamma$ rate. Nevertheless, we can
include them by noting that the full contribution $G^+$ to the Wilson
coefficients is
\begin{eqnarray}\label{C78G}
C_{7(8)}^{(G^+)}=
-(1+\epsilon^{GL}_{ts})(1+\epsilon^{GR}_{tb})F_{7(8)}^{(2)}(x_t).
\end{eqnarray}
As for $\epsilon^{GL}_{ts}=\epsilon^{GR}_{tb}=0$ this contribution is
taken already into account in the SM contribution, one has only to
modify $C_{7(8)}^{(G^+)}$ by
\begin{eqnarray}
\delta^{G^+}C_{7(8)}^{(0)}=-(\epsilon^{GR}_{tb}+\epsilon^{GL}_{ts}
+\epsilon^{GR}_{tb}\epsilon^{GL}_{ts})F_{7(8)}^{(2)}(x_t)\approx
-(\epsilon^{GR}_{tb}+\epsilon^{GL}_{ts})F_{7(8)}^{(2)}(x_t)
\label{eqn:Gplus}
\end{eqnarray}
where we have neglected the higher order term. The prescription
(\ref{eqn:Gplus}) replaces in our approach the recipe given in eq.
(18) of ref. \cite{DEGAGI}.  It formally agrees with the latter if one
drops $\epsilon^{GL}_{ts}$ not considered in \cite{DEGAGI} and
identifies $V_{ts}$ in the $\bar t_R G^+s_L$ vertex used in
\cite{DEGAGI} with $V_{ts}^{\rm eff}$ (see our discussion in section
4) and recalls that $V_{tb}=V_{tb}^{\rm eff}$.  Using then the
substitution rule (\ref{REL3}) in the opposite direction one finds
\begin{eqnarray}
-\epsilon^{GR}_{tb}\rightarrow 1-
{1+\epsilon^\prime_b(t)\tan\beta\over1+\epsilon_b\tan\beta}=
{[\epsilon_b-\epsilon_b^\prime(t)]\tan\beta\over
1+\epsilon_b\tan\beta}
\end{eqnarray}
as in eq. (18) of~\cite{DEGAGI}. We have stressed however, that one
has to be careful in evaluating the corrections $\epsilon^{GR}_{tb}$
and $\epsilon^{GL}_{ts}$, because uncontrolled approximation can
result in overestimating their impact on the $\bar B\to X_s\gamma$
rate.

In the case of $H^+$ contribution, the formula analogous to
(\ref{C78G}) reads
\begin{eqnarray}
C_{7(8)}^{(H^+)}=\left(1-\epsilon^{HL}_{ts}
\right)\left(1-\epsilon^{HR}_{tb}\right)F_{7(8)}^{(2)}(y_t).
\label{eqn:Hplus}
\end{eqnarray}
In the approximation of section \ref{sec:su2xu1}, using eqs. 
(\ref{F1}) and (\ref{PH1}), we have:
\begin{eqnarray}
\left(1-\epsilon^{HL}_{ts}\right)\left(1-\epsilon^{HR}_{tb}\right)
={1\over1+\tilde\epsilon_3\tan\beta}
\left[1-\epsilon_0^\prime\tan\beta+
y^2_by^2_t \frac{\epsilon_Y\epsilon_Y^\prime \tan^2\beta}
{1+\epsilon_0\tan\beta}\right]
\label{eqn:HplusF}
\end{eqnarray}
in agreement with the factor multiplying $F_{7(8)}^{(2)}(y_t)$ in the
formula (87) of \cite{AMGIISST}. However as shown in section \ref{sec:comp}, 
the approximation (\ref{F1A}) is not numerically very accurate and better 
approximation is obtained by using for $\epsilon^{HL}_{ts}$ eq. 
(\ref{eqn:approximation}) with $J=3$, $I=2$ and 
$\epsilon^{HR}_{tb}=\tilde\epsilon_3\tan\beta/(1+\tan\beta)$ obtained from 
(\ref{PH4}) with $\tilde\epsilon_3$ calculated by means of 
(\ref{eqn:epsfromdmd}). Thus, the effect of corrections to the charged 
Higgs boson vertices results in the following modification of the Wilson
coefficients $C_{7(8)}^{(0)}$:
\begin{eqnarray}\label{dC78H}
\delta^{H^+}C_{7(8)}^{(0)}=\left(-\epsilon^{HL}_{ts}
-\epsilon^{HR}_{tb}+\epsilon^{HL}_{ts}
\epsilon^{HR}_{tb}\right)F_{7(8)}^{(2)}(y_t).
\end{eqnarray}
Sticking to the naive approximation of sec. \ref{sec:su2xu1} with the
$\Delta_{JI}$ term in (\ref{F1}) neglected, i.e. inserting in
(\ref{dC78H})
\be
\epsilon_{ts}^{HL}=\epsilon^\prime_t(s)\tan\beta, \qquad
\epsilon_{tb}^{HR}=\frac{\epsilon_b\tan\beta}{1+\epsilon_b\tan\beta}
\ee
one recovers the rule (19) of ref.~\cite{DEGAGI} if, in agreement with
our discussion of sec. 4, one identifies $V_{ts}$ with $V_{ts}^{\rm
  eff}$ in the $\bar t_R H^+s_L$ vertex of \cite{DEGAGI}.

\subsubsection{Chargino Contributions}

The inclusion of the $\tan\beta$ enhanced corrections in the chargino
contributions to Wilson coefficients of higher dimensional operators
involving down quark fields of arbitrary flavour is achieved by means
of the rules (\ref{RLiNew}) and (\ref{RRiNew}) or (\ref{RRiN}). For
the couplings $\tilde t^\dagger c^i_L s_L$ (\ref{RLiNew}) amounts
simply to the replacements
\be
m_t \rightarrow \overline m_t, \qquad V_{ts}\rightarrow V_{ts}^{\rm eff}.
\ee
For the couplings $\tilde t^\dagger c^i_R b_R$ the rule (\ref{RRiNew})
or (\ref{RRiN}) applies implying
\be
V_{tb} m_b \rightarrow V^{\rm eff}_{tb} 
{\overline m_b\over1+\tilde\epsilon_3\tan\beta}~.
\ee
where following the refinements of sec. \ref{sec:comp} 
$\tilde\epsilon_3\equiv\epsilon_0^{(3)}+y_t^2\epsilon_Y^{(33)}$. This
is accidentally equivalent to expressing in the tree level formulae
simply $V_{tb}$ and $m_b$ through $V^{\rm eff}_{tb}$ and $\overline m_b$, 
respectively. The formula~(30) of ref.~\cite{DEGAGI} with
$\epsilon_b=\epsilon_{d_3}\equiv\epsilon_0^{(3)}+y_t^2\epsilon_Y^{(33)}$
is consistent with this rule rule provided $V_{ts}$ and $V_{tb}$ in
this formula are identified with the effective CKM elements.

On the other hand for the Wilson coefficient of the $m_s/m_b$ suppressed 
operator $ m_s \bar s_R \sigma^{\mu\nu} b_L$ the situation is more 
involved as our rules imply
\be
m_t \rightarrow \overline m_t, \qquad V_{tb}\rightarrow V_{tb}^{\rm eff}
\ee
for the vertex involving the $b_L$ quark and
\be\label{xxx}
V_{ts} m_s \rightarrow V^{\rm eff}_{ts} 
\frac{\overline m_s}{1+\tilde\epsilon_3\tan\beta}~,
\ee
for the vertex involving the $s_R$ quark. If the flavour dependent effects
are important $1/(1+\tilde\epsilon_3\tan\beta)$ in (\ref{xxx}) should be 
replaced by the square bracket of the formula (\ref{PH6}) with $I=2$. 
Note that expressing in the tree level formulae simply $V_{ts}$ and $m_s$ 
through 
$V^{\rm eff}_{ts}$ and $\overline m_s$, respectively would give wrong 
$\tan\beta$ dependence.

\subsubsection{Neutral Higgs boson contribution}

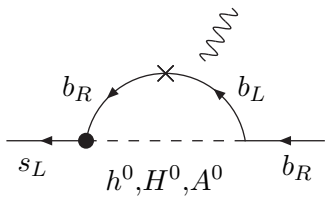
\begin{figure}[htbp]
\begin{center}
\begin{picture}(140,80)(0,0)
\ArrowLine(40,20)(10,20)
\ArrowLine(130,20)(100,20)
\ArrowArc(70,15)(30.4,9,90)
\ArrowArc(70,15)(30.4,90,175)
\DashLine(40,20)(100,20){4}
\Line(67,42.4)(73,48.4)
\Line(67,48.4)(73,42.4)
\Line(67,42.4)(73,48.4)
\Line(67,48.4)(73,42.4)
\Line(67,42.4)(73,48.4)
\Line(67,48.4)(73,42.4)
\Vertex(40,20){3}
\Photon(85,50)(95,70){3}{4}
\Text(70,5)[]{\small $h^0$,$H^0$,$A^0$}
\Text(120,10)[]{\small $b_R$}
\Text(20,10)[]{\small $s_L$}
\Text(37,40)[]{\small $b_R$}
\Text(103,40)[]{\small $b_L$}
\end{picture}
\end{center}
\caption{Additional diagram contributing to $b\to s\gamma$ or $b\to s g$
  transition. The dot and the cross denote the flavour changing
  coupling $\left[X_{LR}\right]^{sb}$ and the helicity flip,
  respectively.}
\label{fig:Hbsg}
\end{figure}

For completeness we include the neutral Higgs contribution to the
Wilson coefficients found in ref. \cite{AMGIISST}.  It is the natural
consequence of the flavour changing neutral Higgs coupling
$\left[X_{LR}\right]^{23}$ which, through the diagram shown in figure
\ref{fig:Hbsg}, leads to the contribution \cite{AMGIISST}
\bea
&&\delta^{H^0}C_{7(8)}^{(0)}={1\over18}{M^2_W\over g^2_2}\sum_{S=A^0,H^0,h^0}
\left[X^S_{LR}\right]^{23}\left[X^S\right]^{33}{\overline m_b^2\over M^2_S}
\nonumber\\
&&\phantom{aaaaaaa}\approx
-{1\over36}{\epsilon_Yy^2_t\tan^3\beta\over(1+\epsilon_0\tan\beta)
(1+\tilde\epsilon_3\tan\beta)^2}{\overline m_b^2\over M^2_A}
\eea
where $\left[X^S\right]^{33}$ and $\left[X^S_{LR}\right]^{23}$ are
given in eqs. (\ref{DIAGF}) and (\ref{BXLRFIN}), respectively.

The last comment concerns the dependence of the $\bar B\rightarrow
X_s\gamma$ amplitude on the CKM matrix elements.  In our scans
presented in sec. \ref{sec:num} for a given set of the MSSM parameters
we use the value of $|V^{\rm eff\ast}_{ts}V^{\rm eff}_{tb}|$
determined consistently from the UT analysis as described in the
preceding subsection.  We do not need therefore to implement the
recipe of ref. \cite{CIDEGAGI2} for correcting the $\bar B\rightarrow
X_s\gamma$ amplitude for new physics effects in $V^{\rm eff}_{ts}$.


\section{Numerical Analysis}
\label{sec:num}

In this section we present numerical analysis of the dependence of
$\Delta M_{d,s}$ and $B^0_{s,d}\to \mu^+\mu^-$ on the parameters of
the MSSM. We will also present the global analysis of these quantities
taking into account available experimental constraints, in particular
the one from the measured rate of the $B\to X_s\gamma$ decay. We
present the results based on our complete approach of section 2 which
includes automatically the $SU(2)\times U(1)$ breaking corrections as
well as the dependence of the flavour changing couplings on the
electroweak gauge couplings. On some plots we compare these results
with the one obtained by using the approximation of sec. 3 based on
$SU(2)\times U(1)$ limit and dominance of $\alpha_s$ and the top and
bottom Yukawa couplings.  The latter describe qualitatively the main
features of the MSSM effects but are not very accurate.

\subsection{The Size of \boldmath{$f_s$} and \boldmath{$f_d$}}

The parameters $f_s$ and $f_d$ introduced in eq. (\ref{DMS0}) and
directly related to the ratio $\Delta M_{s,d}/(\Delta M_{s,d})^{\rm
  SM}$:
\bea
\Delta M_{s,d}/(\Delta M_{s,d})^{\rm SM}\equiv|1+f_{s,d}|\nonumber
\eea
receive contributions from double penguins, charged Higgs boson box
diagrams and chargino box diagrams. As we have already said in sec.
6.1.4, for sparticles heavier than 500 GeV the latter contributions
can be safely neglected. We have also shown in sec. 6.1.3 that for
large $\tan\beta$ the contribution of the $H^\pm$ box diagrams to $f_s$ is
negative and can give up to $\delta^Hf_s\approx-0.1$ for $\mu>0$ and
up to $\delta^Hf_s\approx-0.2$ for $\mu<0$. Except for very small
values of the $A_t$ parameter, for which the double penguin diagram
contribution is small, the charged Higgs boson box diagram contributions are
much smaller than the latter one and do not play any role in the
correlation between $\Delta M_s$ and $BR(B^0_{s,d}\to\mu^+\mu^-)$. The
contribution of the charged Higgs boson box diagrams to $f_d$ is
negligible $-0.02<\delta^Hf_d<0$ whereas the contribution of the
double penguin diagrams to $f_d$ can be of either sign and for very
special values of the parameters can reach $|\delta^{\rm
  DP}f_d|\approx0.2$.

\begin{figure}[htbp]
\begin{center}
\epsfig{file=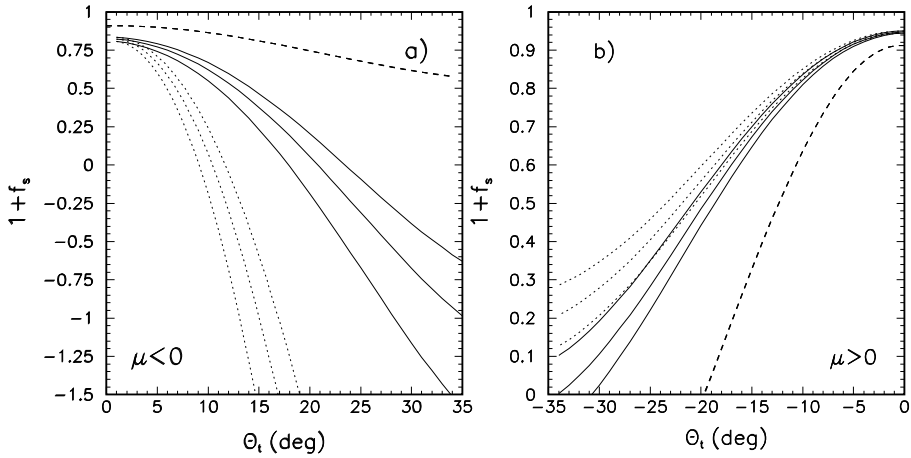,height=6.6cm}
\end{center}
\caption{$(1+f_s)$ in the MSSM for $\tan\beta=50$, $M_{H^+}=200$ GeV,
  $m_{\tilde g}=3M_2$, $M_{\tilde b_R}=800$ GeV and the lighter
  chargino mass $600$ GeV as a function of the stop mixing angle
  $\theta_t$. In panel a) $\mu<0$ and $M_{\tilde t_1}=600$ GeV,
  $M_{\tilde t_2}=750$ GeV. In panel b) $\mu>0$ and $M_{\tilde
    t_1}=500$ GeV, $M_{\tilde t_2}=850$ GeV. Solid lines correspond to
  the complete calculation. Dotted ones - to the approximation 
  based on the formulae of sec. 6.1.2 and the dashed lines to the
  calculation without the resummation of the $\tan\beta$ enhanced
  terms. Consecutive solid and dotted lines correspond to
  $M_2/|\mu|=3/4$, $1$ and $5/4$ counting from left to right (in both
  panels). The results without the resummation do not depend in a
  visible way on $M_2/|\mu|$.}
\label{fig:fsvsteta}
\end{figure}

\begin{figure}[htbp]
\begin{center}
\epsfig{file=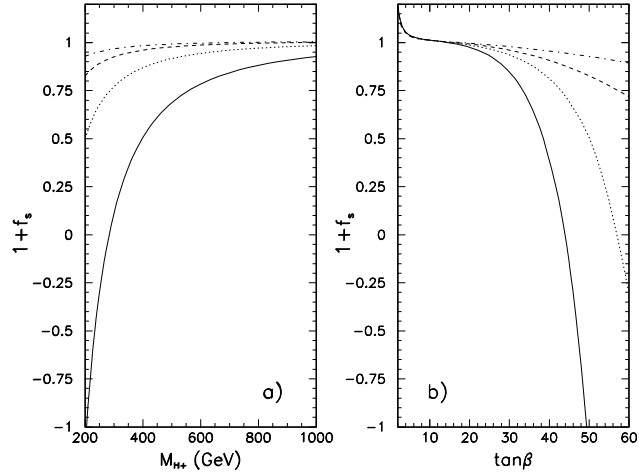,height=6.6cm}
\end{center}
\caption{$(1+f_s)$ in the MSSM for the lighter chargino mass $750$ GeV,
  $|r|\equiv M_2/|\mu|=1$, $m_{\tilde g}=3M_2$ and $M_{\tilde
    b_R}=800$ GeV as a function of $M_{H^+}$ for $\tan\beta=50$ (panel
  a) and as a function of $\tan\beta$ for $M_{H^+}=200$ GeV (panel b).
  Solid and dashed (dotted and dot-dashed) lines correspond to
  $M_{\tilde t_1}=500$ GeV, $M_{\tilde t_2}=850$ GeV (600 and 750
  GeV). Solid and dotted lines, correspond to $\mu<0$ whereas the
  dashed and dot-dashed lines to $\mu>0$.  The stop mixing angle
  $\theta_t=+(-)10^o$ for $\mu<0$ ($\mu>0$).}
\label{fig:fsvsmhtb}
\end{figure}

In figs. \ref{fig:fsvsteta} and \ref{fig:fsvsmhtb} we show the factor
$(1+f_s)$ as a function of the parameters to which its value is most
sensitive, that is $M_{H^+}$, $\tan\beta$ and the mixing angle of the
top squarks $\theta_t$. The sign of the latter is always such that
$\mu A_t$ is positive, allowing for the cancellation of the charged
Higgs boson and chargino contributions to the amplitude of the $\bar
B\to X_s\gamma$ decay.

Strong dependence of the double penguin contribution to $f_s$ on the
left-right mixing of the top squarks is clearly visible in
fig.~\ref{fig:fsvsteta} where we show $1+f_s$ as a function of the
stop mixing angle $\theta_t$. As expected from our considerations of
sec. 5, for $\mu<0$ ($\mu>0$) the resummation of the $\tan\beta$
enhanced terms leads to dramatic increase (decrease \cite{ISRE1}) of
$|f_s|$ due to the appearance of the factors
$(1+\tilde\epsilon_3\tan\beta)(1+\epsilon_0\tan\beta)$ in the
denominators in eqs. (\ref{C2LRA}), (\ref{eqn:C1SLLA}) as compared to
the formulae (4.11) of ref. \cite{BUCHROSL1}.  Curves for different
values of the ratio $M_2/|\mu|$ in fig.~\ref{fig:fsvsteta} illustrate
the impact of the heavier chargino.  Indeed, for shown values of
$M_2/|\mu|$ the lighter chargino is always almost higgsino-like and to
an accuracy of few GeV, $m_{C_1}\approx|\mu|$.  (Without the
resummation, the lines corresponding to different values of $M_2/|\mu|$ are
indistinguishable on the plot.)

Dotted lines in fig.~\ref{fig:fsvsteta} have been obtained by using
the formulae (\ref{C2LRA})-(\ref{eqn:C1SRRA}) and correspond to the
approximation of sec. 3.3. We observe that for $\mu>0$ (which is
favoured by models like minimal SUGRA) when the resummation suppresses
the effects of the double penguin diagrams as compared to the
calculation of \cite{BUCHROSL1}, the approximation of sec. 3.3 works
well for $|\theta_t|\simlt20^o$ but in general gives too strong a
suppression. However, for $\mu<0$, the approximation of sec. 3.3
grossly overestimates the impact of the double penguin diagrams on
$\Delta M_s$ already for $\theta_t\simlt5^o$ and should not be used
for realistic applications.

For the same value of the mixing angle $\theta_t$, larger effects are
obtained for bigger stop mass splitting because in this case the
parameter $|A_t|$ has to be larger. This is clearly seen in
fig.~\ref{fig:fsvsmhtb}.  It should be also stressed that the double
penguin contribution does not vanish when the mass scale of the
sparticles is increased (i.e. when all mass parameters are scaled
uniformly). Thus, large effects decreasing $1+f_s$ below unity can be
present in the MSSM also for the heavy sparticles provided the mass
scale of the MSSM Higgs sector remains low and $\tan\beta$ is large as
illustrated in fig.~\ref{fig:fsvsmhtb}.  Positive contribution to
$1+f_s$ seen in fig.~\ref{fig:fsvsmhtb}b for $\tan\beta<2.5$ and
$M_{H^+}=200$ GeV is due to the ordinary charged Higgs boson box
diagrams which contribute to the universal part of $f_s$, $f_d$ and
$f_\varepsilon$ through the Wilson coefficient of the standard
$Q_1^{\rm VLL}$ operator.  For lighter $H^+$ and light charginos
$1+f_s$ can reach values $\sim2$ \cite{BUGAGOJASI}. Such high values
of $1+f_s$ will be definitely excluded if the measured $\Delta M_s$ is
below 40/ps. 

The lines corresponding to $M_{\tilde t_1}=500$ GeV, $M_{\tilde t_2}=850$ GeV 
in fig. \ref{fig:fsvsmhtb} can be directly compared with their counterparts 
in fig. 13 of ref. \cite{BUCHROSL1}. As can be read off, for $\mu=-750$ GeV 
($\mu=750$ GeV) the resummation of the $\tan\beta$ enhanced terms increases 
(decreases) $f_s$ roughly by a factor of $1.5$ ($2.4$).

\begin{figure}[htbp]
\begin{center}
\epsfig{file=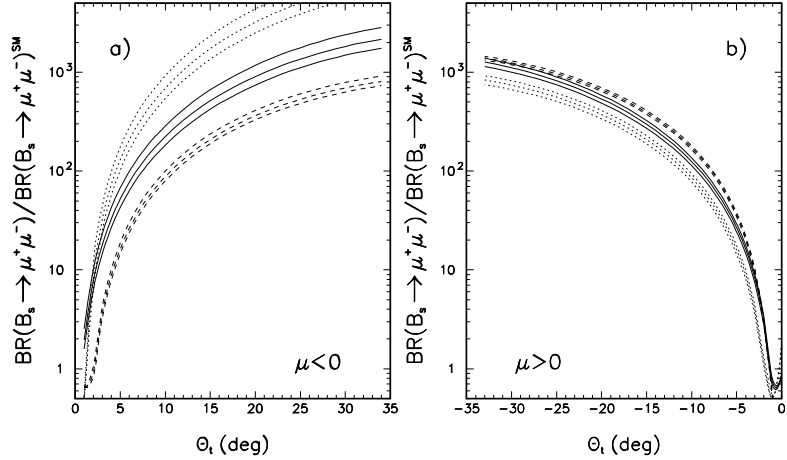,height=6.6cm}
\end{center}
\caption{The ratio $BR(B^0_s\to\mu^+\mu^-)/BR(B^0_s\to\mu^+\mu^-)^{\rm SM}$
  in the MSSM for $\tan\beta=50$, $M_{H^+}=200$, $m_{\tilde g}=3M_2$,
  $M_{\tilde b_R}=800$ GeV and the lighter chargino mass $600$ GeV as
  a function of the stop mixing angle $\theta_t$. In panel a) $\mu<0$
  and $M_{\tilde t_1}=600$ GeV, $M_{\tilde t_2}=750$ GeV. In panel b)
  $\mu>0$ and $M_{\tilde t_1}=500$ GeV, $M_{\tilde t_2}=850$ GeV.
  Solid lines correspond to the complete calculation. Dotted ones - to
  the approximation of based on the formulae of sec. 6.2 and the
  dashed lines to the calculation without the resummation of the
  $\tan\beta$ enhanced terms. Consecutive lines of each type
  correspond to $M_2/|\mu|=3/4$, $1$ and $5/4$ counting from left to
  right (in both panels).}
\label{fig:BRvsteta}
\end{figure}

\subsection{Predictions for \boldmath{$B^0_{s,d}\to \mu^+\mu^-$}}

In \cite{CHSL} and \cite{BOEWKRUR} the branching ratios $B^0_{s,d}\to
\mu^+\mu^-$ have been calculated diagrammatically with one loop
accuracy. The resummation of large $\tan\beta$ effects have not been
done in these papers. The latter effects have been included in the
$SU(2)\times U(1)$ limit first in \cite{BAKO} and their importance has
been subsequently emphasized in \cite{ISRE1}. Here we present for the
first time results for $B^0_{s,d}\to \mu^+\mu^-$ in the approach that
combines the full one loop diagrammatic calculation of refs.
\cite{CHSL} and \cite{BOEWKRUR} with the resummation of large
$\tan\beta$ enhanced terms. This includes both, $SU(2)\times U(1)$
breaking corrections to the calculations of refs.  \cite{BAKO,ISRE1}
and the effects of the electroweak gauge couplings. We will asses the
importance of these improvements with respect to the approaches of
refs. \cite{BAKO,ISRE1}.

In figure \ref{fig:BRvsteta} we show the ratio of the branching ratios
$BR(B^0_s\to\mu^+\mu^-)$ predicted in the SM and in the MSSM with
$\tan\beta=50$ and $M_{H^+}=200$ GeV as a function of the mixing angle
of the top squark. Strong dependence on the latter parameter is
evident.  As in figure \ref{fig:fsvsteta}, curves of a given type
corresponding to smaller values of $M_2/|\mu|$ illustrate the effects
of lowering the mass of the heavier chargino ($m_{C_1}\approx|\mu|$,
$m_{C_2}\approx M_2$ for all these curves) and of the increasing
deviation of the lighter chargino from pure higgsino. Dotted lines in
figure \ref{fig:BRvsteta} have been obtained by using the formulae
(\ref{CS}) and (\ref{CP}) of sec. 6.2 which correspond to the approach
of ref. \cite{ISRE1} that is, to the one of sec. 3 with $\epsilon_0$
and $\epsilon_Y$ computed from the formulae (\ref{eqn:e}) and
(\ref{eqn:eY}), respectively. Dashed lines show the prediction of the
MSSM without the resummation of the $\tan\beta$ enhanced terms.
$SU(2)\times U(1)$ symmetry limit (secs. 3.2-3.4) of ref.
\cite{ISRE1}.

\begin{figure}[htbp]
\begin{center}
\epsfig{file=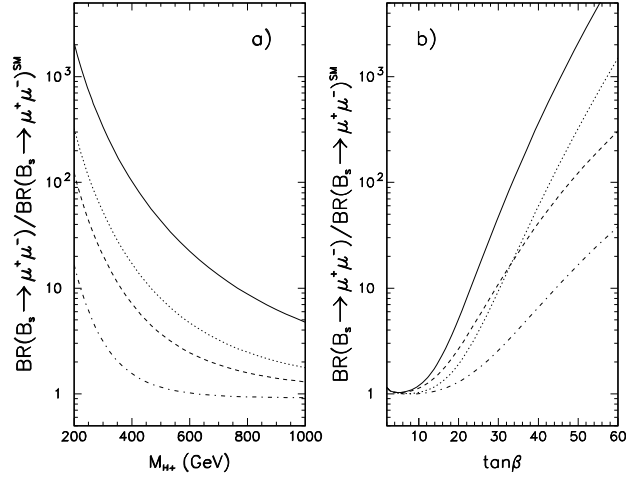,height=6.6cm}
\end{center}
\caption{The ratio $BR(B^0_s\to\mu^+\mu^-)/BR(B^0_s\to\mu^+\mu^-)^{\rm SM}$ 
  in the MSSM for the lighter chargino mass $750$ GeV, $|r|\equiv
  M_2/|\mu|=1$, $m_{\tilde g}=3M_2$ and $M_{\tilde b_R}=800$ GeV as a
  function of $M_{H^+}$ for $\tan\beta=50$ (panel a) and as a function
  of $\tan\beta$ for $M_{H^+}=200$ GeV (panel b). Solid and dashed
  (dotted and dot-dashed) lines correspond to $M_{\tilde t_1}=500$
  GeV, $M_{\tilde t_2}=850$ GeV (600 and 750 GeV). Solid and dotted
  lines, correspond to $\mu<0$ whereas the dashed and dot-dashed lines
  to $\mu>0$.  The stop mixing angle $\theta_t=+(-)10^o$ for $\mu<0$
  ($\mu>0$).}
\label{fig:BRvsmhtb}
\end{figure}

As it could be expected, for $\mu>0$ ($\mu<0$) the resummation decreases 
(increases) the branching ratio predicted in the MSSM. Similarly as in the 
case of $f_s$, for $\mu>0$ the approximation of sec. 3 gives the branching 
ratio smaller than the complete calculation, while for $\mu<0$ it 
overestimates this rate significantly. For example, for $\mu>0$, 
$M_2/|\mu|=1$ and $\theta_t=-15^o$ and other parametrs as in fig. 
\ref{fig:BRvsteta},  
$BR(B^0_s\to\mu^+\mu^-)$ obtained in the 
approximation of sec. 6.2  is smaller 
than the one obtained using the complete approach by  a factor of 
$1.5$, while it is bigger by a factor of $2$ if no
resummation of $\tan\beta$ enhanced terms is performed.
For $\mu<0$ and $\theta_t=15^o$, instead, the 
formulae of sec. 6.2  give 
$BR(B^0_s\to\mu^+\mu^-)$ approximately $3$
times bigger  than the complete calculation, while neglecting the resummation
results in the branching  ratio that is smaller by 
a factor of $1.3$ than obtained in 
the full approach.

In figures \ref{fig:BRvsmhtb}a and \ref{fig:BRvsmhtb}b we show the
dependence of the ratio of the branching ratios
$BR(B^0_s\to\mu^+\mu^-)$ predicted in the SM and in the MSSM as a
functions of the charged Higgs boson mass and $\tan\beta$,
respectively for two different choices of the top squark masses and
both signs of $\mu$. Here we show only the results of our complete
calculation based on the approach of sec. 2.

\begin{figure}[htbp]
\begin{center}
\epsfig{file=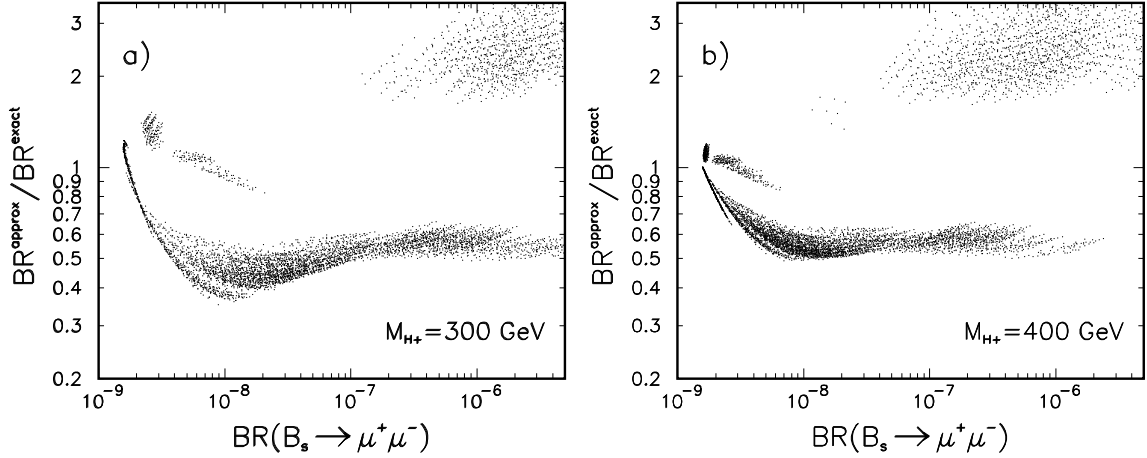,width=\linewidth}
\end{center}
\caption{The ratio 
  $BR(B^0_s\to\mu^+\mu^-)^{\rm approx}/BR(B^0_s\to\mu^+\mu^-)^{\rm exact}$ 
 in the MSSM for $\tan\beta=50$ and $M_{H^+}=300$ GeV
  (panel a) and $400$ GeV (panel b). Points corresponding to
  unacceptable $\bar B\to X_s\gamma$ rate have been rejected.}
\label{fig:ratiovsBR}
\end{figure}

The values of the ratio 
$BR(B^0_s\to\mu^+\mu^-)/BR(B^0_s\to\mu^+\mu^-)^{\rm SM}$ shown in figures 
\ref{fig:BRvsteta} and \ref{fig:BRvsmhtb} are also representative for the 
ratio $BR(B^0_d\to\mu^+\mu^-)/BR(B^0_d\to\mu^+\mu^-)^{\rm SM}$ if one
neglects the small variation of 
$|(V^{\rm eff}_{td})^{\rm MSSM}/(V^{\rm eff}_{td})^{\rm SM}|^2$ with the 
supersymmetric parameters which we have discussed in sec. 6.4.

Finally, in figure \ref{fig:ratiovsBR} we show the scatter plot of the
ratio of the $B^0_s\to\mu^+\mu^-$ rates computed by using the formulae
(\ref{CS}) and (\ref{CP}) of sec. 6.2 and computed using our complete
approach of sec.  2 for the charged Higgs boson mass equal 300 and 400
GeV.  Large, up to 50\%, deviations of the approximate method with
respect to the full calculation are typical for most of the
points. We observe that the $\bar B\to X_s\gamma$ constraint allow the
approximate predictions of the branching ratio to be both, smaller and
bigger than the one based on the complete calculation. Approximate
predictions overestimating the exact ones correspond to very big
negative contributions to $\Delta M_s$ ($1+f_s<0$) but, as found in
\cite{BUCHROSL2}, are not excluded by the lower experimental limit
$\Delta M_s>15/$ps provided $1+f_s\simlt-0.5$. However, for $M_{H+}=200$ 
GeV such points give $BR(B^0_s\to\mu^+\mu^-)$ above the CDF bound
(\ref{eqn:CDFlimit}).

\subsection{Correlation between \boldmath{$\Delta M_s$} and 
\boldmath{$B^0_{s,d}\to \mu^+\mu^-$}}

As seen from eq. (\ref{CORRMAIN}), important in the study of the correlation
in question is the ratio $(\Delta M_s)^{\rm exp}/(\Delta M_s)^{\rm SM}$
\cite{BUCHROSL2}. In order to find $(\Delta M_s)^{\rm exp}/(\Delta M_s)^{\rm
  SM}$ one has to deal with the non-perturbative uncertainties contained in 
the evaluation of $(\Delta M_s)^{\rm SM}$.  The allowed range for 
$(\Delta M_s)^{\rm exp}/(\Delta M_s)^{\rm SM}$ can be obtained by varying 
all relevant SM parameters like $\overline m_t$, $V^{\rm eff}_{ts}$ and 
$F_{B_s}\sqrt{\hat B_{B_s}}$.  A conservative scanning of these parameters 
performed in~\cite{BUCHROSL1} resulted in
\be\label{BOUND}
a \left[\frac{(\Delta M_s)^{\rm exp}}{15/ps}\right]\le 
\frac{(\Delta M_s)^{\rm exp}}{(\Delta M_s)^{\rm SM}}
\le b \left[\frac{(\Delta M_s)^{\rm exp}}{15/ps}\right]
\ee
with $a=0.52$ and $b=1.29$.  It is however clear that the numerical
values of the parameters $a$ and $b$ depend on the error analysis and
the difference $b-a$ should  become smaller as the uncertainties
in the parameters $\overline m_t$, $V^{\rm eff}_{ts}$ and in particular 
in $F_{B_s}\sqrt{\hat B_{B_s}}$ are reduced with time. For example, the 
very recent analysis which uses the Bayesian approach gives $a=0.71$ and
$b=1.0$~\cite{BUPAST} that correspond to the $95\%$ probability range
$15.1/ps\le(\Delta M_s)^{\rm SM}\le 21.0/ps$.

\begin{figure}[htbp]
\begin{center}
\epsfig{file=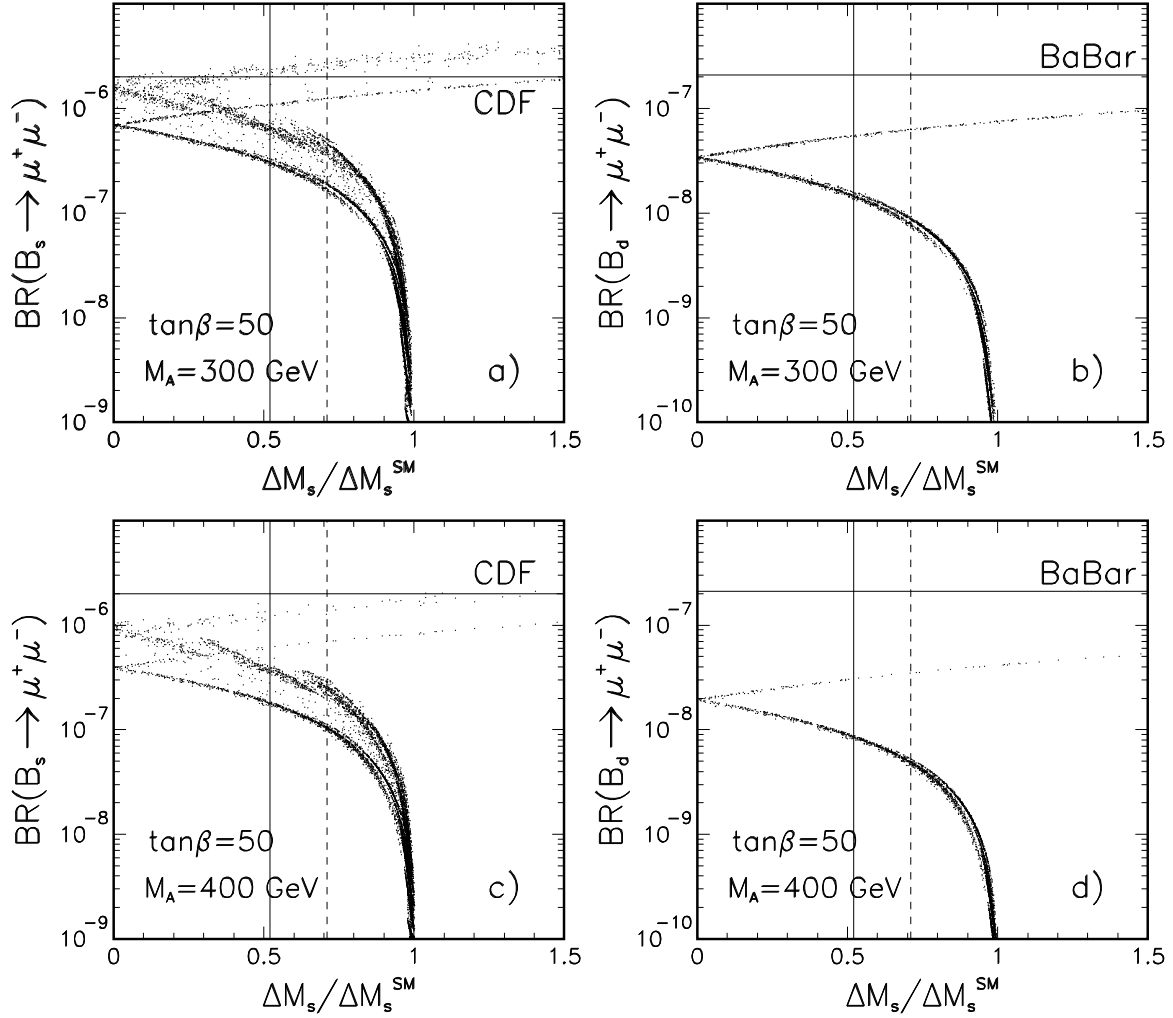,width=\linewidth}
\vspace{-10mm}
\caption{\protect Correlation between $\Delta M_s/(\Delta M_s)^{\rm SM}$ 
and $B^0_{s,d}\to\mu^+\mu^-$ in the MSSM with flavour violation ruled
by the CKM matrix. Lower (upper) branches of points correspond to
$0<1+f_s<1$ ($1+f_s<0$).  Current experimental bounds: $BR(B^0_s\ra
\mu^+\mu^-)<2\cdot10^{-6}$ (CDF)~\cite{CDF} and $BR(B^0_d\ra
\mu^+\mu^-)<2.1\cdot10^{-7}$ (BaBar)~\cite{ALEXAN} are shown by the
horizontal solid lines. Solid (dashed) vertical lines show the lower limit
on $\Delta M_s/(\Delta M_s)^{\rm SM}$ following from eq. (\ref{BOUND})
with $a=0.52$ as in ref. \cite{BUCHROSL1} ($a=0.71$ as in \cite{BUPAST}).
\label{fig:bcorr}}
\end{center}
\end{figure}

We illustrate the correlations between $\Delta M_s$ and 
$BR(B^0_{s,d}\to \mu^+\mu^-)$ in fig.~\ref{fig:bcorr} where 
we plot $BR(B^0_{s,d}\to\mu^+\mu^-)$ as functions of 
$(\Delta M_s)^{\rm exp}/(\Delta M_s)^{\rm SM}$ for $\tan\beta=50$ 
and $M_A=300$ and 400 GeV by scanning the other MSSM parameters 
with the restriction that sparticles are heavier than 500 GeV
and the $\bar B\to X_s\gamma$ constraint is satisfied. For each point 
in the MSSM parameter space we satisfy the experimental constraints on 
$\epsilon_K$, $V_{ub}/V_{cb}$, $a_{\psi K_S}$ and $\Delta M_d$ by
appropriately adjusting the Wolfenstein and nonperturbative $\hat B_K$, 
$F_{B_d}$ and $\xi$ parameters, that is, performing the unitarity triangle 
analysis \cite{BUCHROSL1,CHRO,BUPAST} described in sec. 6.4 (points for 
which these constraints cannot be satisfed are rejected).
$(\Delta M_d)^{\rm exp}$ and the parameter $\varepsilon_K$ do not 
constrain our scan significantly as the corresponding quantities 
$f_d$ and $f_\varepsilon$ are small in most of the parameter space.
Since the element $V_{td}^{\rm eff}$ is well constrained by
$\epsilon_K$, $V_{ub}/V_{cb}$, $a_{\psi K_S}$, fitting the 
experimental value of $\Delta M_d$ practically fixes $F_{B_d}$. 
For this reason, the correlation of $BR(B^0_d\to\mu^+\mu^-)$ with 
$\Delta M_s/\Delta M^{\rm exp}_s$ clearly seen in figures 
\ref{fig:bcorr}b and \ref{fig:bcorr}d is much tighter than the analogous 
correlation of $BR(B^0_s\to\mu^+\mu^-)$ (figures \ref{fig:bcorr}a,c)
where the $F_{B_s}\approx\xi F_{B_d}$ can still be varied independently
(some structure in the density of points visible in figs.~\ref{fig:bcorr}a,c
is an artifact of the scanning method used to produce the figure, preferring
the points concentrated around the minimal or the maximal allowed value 
for $BR(B^0_s\to\mu^+\mu^-)$).
Vertical lines shown in figures \ref{fig:bcorr} correspond to the lower 
limit on $(\Delta M_s)^{\rm exp}/(\Delta M_s)^{\rm SM}$ following from
(\ref{BOUND}) with $a$ taken from ref. \cite{BUCHROSL1} (solid) and
\cite{BUPAST} (dashed).

For $M_A=200$ GeV and $\tan\beta=50$ shown in the plots of \cite{BUCHROSL2} 
all points corresponding to the rather unlikely scenario with $1+f_s<0$
were eliminated by the combination of the lower limit~(\ref{BOUND}) and
the CDF upper bound $BR(B^0_s\to\mu^+\mu^-)<2\times10^{-6}$~\cite{CDF}. 
This is not the case for heavier $A^0$ and/or smaller $\tan\beta$ values.

Therefore for such points we can only use~(\ref{Bdmumu}) to find 
\begin{eqnarray}
BR(B^0_d\ra \mu^+\mu^-)<3.6~(3.1)\cdot10^{-8}
\left[{1.15\over F_{B_s}/F_{B_d}}\right]^2
\left[{BR(B^0_s\ra \mu^+\mu^-)^{\rm exp}\over 10^{-6}}\right]
\label{eqn:BOUNDFF}
\end{eqnarray}
with the numerical factor corresponding to the analyses
in~\cite{BUCHROSL1} and~\cite{BUPAST}, respectively.  With the current
CDF bound one has the upper bound $BR(B^0_d\ra \mu^+\mu^-) <
8~(7)\cdot10^{-8}$ which is still lower than the current BaBar
bound~\cite{ALEXAN}.

For a more likely situation of $0<1+f_s<1$ and $(\Delta M_s)^{\rm exp}$ 
satisfying~(\ref{BOUND}) we get upper bounds on both branching ratios:
\begin{eqnarray}
&&BR(B^0_s\ra \mu^+\mu^-)\simlt1.2\cdot10^{-6}~(8\cdot10^{-7})
\phantom{aaa}{\rm for}\phantom{aa} a=0.52~(0.71) ,\nonumber\\
&&BR(B^0_d\ra \mu^+\mu^-)\simlt3\cdot10^{-8}~(2\cdot10^{-8})
\phantom{aaaa}{\rm for}\phantom{aa} a=0.52~(0.71) .   
\label{eqn:BOUNDF}
\end{eqnarray}
where the two values for the parameter $a$ correspond to the analyses
in~\cite{BUCHROSL1} and~\cite{BUPAST}, respectively. This should be
compared with the SM values that are in the ballpark of $3\cdot10^{-9}$ 
and $1\cdot 10^{-10}$, respectively. On the basis of our discussion of 
the contribution $(\Delta M_s)^{\chi^\pm}$ in sec. 6, we would like 
to emphasize that the upper limits on $BR(B^0_{s,d}\to\mu^+\mu^-)$ 
obtained here for heavy sparticle spectrum cannot be significantly 
altered by lowering the sparticle masses.

The upper bounds (\ref{eqn:BOUNDF}) and (\ref{eqn:BOUNDFF}) are very 
impotrant because, as has been pointed out in ref. \cite{CHRO} and recently 
confirmed in \cite{ISRE2}, they do not apply in the MSSM in which flavour
violation originates from the squark mass matrices. The reason is that 
in this case leading supersymmetric corrections to $\Delta M_s$ and to
$BR(B^0_d\to\mu^+\mu^-)$ are governed by different off diagonal elements 
of the squark mass matrices and the correlation between these two quantities
does not hold. One concludes therefore, that within the supersymmetric
framework, observation of the $B^0_d\to\mu^+\mu^-$ decay at the level higher
than $\approx3\times10^{-8}$, apart from implying that the scale of the Higgs
boson sector is not far from the electroweak scale, would be also a very
strong evidence for non-minimal flavour violation in the quark sector.

In principle, if the measured values of $BR(B^0_{s,d}\to\mu^+\mu^-)$ 
turn out to be significantly bigger than predicted by the SM but do not 
respect the correlation with the measured value of $\Delta M_s$, this could 
also be a sign of nonminimal flavour violation. In practice, however, 
usefulness of this test is limited by the precision with which the
nonperturbative parameters $F_{B_s}$, $B_{B_s}$ etc. are known. 

\section{Summary and Conclusions}
\setcounter{equation}{0}

In this paper we have presented an effective Lagrangian formalism for
the calculation of flavour changing neutral and charged currents that
is suited for theories containing a sector of heavy fields. While our
approach is rather general, we have discussed it explicitly in the
context of the MSSM with the CKM matrix as the only source of flavour
violation, heavy supersymmetric particles and neutral and charged
Higgs masses below $500$~GeV.

Recently a number of analyses of large $\tan\beta$ effects in weak
decays appeared in the literature. The present analysis is the first
one that includes simultaneously
\begin{itemize}
\item Resummed large $\tan\beta$ effects,
\item $SU(2)\times U(1)$ gauge symmetry breaking effects,
\item The contributions of electroweak couplings $g_1$ and $g_2$ to
  the effective parameters in the low energy Lagrangian.
\end{itemize}
Our approach is therefore equivalent to the full diagramatic one loop
calculation supplemented by the resummation of the $\tan\beta$ enhanced 
terms. We have analyzed the importance of these effects demonstrating 
that for reliable quantitative predictions all of them have to be taken 
into account. Thus, we have generalized previous analyses where either 
the resummed large $\tan\beta$ effects have been calculated in the 
$SU(2)\times U(1)$ symmetry limit \cite{BAKO,ISRE1,AMGIISST} or the 
complete one loop calculations have been performed leaving out, however, 
the resummation of large $\tan\beta$ effects \cite{CHSL,BOEWKRUR,HULIYAZH}.

To compare our approach with the one of \cite{ISRE1,AMGIISST} we have
also derived analytic formulae for the neutral and charged Higgs boson
couplings to quarks in the $SU(2)\times U(1)$ symmetry limit resumming
large $\tan\beta$ terms. Our results agree here with those presented
in \cite{ISRE1,AMGIISST}. These transparent formulae, which make all
large $\tan\beta$ effects explicit, depend only on four universal
flavour independent parameters $\epsilon_0$, $\epsilon_Y$,
$\epsilon_0^\prime$ and $\epsilon_Y^\prime$. However, the comparison
of the results obtained in this approximation with the ones following
from the complete calculation shows clearly that the former can only
be used for a semi-quantitative analysis. We have also shown that the
accuracy of the approximation can be greatly improved while
maintaining the transparency of the resulting expressions. This is
achieved by introducing flavour dependence into the parameters
$\epsilon_0$, $\epsilon_Y$, $\epsilon_0^\prime$ and
$\epsilon_Y^\prime$ by extracting them directly from the complete (but
relatively simple to compute) correction to the down quark mass
matrix.  In this manner we have obtained analytic formulae for the
neutral and charged Higgs couplings to quarks that still make the
large $\tan\beta$ effects explicit but include automatically the
effects of electroweak couplings $g_1$, $g_2$ and of the, slightly
less important, $SU(2)\times U(1)$ symmetry breaking effects. Our
analytic formulae reproduce within $5-10\%$ the results of the full
approach to be compared with only $20-40\%$ accuracy of the formulae
obtained in the approach of refs. \cite{ISRE1,AMGIISST}.

The main message of our analysis is that the simple approach adopted
in \cite{BAKO,ISRE1,AMGIISST} overestimates the effects of the
resummation of large $\tan\beta$ contributions. This is clearly seen
in figs.  \ref{fig:fsvsteta}, \ref{fig:BRvsteta}.

The most spectacular effects in weak decays predicted by the MSSM with
large $\tan\beta$ are the huge increase of the branching ratios
$BR(B^0_{s,d}\to \mu^+\mu^-)$ and a significant suppression of $\Delta
M_s$.  As pointed out in \cite{BUCHROSL2} these effects are tightly
correlated so that for most probable ranges of supersymmetric
parameters the experimental lower bound on $\Delta M_s$ implies an
upper bound on $BR(B^0_{s,d}\to\mu^+\mu^-)$. In \cite{BUCHROSL2} and
in the present paper both these quantities, as well as their
correlation have been calculated including for the first time all the
effects listed above. Simultaneously we have clarified some points in
the calculation of large $\tan\beta$ effects in the $B\to X_s\gamma$
rate that we have used to constrain the allowed range of
supersymmetric parameters.

For the same set of parameters our results for $BR(B^0_{s,d}\to\mu^+\mu^-)$ 
without the inclusion of the $\Delta M_s$ constraint are typically between 
the values presented in \cite{ISRE1,AMGIISST} and in 
\cite{CHSL,HULIYAZH,BOEWKRUR}. We also find that the impact of the resummed 
large $\tan\beta$ corrections on our analysis of $\Delta M_s$ in 
\cite{BUCHROSL1} is smaller than obtained in \cite{ISRE1,AMGIISST}. Moreover 
we have shown that large $\tan\beta$ effects in $\Delta M_s$ are dominated 
by double neutral Higgs penguin diagrams and receive significant 
contributions from box diagrams with charged Higgs exchanges.

For the near future of particular interest are the upper bounds on
$BR(B^0_{s,d}\to\mu^+\mu^-)$ as functions of the ratio $(\Delta
M_s)^{\rm exp}/ (\Delta M_s)^{\rm SM}$ that we have already presented
in \cite{BUCHROSL2}. In the most likely scenario with $0<(1+f_s)<1$
these bounds are becoming very strong when this ratio approaches
unity. This is illustrated in fig.~\ref{fig:bcorr}. For 
$(\Delta M_s)^{\rm exp} \ge(\Delta M_s)^{\rm SM}$ substantial enhancements 
of $BR(B^0_{s,d}\to\mu^+\mu^-)$ with respect to the values obtained in
the SM are not possible within the MSSM scenario considered here.
Violation of our bounds would either rule out this version of MSSM
and/or signal new sources of flavour violation~\cite{CHRO}.

As the upper bounds on $BR(B^0_{s,d}\to \mu^+\mu^-)$ discussed here
are sensitive functions of the ratio $(\Delta M_s)^{\rm exp}/ (\Delta
M_s)^{\rm SM}$, their quantitative usefulness will depend on the value
of $(\Delta M_s)^{\rm exp}$ and on the accuracy with which $(\Delta
M_s)^{\rm SM}$ can be calculated.  In this respect the present efforts
of experimentalists to measure $BR(B^0_{s,d}\to \mu^+\mu^-)$ and
$\Delta M_s$ and of theorists to calculate $F_{B_{d,s}}$ and the
parameters $B_{d,s}$ are very important.
\vskip0.5cm

\noindent{\it Note added}: Very recently a new analysis of resummed large 
$\tan\beta$ effects in FCNC 
transitions has been presented in \cite{DEPI}. It goes beyond the framework 
of minimal flavour and CP violation and discusses the ratio 
$\varepsilon^\prime/\varepsilon$ and CP asymmetries not considered 
in our paper as the new physics effects in these quantities in our scenario 
are tiny. However these authors, similarly to \cite{ISRE1,AMGIISST}, work
in the $SU(2)\times U(1)$ limit and do not take into account the 
refinements discussed in section 5 nor they included the constraint from 
$B\to X_s\gamma$ decay on the supersymmetric parameters. On the other hand
they stress that the inclusion of light quark contributions in the 
formalism of \cite{ISRE1,AMGIISST} could have an impact for certain very 
special values of supersymmetric parameters for which $\tilde\epsilon_3$ 
approaches $-1$. We would like to emphasis that in our full approach of 
section 2 and in the improved approximation presented in section 5 the 
contributions of light quarks are automatically included. Moreover when the 
experimental constraints from $B\to X_s\gamma$ decay, $\Delta M_{s,d}$ and 
$B\to\mu^+\mu^-$ are taken into account and the Higgs masses are kept below 
$500$ GeV, $\tilde\epsilon_3$ never approaches $-1$ and the light quarks 
except for assuring the GIM mechanism are unimportant in the resummed large 
$\tan\beta$ corrections.

\section*{Acknowledgements}

A.J.B. would like to thank A. Dedes, G. Isidori, F. Kr\"uger and J. Urban 
for their interest and discussions. P.H.Ch. would like to thank G. Isidori 
for the discussions and the CERN Theory group for hospitality during the 
completion of this paper. The work of A.J.B. and J.R. has been supported 
in part by the German Bundesministerium f\"ur Bildung und Forschung under 
the contract 05HT1WOA3 and the DFG Project Bu. 706/1-1. J.R. and \L .S.
have been supported by the Polish State Committee for Scientific 
Research grant 2~P03B~040~24 for 2003-2005. The work of P.H.Ch. 
has been partly supported by the Polish State Committee for Scientific 
Research grant 2~P03B~129~24 for 2003-2004 and by the EC Contract 
HPRN-CT-2000-00148 for years 2000-2004. 

\newpage

\def\theequation{\thesection.\arabic{equation}}

\newcounter{lp}
\setcounter{lp}{0}
\newcommand{\lpv}{\addtocounter{lp}{1}\noindent\arabic{lp}.~}

\setcounter{section}{0}
\renewcommand{\thesection}{Appendix}

\setcounter{subsection}{0}
\renewcommand{\thesubsection}{\Alph{section}.\arabic{subsection}}

\setcounter{equation}{0}
\renewcommand{\theequation}{\Alph{section}.\arabic{equation}}

\section{Feynman rules and Green's functions}

For completeness we collect here some of the the Feynman
rules given in \cite{ROS} and formulae for the 2-, 3- and 4-point
Green's functions used in the paper.

\subsection{Feynman rules}

We write the fermion-fermion-scalar vertices as $i(V_L P_L + V_R
P_R)$. The couplings $V_L, V_R$ are listed  below:\\[2mm]

\begin{tabular}{ll}
\begin{picture}(140,70)(0,-10)
\DashArrowLine(10,0)(60,0){5}
\Text(0,0)[c]{$U_i^-$}
\ArrowLine(60,0)(110,0)
\Text(120,0)[c]{$N_j$}
\ArrowLine(60,50)(60,0)
\Text(65,45)[l]{$u^J$}
\Vertex(60,0){2}
\end{picture}
&
\raisebox{30\unitlength}{
\begin{minipage}{5cm}
\lefteqn{
V_{uUN}^{LJij} = {-1\over\sqrt2} Z^{Ji\star}_U
({g_1\over3}Z^{1j}_N  + g_2Z^{2j}_N ) 
- y_{u_J}Z_U^{(J+3)i\star} Z_N^{4j} 
}
\lefteqn{
V_{uUN}^{RJij} = {2\sqrt2\over3}g_1 Z_U^{(J+3)i\star}
Z_N^{1j\star} - y_{u_J} Z_U^{Ji\star} Z_N^{4j\star} 
}
\end{minipage}
}
\end{tabular}

\begin{tabular}{ll}
\begin{picture}(140,70)(0,-10)
\DashArrowLine(10,0)(60,0){5}
\Text(0,0)[c]{$D_i^+$}
\ArrowLine(60,0)(110,0)
\Text(120,0)[c]{$N_j$}
\ArrowLine(60,50)(60,0)
\Text(65,45)[l]{$d^I$}
\Vertex(60,0){2}
\end{picture}
&
\raisebox{30\unitlength}{
\begin{minipage}{5cm}
\lefteqn{
V_{dDN}^{LIij} = {-1\over\sqrt2} Z^{Ii}_D
({g_1\over3}Z^{1j}_N - g_2Z^{2j}_N) + y_{d_I}Z_D^{(I+3)i}Z_N^{3j}  
}
\lefteqn{
V_{dDN}^{RIij} = {-\sqrt2\over3}g_1 Z_D^{(I+3)i}Z_N^{1j\star}
+ y_{d_I} Z_D^{Ii} Z_N^{3j\star} 
}
\end{minipage}
}
\end{tabular}

\begin{tabular}{ll}
\begin{picture}(140,70)(0,-10)
\DashArrowLine(10,0)(60,0){5}
\Text(0,0)[c]{$U_i^-$}
\ArrowLine(60,0)(110,0)
\Text(120,0)[c]{$C_j^-$}
\ArrowLine(60,50)(60,0)
\Text(65,45)[l]{$d^I$}
\Vertex(60,0){2}
\end{picture}
&
\raisebox{30\unitlength}{
\begin{minipage}{5cm}
\lefteqn{
V_{dUC}^{LIij} = (-g_2Z_U^{Ji\star} Z_+^{1j} + y_{u_J}
Z_U^{(J+3)i\star} Z_+^{2j}) V^{JI}
}
\lefteqn{
V_{dUC}^{RIij} = - y_{d_I} Z_U^{Ji\star} Z_-^{2j\star} V^{JI} 
}
\end{minipage}
}
\end{tabular}

\begin{tabular}{ll}
\begin{picture}(140,70)(0,-10)
\DashArrowLine(10,0)(60,0){5}
\Text(0,0)[c]{$D_i^+$}
\ArrowLine(60,0)(110,0)
\Text(120,0)[c]{$C_j^+$}
\ArrowLine(60,50)(60,0)
\Text(65,45)[l]{$u^J$}
\Vertex(60,0){2}
\end{picture}
&
\raisebox{30\unitlength}{
\begin{minipage}{5cm}
\lefteqn{
V_{uDC}^{LJij} = -(g_2 Z_D^{Ii} Z_-^{1j} + y_{d_I}
Z_D^{(I+3)i} Z_-^{2j}) V^{JI\star}
}
\lefteqn{
V_{uDC}^{RJij} = y_{u_J} Z_D^{Ii} Z_+^{2j\star} V^{JI\star}
}
\end{minipage}
}
\end{tabular}

\begin{tabular}{ll}
\begin{picture}(150,70)(0,-10)
\DashLine(60,0)(10,0){5}
\Text(0,0)[c]{$H^0_k$}
\ArrowLine(60,0)(110,0)
\Text(120,0)[c]{$N^0_i$}
\ArrowLine(60,50)(60,0)
\Text(65,45)[l]{$N^0_j$}
\Vertex(60,0){2}
\end{picture}
&
\raisebox{25\unitlength}{
\begin{minipage}{5cm}
\lefteqn{
V_{NNS}^{Lijk} = V_{NNS}^{Rijk\star}={1\over2}\left[
(Z_R^{1k} Z_N^{3j} - Z_R^{2k} Z_N^{4j})\right.
}
\lefteqn{\left.
\phantom{aaaaaaaaaaaaaa}\times
(g_1 Z_N^{1i}-g_2 Z_N^{2i})+(j\leftrightarrow i)\right]
}
\end{minipage}
}
\end{tabular}

\begin{tabular}{ll}
\begin{picture}(150,70)(0,-10)
\DashLine(60,0)(10,0){5}
\Text(0,0)[c]{$A^0_k$}
\ArrowLine(60,0)(110,0)
\Text(120,0)[c]{$N^0_i$}
\ArrowLine(60,50)(60,0)
\Text(65,45)[l]{$N^0_j$}
\Vertex(60,0){2}
\end{picture}
&
\raisebox{25\unitlength}{
\begin{minipage}{5cm}
\lefteqn{
V_{NNP}^{Lijk} = - V_{NNP}^{Rjik\star} ={-i\over2}\left[
(Z_H^{1k} Z_N^{3j} - Z_H^{2k} Z_N^{4j})\right.
}
\lefteqn{\left.
\phantom{aaaaaaaaaaaaaa}\times
(g_1 Z_N^{1i}-g_2 Z_N^{2i})+(j\leftrightarrow i)\right]
}
\end{minipage}
}
\end{tabular}

\begin{tabular}{ll}
\begin{picture}(150,70)(0,-10)
\DashLine(60,0)(10,0){5}
\Text(0,0)[c]{$H^0_k$}
\ArrowLine(60,0)(110,0)
\Text(120,0)[c]{$C^+_i$}
\ArrowLine(60,50)(60,0)
\Text(65,45)[l]{$C^+_j$}
\Vertex(60,0){2}
\end{picture}
&
\raisebox{25\unitlength}{
\begin{minipage}{5cm}
\lefteqn{
V_{CCS}^{Lijk} = V_{CCS}^{Rjik\star} = - {g_2\over\sqrt2}
(Z_R^{1k} Z_-^{2i} Z_+^{1j} + Z_R^{2k} Z_-^{1i} Z_+^{2j}) 
}
\end{minipage}
}
\end{tabular}

\begin{tabular}{ll}
\begin{picture}(150,70)(0,-10)
\DashLine(60,0)(10,0){5}
\Text(0,0)[c]{$A^0_k$}
\ArrowLine(60,0)(110,0)
\Text(120,0)[c]{$C^+_i$}
\ArrowLine(60,50)(60,0)
\Text(65,45)[l]{$C^+_j$}
\Vertex(60,0){2}
\end{picture}
&
\raisebox{25\unitlength}{
\begin{minipage}{5cm}
\lefteqn{
V_{CCP}^{Lijk} = - V_{CCP}^{Rjik\star} = {ig_2\over\sqrt2} 
(Z_H^{1k}Z_-^{2i} Z_+^{1j} + Z_H^{2k} Z_-^{1i} Z_+^{2j}) 
}
\end{minipage}
}
\end{tabular}

\begin{tabular}{ll}
\begin{picture}(150,70)(0,-10)
\DashArrowLine(10,0)(60,0){5}
\Text(0,0)[c]{$H^+_k$}
\ArrowLine(60,0)(110,0)
\Text(120,0)[c]{$C^+_j$}
\ArrowLine(60,50)(60,0)
\Text(65,45)[l]{$N_i$}
\Vertex(60,0){2}
\end{picture}
&
\raisebox{30\unitlength}{
\begin{minipage}{5cm}
\lefteqn{
V_{NCH}^{Lijk} = Z_H^{1k} \left({1\over\sqrt2} Z_-^{2j}
(g_1Z_N^{1i}  + g_2Z_N^{2i}) - Z_-^{1j} Z_N^{3i} g_2\right)
}
\lefteqn{
V_{NCH}^{Rijk} = - Z_H^{2k} \left({1\over\sqrt2}
Z_+^{2j\star} (g_1Z_N^{1i\star}  + g_2Z_N^{2i\star}) + Z_+^{1j\star}
Z_N^{4i\star}g_2\right)
}
\end{minipage}
}
\end{tabular}

\vskip 2mm

The three-scalar  vertices are defined as $iV$, with $V$ given by:
\\[2mm]

\begin{tabular}{ll}
\begin{picture}(150,70)(0,-10)
\DashLine(10,0)(60,0){5}
\Text(0,0)[c]{$H^0_k$}
\DashArrowLine(60,0)(110,0){5}
\Text(120,0)[c]{$D^+_j$}
\DashArrowLine(60,50)(60,0){5}
\Text(65,45)[l]{$D^+_i$}
\Vertex(60,0){2}
\end{picture}
&
\raisebox{25\unitlength}{
\begin{minipage}{5cm}
\lefteqn{
V_{DDS}^{ijk}=\left({g_1^2\over6}B_R^k \left(\delta^{ji} +
{3-4s_W^2 \over 4s_W^2} Z_D^{Ij\star} Z_D^{Ii}\right)\right.  
}
\lefteqn{
-v_d y_{d_I}^2Z_R^{1k}(Z_D^{Ij\star}Z_D^{Ii}+Z_D^{(I+3)j\star}Z_D^{(I+3)i})
}
\lefteqn{
- {1\over\sqrt2} Z_R^{1k} (A_d^{IJ\star} Z_D^{Ii} Z_D^{(J+3)j\star}
+ A_d^{IJ} Z_D^{Ij\star} Z_D^{(J+3)i})
}
\lefteqn{
\left. - {1\over\sqrt2} y_{d_I} Z_R^{2k} (\mu^{\star} Z_D^{Ij\star} 
Z_D^{(I+3)i} + \mu Z_D^{Ii} Z_D^{(I+3)j\star} )\right)
}
\end{minipage}
}
\end{tabular}

\begin{tabular}{ll}
\begin{picture}(150,70)(0,-10)
\DashLine(10,0)(60,0){5}
\Text(0,0)[c]{$A^0_k$}
\DashArrowLine(60,0)(110,0){5}
\Text(120,0)[c]{$D^+_j$}
\DashArrowLine(60,50)(60,0){5}
\Text(65,45)[l]{$D^+_i$}
\Vertex(60,0){2}
\end{picture}
&
\raisebox{30\unitlength}{
\begin{minipage}{5cm}
\lefteqn{
V_{DDP}^{ijk}={i\over\sqrt2}\left(y_{d_I}(\mu^\star Z_D^{Ij\star}Z_D^{(I+3)i} 
-\mu Z_D^{Ii} Z_D^{(I+3)j\star}) Z_H^{2k}\right.
} 
\lefteqn{
\left.+(A_d^{IJ\star}Z_D^{Ii} Z_D^{(J+3)j\star} - A_d^{IJ} Z_D^{Ij\star}
Z_U^{(J+3)i}) Z_H^{1k} \right)
}
\end{minipage}
}
\end{tabular}

\begin{tabular}{ll}
\begin{picture}(150,70)(0,-10)
\DashLine(10,0)(60,0){5}
\Text(0,0)[c]{$H^0_k$}
\DashArrowLine(60,0)(110,0){5}
\Text(120,0)[c]{$U^-_j$}
\DashArrowLine(60,50)(60,0){5}
\Text(65,45)[l]{$U^-_i$}
\Vertex(60,0){2}
\end{picture}
&
\raisebox{25\unitlength}{
\begin{minipage}{5cm}
\lefteqn{
V_{UUS}^{ijk}=\left(-{g^2_1\over3} B_R^k \left(\delta^{ij} +
{3-8s_W^2 \over 4s_W^2} Z_U^{Ii\star} Z_U^{Ij}\right)\right.  
}
\lefteqn{
 - v_u y_{u_I}^2 Z_R^{2k} (Z_U^{Ii\star} Z_U^{Ij} + Z_U^{(I+3)i\star}
 Z_U^{(I+3)j})
}
\lefteqn{
+ {1\over\sqrt2} Z_R^{2k} (A_u^{IJ\star} Z_U^{Ii\star} Z_U^{(J+3)j}
+ A_u^{IJ} Z_U^{Ij} Z_U^{(J+3)i\star})
}
\lefteqn{
\left.  + {1\over\sqrt2} y_{u_I} Z_R^{1k} (\mu^{\star} Z_U^{Ij} 
Z_U^{(I+3)i\star} + \mu Z_U^{Ii\star} Z_U^{(I+3)j} )\right)
}
\end{minipage}
}
\end{tabular}

\begin{tabular}{ll}
\begin{picture}(150,70)(0,-10)
\DashLine(10,0)(60,0){5}
\Text(0,0)[c]{$A^0_k$}
\DashArrowLine(60,0)(110,0){5}
\Text(120,0)[c]{$U^-_j$}
\DashArrowLine(60,50)(60,0){5}
\Text(65,45)[l]{$U^-_i$}
\Vertex(60,0){2}
\end{picture}
&
\raisebox{30\unitlength}{
\begin{minipage}{5cm}
\lefteqn{
V_{UUP}^{ijk} = {i\over\sqrt2} \left(y_{u_I} (\mu Z_U^{Ii\star} Z_U^{(I+3)j} -
\mu^{\star} Z_U^{Ij} Z_U^{(I+3)i\star}) Z_H^{1k}\right.
} 
\lefteqn{
\left. + (A_u^{IJ} Z_U^{Ij} Z_U^{(J+3)i\star} - A_u^{IJ\star} Z_U^{Ii\star}
Z_U^{(J+3)j}) Z_H^{2k} \right)
}
\end{minipage}
}
\end{tabular}

\begin{tabular}{ll}
\begin{picture}(150,70)(0,-10)
\DashArrowLine(10,0)(60,0){5}
\Text(0,0)[c]{$H^+_k$}
\DashArrowLine(60,0)(110,0){5}
\Text(120,0)[c]{$D^+_j$}
\DashArrowLine(60,50)(60,0){5}
\Text(65,45)[l]{$U^-_i$}
\Vertex(60,0){2}
\end{picture}
&
\raisebox{25\unitlength}{
\begin{minipage}{5cm}
\lefteqn{
V_{UDH}^{ijk} = \left[-{\sqrt2M_W \over g_2} y_{u_J}y_{d_I} 
V^{JI} Z_D^{(I+3)j\star} Z_U^{(J+3)i\star}\delta^{1k} \right.  
}
\lefteqn{
+{1\over \sqrt{2}} \left( {-g_2^2 \over 2} v_l Z_H^{lk} + v_d
y_{d_I}^2 Z_H^{1k} + v_u y_{u_J}^2 Z_H^{2k} \right) V^{JI}
Z_D^{Ij\star} Z_U^{Ji\star} 
}
\lefteqn{
+ (Z_H^{1k} \mu^{\star} y_{u_J} V^{JI} - Z_H^{2k} A_u^{KJ} V^{KI})
Z_U^{(J+3)i\star} Z_D^{Ij\star} 
}
\lefteqn{
\left. + (Z_H^{1k} A_d^{KI\star} V^{JK} 
- Z_H^{2k} \mu y_{d_I} V^{JI}) Z_U^{Ji\star} Z_D^{(I+3)j\star}
\right]
}
\end{minipage}
}
\end{tabular}

\vskip0.5cm
As explained in sec. 2, the CKM matrix $V$ in the above rules is
related to the ``measured'' CKM matrix $V_{\rm eff}$ as in
(\ref{eqn:CKMeff}) or, in the approximation of sec.~\ref{sec:su2xu1},
as in (\ref{eqn:CKMcorr}). The Yukawa couplings are related to the
physical quark masses as:
\begin{eqnarray}
y_{d_I} = -{\sqrt2\over v_d}{\overline{m}_{d_I}
\over(1+\tilde\epsilon_I\tan\bar\beta)}, \qquad  
y_{u_J} = {\sqrt2\over v_u}\overline{m}_{u_J}\label{appbasic}
\end{eqnarray} 
Definitions of the mixing matrices $Z_U$, $Z_U$, $Z_\pm$ and $Z_N$ can be
found in~\cite{ROS}. Note that it is $V$, not $V_{\rm eff}$ that
enters the squark mass squared matrices.  It should be also remembered
that computing Wilson coefficients of the higher dimension operators
generated by charginos one should modify their vertices according to
the rules formulated in secs.~\ref{subsec:cqvert}
and~\ref{subsec:cqvert1}.

\subsection{Quark self-energies}

Definitions of the formfactors are given in fig.~\ref{fig:Sigcorr}.
Also,
$\left(\Sigma^q_{mR}\right)^{IJ}=\left(\Sigma^q_{mL}\right)^{JI\star}$\\[2mm]
\begin{eqnarray}
&&\left(4\pi\right)^2\left(\Sigma^d_{VL}\right)^{JI}= 
{32\pi\alpha_s\over3}\sum_{k=1}^6 Z_D^{Jk\star} Z_D^{Ik}
B_1(s,m^2_{\tilde g},m^2_{D_k})\nonumber\\
&&\hskip 2mm + \sum_{k=1}^6\sum_{l=1}^4 V_{dDN}^{LJkl\star} V_{dDN}^{LIkl}
B_1(s,m^2_{N_l},m^2_{D_k})
+ \sum_{k=1}^6\sum_{l=1}^2 V_{dUC}^{LJkl\star} V_{dUC}^{LIkl}
B_1(s,m^2_{C_l},m^2_{U_k})\nonumber\\
&&\left(4\pi\right)^2\left(\Sigma^d_{VR}\right)^{JI}=
{32 \pi \alpha_s\over 3}\sum_{k=1}^6 Z_D^{(J+3)k\star} Z_D^{(I+3)k}
B_1(s,m^2_{\tilde g},m^2_{D_k})\nonumber\\
&&\hskip 2mm+ \sum_{k=1}^6\sum_{l=1}^4 V_{dDN}^{RJkl\star} V_{dDN}^{RIkl}
B_1(s,m^2_{N_l},m^2_{D_k})
+ \sum_{k=1}^6\sum_{l=1}^2 V_{dUC}^{RJkl\star} V_{dUC}^{RIkl}
B_1(s,m^2_{C_l},m^2_{U_k})\nonumber\\
&&\left(4\pi\right)^2\left(\Sigma^d_{mL}\right)^{JI}=
-{32\pi\alpha_s\over 3} m_{\tilde g} \sum_{k=1}^6 Z_D^{(J+3)k\star}
Z_D^{Ik} B_0(s,m^2_{\tilde g},m^2_{D_k})\nonumber\\
&&\hskip 2mm +\sum_{k=1}^6\sum_{l=1}^4 V_{dDN}^{RJkl\star} 
V_{dDN}^{LIkl} m_{N_l} B_0(s,m^2_{N_l},m^2_{D_k})\nonumber\\
&&\hskip 2mm +\sum_{k=1}^6\sum_{l=1}^2V_{dUC}^{RJkl\star} 
V_{dUC}^{LIkl} m_{C_l} B_0(s,m^2_{C_l},m^2_{U_k})
\end{eqnarray}

\begin{eqnarray}
&&\left(4\pi\right)^2\left(\Sigma^u_{VL}\right)^{JI}=
{32\pi\alpha_s\over 3}\sum_{k=1}^6 Z_U^{Jk} Z_U^{Ik\star}
B_1(s,m^2_{\tilde g},m^2_{U_k})\nonumber\\
&&\hskip 2mm + \sum_{k=1}^6\sum_{l=1}^4 V_{uUN}^{LJkl\star} V_{uUN}^{LIkl}
B_1(s,m^2_{N_l},m^2_{U_k})
+ \sum_{k=1}^6\sum_{l=1}^2 V_{uDC}^{LJkl\star} V_{uDC}^{LIkl}
B_1(s,m^2_{C_l},m^2_{D_k})\nonumber\\
&&\left(4\pi\right)^2\left(\Sigma^u_{VR}\right)^{JI} =
{32\pi\alpha_s\over 3}\sum_{k=1}^6 Z_U^{(J+3)k} Z_U^{(I+3)k\star}
B_1(s,m^2_{\tilde g},m^2_{U_k})\nonumber\\
&&\hskip 2mm+ \sum_{k=1}^6\sum_{l=1}^4 V_{uUN}^{RJkl\star} V_{uUN}^{RIkl}
B_1(s,m^2_{N_l},m^2_{U_k})
+ \sum_{k=1}^6\sum_{l=1}^2 V_{uDC}^{RJkl\star} V_{uDC}^{RIkl}
B_1(s,m^2_{C_l},m^2_{D_k})\nonumber\\
&&\left(4\pi\right)^2\left(\Sigma^u_{mL}\right)^{JI}=
-{32\pi\alpha_s\over 3}m_{\tilde g} \sum_{k=1}^6 Z_U^{(J+3)k}
Z_U^{Ik\star} B_0(s,m^2_{\tilde g},m^2_{U_k})\nonumber\\
&&\hskip 2mm  +\sum_{k=1}^6\sum_{l=1}^4 V_{uUN}^{RJkl\star}
 V_{uUN}^{LIkl} m_{N_l} B_0(s,m^2_{N_l},m^2_{U_k})\nonumber\\
&&\hskip 2mm +\sum_{k=1}^6\sum_{l=1}^2 V_{uDC}^{RJkl\star}  
V_{uDC}^{LIkl} m_{C_l} B_0(s,m^2_{C_l},m^2_{D_k})
\end{eqnarray}

\subsection{Higgs-fermion effectve couplings}

The effective Higgs-fermion vertex is defined in fig.~\ref{fig:NHiggscorr}.
The formulae given below are completely general and are valid also in
the MSSM with flavour violation originating in the sfermion mass matrices.

\noindent {\bf i) CP-even neutral Higgs boson vertex $H^0_k \bar d_J d_I$.}  
Using the general notation of eq.~(\ref{eqn:Sverts}) with $S=H^0_k$ 
($H^0_1 = H^0, H^0_2 = h^0$) we find:

\begin{eqnarray}
\left(4\pi\right)^2\left (\Delta F^{dH^0_k}_{L}\right)^{JI}&=& 
{32 \pi \alpha_s\over 3} m_{\tilde g} \sum_{l,m=1}^6 V_{DDS}^{lmk}  
Z_D^{(J+3)l\star} Z_D^{Im}  C_0(m_{\tilde g}^2,m_{D_l}^2,m_{D_m}^2)
\nonumber\\
&-&  \sum_{l,m=1}^6\sum_{n=1}^4 V_{DDS}^{lmk} V_{dDN}^{RJln\star} 
V_{dDN}^{LImn} m_{N_n} C_0(m_{N_n}^2,m_{D_l}^2,m_{D_m}^2)
\nonumber\\
&-& \sum_{l,m=1}^4\sum_{n=1}^6 V_{dDN}^{RJnl\star} V_{dDN}^{LInm} 
\left[V_{NNS}^{Rlmk} C_2(m_{D_n}^2,m_{N_l}^2,m_{N_m}^2)\right.\nonumber\\
&&\phantom{aaaaaaaaaa}+
\left. V_{NNS}^{Llmk} m_{N_l} m_{N_m} C_0(m_{D_n}^2,m_{N_l}^2,m_{N_m}^2) 
\right]  \nonumber\\
&-& \sum_{l,m=1}^6\sum_{n=1}^2 V_{UUS}^{lmk} V_{dUC}^{RJln\star}
V_{dUC}^{LImn} m_{C_n} C_0(m_{C_n}^2,m_{U_l}^2,m_{U_m}^2)
\nonumber\\
&-& \sum_{l,m=1}^2\sum_{n=1}^6 V_{dUC}^{RJnl\star} V_{dUC}^{LInm}
\left[ V_{CCS}^{Rmlk} C_2(m_{U_n}^2,m_{C_l}^2,m_{C_m}^2)\right.\nonumber\\
&&\phantom{aaaaaaaaaa}
+\left. V_{CCS}^{Lmlk} 
m_{C_l} m_{C_m} C_0(m_{U_n}^2,m_{C_l}^2,m_{C_m}^2) \right]
\end{eqnarray}

\begin{eqnarray}
\left(4\pi\right)^2\left (\Delta F^{dH^0_k}_R\right)^{JI}&=& 
{32 \pi \alpha_s\over 3} m_{\tilde g} \sum_{l,m=1}^6 V_{DDS}^{lmk}  
Z_D^{Jl\star} Z_D^{(I+3)m}  C_0(m_{\tilde g}^2,m_{D_l}^2,m_{D_m}^2)
\nonumber\\
&-& \sum_{l,m=1}^6\sum_{n=1}^4 V_{DDS}^{lmk} V_{dDN}^{LJln\star} 
V_{dDN}^{RImn} m_{N_n} C_0(m_{N_n}^2,m_{D_l}^2,m_{D_m}^2)
\nonumber\\
&-&  \sum_{l,m=1}^4\sum_{n=1}^6 V_{dDN}^{LJnl\star} V_{dDN}^{RInm} 
\left[ V_{NNS}^{Llmk} C_2(m_{D_n}^2,m_{N_l}^2,m_{N_m}^2)\right.\nonumber\\
&&\phantom{aaaaaaaaaa}
+\left. V_{NNS}^{Rlmk} m_{N_l} m_{N_m} C_0(m_{D_n}^2,m_{N_l}^2,m_{N_m}^2) 
\right] \nonumber\\
&-& \sum_{l,m=1}^6\sum_{n=1}^2 V_{UUS}^{lmk} V_{dUC}^{LJln\star}
V_{dUC}^{RImn} m_{C_n} C_0(m_{C_n}^2,m_{U_l}^2,m_{U_m}^2)
\nonumber\\
&-& \sum_{l,m=1}^2\sum_{n=1}^6 V_{dUC}^{LJnl\star} V_{dUC}^{RInm}
\left[V_{CCS}^{Lmlk}C_2(m_{U_n}^2,m_{C_l}^2,m_{C_m}^2)\right.\nonumber\\
&&\phantom{aaaaaaaaaa}
+\left. V_{CCS}^{Rmlk} 
m_{C_l} m_{C_m} C_0(m_{U_n}^2,m_{C_l}^2,m_{C_m}^2) \right)]
\end{eqnarray}

\vskip 5mm

\noindent {\bf ii)  CP-odd neutral Higgs boson vertex $A^0_k\bar d_J d_I$}.  
In this case $S=A^0_k$ ($A^0_1 = A^0$, $A^0_2 = G^0$) and

\begin{eqnarray}
\left(4\pi\right)^2\left (\Delta F^{dA^0_k}_L\right)^{JI}&=& 
{32\pi\alpha_s\over3} m_{\tilde g} \sum_{l,m=1}^6 V_{DDP}^{lmk}  
Z_D^{(J+3)l\star} Z_D^{Im}  C_0(m_{\tilde g}^2,m_{D_l}^2,m_{D_m}^2)
\nonumber\\
&-& \sum_{l,m=1}^6\sum_{n=1}^4 V_{DDP}^{lmk} V_{dDN}^{RJln\star} 
V_{dDN}^{LImn} m_{N_n} C_0(m_{N_n}^2,m_{D_l}^2,m_{D_m}^2)
\nonumber\\
&-& \sum_{l,m=1}^4\sum_{n=1}^6 V_{dDN}^{RJnl\star} V_{dDN}^{LInm}
\left[V_{NNS}^{Rlmk}C_2(m_{D_n}^2,m_{N_l}^2,m_{N_m}^2)\right.\nonumber\\
&&\phantom{aaaaaaaaaa}
+\left. V_{NNS}^{Llmk}m_{N_l} m_{N_m} C_0(m_{D_n}^2,m_{N_l}^2,m_{N_m}^2) 
\right]  \nonumber\\
&-& \sum_{l,m=1}^6\sum_{n=1}^2 V_{UUP}^{lmk} V_{dUC}^{RJln\star}
V_{dUC}^{LImn} m_{C_n} C_0(m_{C_n}^2,m_{U_l}^2,m_{U_m}^2)
\nonumber\\
&-&  \sum_{l,m=1}^2\sum_{n=1}^6 V_{dUC}^{RJnl\star} V_{dUC}^{LInm} 
\left[V_{CCP}^{Rmlk} C_2(m_{U_n}^2,m_{C_l}^2,m_{C_m}^2)\right.\nonumber\\
&&\phantom{aaaaaaaaaa}+\left. V_{CCP}^{Lmlk} 
m_{C_l} m_{C_m} C_0(m_{U_n}^2,m_{C_l}^2,m_{C_m}^2) \right]
\end{eqnarray}

\begin{eqnarray}
\left(4\pi\right)^2\left (\Delta F^{dA^0_k}_R\right)^{JI}&=& 
{32\pi\alpha_s\over 3} m_{\tilde g} \sum_{l,m=1}^6 V_{DDP}^{lmk}  
Z_D^{Jl\star} Z_D^{(I+3)m}  C_0(m_{\tilde g}^2,m_{D_l}^2,m_{D_m}^2)
\nonumber\\
&-&\sum_{l,m=1}^6\sum_{n=1}^4 V_{DDP}^{lmk} V_{dDN}^{LJln\star} 
V_{dDN}^{RImn} m_{N_n} C_0(m_{N_n}^2,m_{D_l}^2,m_{D_m}^2)
\nonumber\\
&-&  i\sum_{l,m=1}^4\sum_{n=1}^6 V_{dDN}^{LJnl\star} V_{dDN}^{RInm} 
\left[V_{NNP}^{Llmk} C_2(m_{D_n}^2,m_{N_l}^2,m_{N_m}^2)\right.\nonumber\\
&&\phantom{aaaaaaaaaa}
+\left. V_{NNP}^{Rlmk} 
m_{N_l} m_{N_m} C_0(m_{D_n}^2,m_{N_l}^2,m_{N_m}^2) 
\right) \nonumber\\
&-& \sum_{l,m=1}^6\sum_{n=1}^2 V_{UUP}^{lmk} V_{dUC}^{LJln\star} 
V_{dUC}^{RImn} m_{C_n} C_0(m_{C_n}^2,m_{U_l}^2,m_{U_m}^2)
\nonumber\\
&-& \sum_{l,m=1}^2\sum_{n=1}^6 V_{dUC}^{LJnl\star} V_{dUC}^{RInm} 
\left[V_{CCP}^{Lmlk} 
C_2(m_{U_n}^2,m_{C_l}^2,m_{C_m}^2)\right.\nonumber\\
&&\phantom{aaaaaaaaaa}+\left. V_{CCP}^{Rmlk} 
m_{C_l} m_{C_m} C_0(m_{U_n}^2,m_{C_l}^2,m_{C_m}^2) \right]
\end{eqnarray}

\vskip 1cm

\noindent {\bf iii)  Charged Higgs boson vertex $H^+_k\bar u_J d_I$} 
where $H^+1=H^+$ and $H^+_2=G^+$ (see fig~\ref{fig:CHiggscorr}).

\begin{eqnarray}
\left(4\pi\right)^2\left (\Delta F^k_L\right)^{JI}&=& -
{32 \pi \alpha_s\over 3} m_{\tilde g} \sum_{l,m=1}^6 V_{UDH}^{mlk}
Z_U^{(J+3)m} Z_D^{Il} C_0(m_{\tilde g}^2,m_{U_m}^2,m_{D_l}^2)
\nonumber\\
&+&  \sum_{l,m=1}^6\sum_{n=1}^4 V_{UDH}^{mlk} V_{uUN}^{RJmn\star} 
V_{dDN}^{LIln} m_{N_n} C_0(m_{N_n}^2,m_{U_m}^2,m_{D_l}^2)
\nonumber\\
&+& \sum_{l=1}^4\sum_{m=1}^2\sum_{n=1}^6 V_{uDC}^{RJnm\star}V_{dDN}^{LInl}
\left[V_{NCH}^{Rlmk} C_2(m_{D_n}^2,m_{C_m}^2,m_{N_l}^2)\right.\nonumber\\
&&\phantom{aaaaaaaaa}
+\left. V_{NCH}^{Llmk} m_{C_m} m_{N_l} C_0(m_{D_n}^2,m_{C_m}^2,m_{N_l}^2)
\right] \nonumber\\
&+& \sum_{l=1}^4\sum_{m=1}^2\sum_{n=1}^6 V_{uUN}^{RJnl\star} V_{dUC}^{LInm} 
\left[V_{NCH}^{Rlmk} C_2(m_{U_n}^2,m_{N_l}^2,m_{C_m}^2)\right.\nonumber\\
&&\phantom{aaaaaaaaa}
+\left. V_{NCH}^{Llmk} m_{N_l} m_{C_m} C_0(m_{U_n}^2,m_{N_l}^2,m_{C_m}^2)
\right]
\end{eqnarray}

\begin{eqnarray}
\left(4\pi\right)^2\left (\Delta F^k_R\right)^{JI}&=& -
{32 \pi \alpha_s\over 3} m_{\tilde g} \sum_{l,m=1}^6 V_{UDH}^{mlk}
Z_U^{Jm} Z_D^{(I+3)l} C_0(m_{\tilde g}^2,m_{U_m}^2,m_{D_l}^2)
\nonumber\\
&+& \sum_{l,m=1}^6\sum_{n=1}^4 V_{UDH}^{mlk} V_{uUN}^{LJmn\star} 
V_{dDN}^{RIln} m_{N_n} C_0(m_{N_n}^2,m_{U_m}^2,m_{D_l}^2)
\nonumber\\
&+& \sum_{l=1}^4\sum_{m=1}^2\sum_{n=1}^6 V_{uDC}^{LJnm\star} V_{dDN}^{RInl} 
\left[V_{NCH}^{Llmk} C_2(m_{D_n}^2,m_{C_m}^2,m_{N_l}^2)\right.
\nonumber\\
&&\phantom{aaaaaaaaa}
+\left. V_{NCH}^{Rlmk} 
m_{C_m} m_{N_l} C_0(m_{D_n}^2,m_{C_m}^2,m_{N_l}^2)\right] 
\nonumber\\
&+& \sum_{l=1}^4\sum_{m=1}^2\sum_{n=1}^6 V_{uUN}^{LJnl\star} V_{dUC}^{RInm}
\left[V_{NCH}^{Llmk} C_2(m_{U_n}^2,m_{N_l}^2,m_{C_m}^2) \right.\nonumber\\
&&\phantom{aaaaaaaaa}
+\left. V_{NCH}^{Rlmk} m_{N_l} m_{C_m} C_0(m_{U_n}^2,m_{N_l}^2,m_{C_m}^2)
\right]
\end{eqnarray}

\subsection{Box diagram contributions} 

Box contributions to $\Delta F=2$ processes as given below are related
to the Wilson coefficients $C_i$ defined in eq.~(\ref{eqn:heff_dF2}) through
\begin{eqnarray}
B_i= G_F^2M^2_W (V_{\rm eff}^{tb\ast}V_{\rm eff}^{ts})^2 C_i(M_{SUSY})
\label{eqn:boxwil}
\end{eqnarray}

\noindent {\bf i) Charged Higgs contributions} (including the mixed 
$H^{\pm}W^{\mp}$ diagram). In order to simplify the 
notation, we denote $P_{LR}^{JI1}\equiv [P^H_{LR}]^{JI}$, 
$P_{LR}^{JI2}\equiv [P^G_{LR}]^{JI}$ and similarly for the $RL$ couplings 
($[P^H],[P^G]$ are defined in eqs.~(\ref{LH}), (\ref{LG})). If not written 
down explicitly, the arguments of the loop integrals $D_0$ and $D_2$ are
$(m^2_{H^+_k},m^2_{H^+_l},\overline{m}^2_{u_M},\overline{m}^2_{u_N})$.

\begin{eqnarray}
\left(B^{\rm VLL}\right)_{hg} &=&-{g^2_2\over2} \sum_{M,N=1}^3 
V_{\rm eff}^{MJ\star} V_{\rm eff}^{NI} P_{RL}^{NJ1\star} P_{RL}^{MI1}   
\overline{m}_{u_M} \overline{m}_{u_N} 
D_0(M_W^2,m_{H^+_1}^2,\overline{m}_{u_M}^2,\overline{m}_{u_N}^2)
\nonumber\\
&+&{1\over8} \sum_{M,N=1}^3 P_{RL}^{MJ1\star} P_{RL}^{NJ1\star} 
P_{RL}^{MI1} P_{RL}^{NI1} 
D_2(m_{H^+_1}^2,m_{H^+_1}^2,\overline{m}_{u_M}^2, \overline{m}_{u_N}^2)
\nonumber\\
&+&{1\over4} \sum_{M,N=1}^3 P_{RL}^{MJ2\star} P_{RL}^{NJ1\star}  
P_{RL}^{MI1} P_{RL}^{NI2} 
D_2(M_W^2,m_{H^+_1}^2,\overline{m}_{u_M}^2, \overline{m}_{u_N}^2)
\nonumber\\ 
\left(B^{\rm VRR}\right)_{hg} &=&{1\over8}\sum_{k,l=1}^2 \sum_{M,N=1}^3 
P_{LR}^{NJk\star}P_{LR}^{MJl\star} P_{LR}^{MIk}P_{LR}^{NIl} D_2
\nonumber\\
\left(B^{\rm SLL}_1\right)_{hg}&=&{1\over2}\sum_{k,l=1}^2 \sum_{M,N=1}^3 
P_{LR}^{MJl\star} P_{LR}^{NJk\star} P_{RL}^{MIk} P_{RL}^{NIl} 
\overline{m}_{u_M} \overline{m}_{u_N} D_0\nonumber\\
\left(B^{\rm SLL}_2\right)_{hg} &=& 0 \nonumber\\
\left(B^{\rm SRR}_1\right)_{hg} &=&{1\over2} \sum_{k,l=1}^2 \sum_{M,N=1}^3 
P_{RL}^{MJl\star} P_{RL}^{NJk\star} P_{LR}^{MIk} P_{LR}^{NIl} 
\overline{m}_{u_M} \overline{m}_{u_N} D_0
\nonumber\\
\left(B^{\rm SRR}_2\right)_{hg} &=& 0\nonumber\\
\left(B^{\rm LR}_1\right)_{hg} &=& {1\over4} \sum_{k,l=1}^2 \sum_{M,N=1}^3 
P_{RL}^{MJl\star}P_{LR}^{NJk\star} P_{RL}^{MIk} P_{LR}^{NIl} D_2\nonumber\\
\left(B^{\rm LR}_2\right)_{hg} &=& 
-{g^2_2\over2} \sum_{k=1}^2 \sum_{M,N=1}^3 
V_{\rm eff}^{MJ\star} V_{\rm eff}^{NI} P_{LR}^{NJk\star} P_{LR}^{MIk} 
D_2(M_W^2,m_{H^+_k}^2,\overline{m}_{u_M}^2,\overline{m}_{u_N}^2)\nonumber\\
&+& \sum_{k,l=1}^2 \sum_{M,N=1}^3 
P_{LR}^{MJl\star}P_{RL}^{NJk\star} P_{RL}^{MIk} P_{LR}^{NIl} 
\overline{m}_{u_M} \overline{m}_{u_N} D_0
\end{eqnarray}

\noindent {\bf Chargino contributions.}   Arguments of the loop integrals are
$(m^2_{C_m},m^2_{C_n},m^2_{U_k},m^2_{U_l})$.

\begin{eqnarray}
\left(B^{\rm VLL}\right)_C &=&{1\over8} \sum_{k,l=1}^6 
\sum_{m,n=1}^2  V_{dUC}^{LJkn\star} V_{dUC}^{LJlm\star} 
V_{dUC}^{LIkm} V_{dUC}^{LIln} D_2  \nonumber\\
\left(B^{\rm VRR}\right)_C &=& {1\over8} \sum_{k,l=1}^6 
\sum_{m,n=1}^2  V_{dUC}^{RJkn\star} V_{dUC}^{RJlm\star} 
V_{dUC}^{RIkm} V_{dUC}^{RIln} D_2 \nonumber\\
\left(B^{\rm SLL}_1\right)_C &=&-{1\over4} \sum_{k,l=1}^6 
\sum_{m,n=1}^2 V_{dUC}^{RJkn\star} V_{dUC}^{RJlm\star} 
V_{dUC}^{LIkm} V_{dUC}^{LIln} m_{C_m} m_{C_n} D_0 \nonumber\\
\left(B^{\rm SLL}_2\right)_C &=& {1\over16} \sum_{k,l=1}^6 
\sum_{m,n=1}^2 V_{dUC}^{RJkn\star} V_{dUC}^{RJlm\star} 
V_{dUC}^{LIkm} V_{dUC}^{LIln} m_{C_m} m_{C_n} D_0 \nonumber\\
\left(B^{\rm SRR}_1\right)_C &=& -{1\over4} \sum_{k,l=1}^6 
\sum_{m,n=1}^2 V_{dUC}^{LJkn\star} V_{dUC}^{LJlm\star} 
V_{dUC}^{RIkm} V_{dUC}^{RIln} m_{C_m} m_{C_n} D_0 \nonumber\\
\left(B^{\rm SRR}_2\right)_C &=& {1\over16} \sum_{k,l=1}^6 
\sum_{m,n=1}^2 V_{dUC}^{LJkn\star} V_{dUC}^{LJlm\star} 
V_{dUC}^{RIkm} V_{dUC}^{RIln} m_{C_m} m_{C_n} D_0 \nonumber\\
\left(B^{\rm LR}_1\right)_C &=& -{1\over2} \sum_{k,l=1}^6 
\sum_{m,n=1}^2  V_{dUC}^{LJkn\star} V_{dUC}^{RJlm\star} 
V_{dUC}^{LIkm} V_{dUC}^{RIln}  m_{C_m} m_{C_n} D_0 \nonumber\\
\left(B^{\rm LR}_2\right)_C &=& -{1\over2} \sum_{k,l=1}^6 
\sum_{m,n=1}^2 V_{dUC}^{LJlm\star} V_{dUC}^{RJkn\star} 
V_{dUC}^{LIkm} V_{dUC}^{RIln} D_2
\end{eqnarray}

\subsection{Loop integrals}

We define three- and four-point loop integrals at vanishing external 
momenta as:
\begin{eqnarray}
{1\over(4\pi)^2}C_{2n}(m_1^2,m_2^2,m_3^2) = \int{d^d k\over(2\pi)^d} 
{i ~k^{2n}\over (k^2 - m_1^2)(k^2-m_2^2)(k^2 - m_3^2)}~,\phantom{aaaaaa}
\label{eq:cndef}\\
{1\over(4\pi)^2}D_{2n}(m_1^2,m_2^2,m_3^2,m_4^2) =
\int{d^4 k\over(2\pi)^d} {i ~k^{2n}\over (k^2 - m_1^2)(k^2-m_2^2)
(k^2 - m_3^2)(k^2 - m_4^2)}~.\label{eq:dndef}
\end{eqnarray}
Explicit formulae for the $C_0$, $C_2$, $D_0$ and $D_2$ functions 
are as follows
\begin{eqnarray}
&&C_0(x,y,z)= {y\over (x-y)(z-y)}\log{y\over x} +{z\over
(x-z)(y-z)}\log{z\over x}~,\label{c0gen} \\
&&C_2(x,y,z)=\Delta + \log{x\over\mu^2} 
+{y^2\over(x-y)(z-y)}\log{y\over x} 
+{z^2\over(x-z)(y-z)}\log{z\over x}~.\phantom{aa}
\label{c2gen}
\end{eqnarray}
where $\Delta={2\over d-4}+\log4\pi\gamma_E-1$ and $\mu$ is the
renormalization scale. In the flavour changing penguin diagrams
$\Delta$ and $\mu$ dependence always cancels out after summation 
over squark, chargino and neutralino mixing matrices.
\begin{eqnarray}
D_0(x,y,z,t) &=& {y\over (y-x)(y-z)(y-t)}\log{y\over x}
+{z\over (z-x)(z-y)(z-t)}\log{z\over x}\nonumber\\
&+&{t\over (t-x)(t-y)(t-z)}\log{t\over x}~,\label{d0gen}\\
D_2(x,y,z,t) &=& {y^2\over (y-x)(y-z)(y-t)}\log{y\over x}
+{z^2\over (z-x)(z-y)(z-t)}\log{z\over x}\nonumber\\
&+&{t^2\over (t-x)(t-y)(t-z)}\log{t\over x}~. \label{d2gen}
\end{eqnarray}
The two-point loop integrals $B_0$ and $B_1$ at $s=0$ can be expressed as
\begin{eqnarray}
B_0(0,x,y)=\Delta + {x\over x-y}\log{x\over\mu^2} 
                  + {y\over y-x}\log{y\over\mu^2}\equiv
\Delta + \log{x\over\mu^2}+{y\over x-y}\log{x\over y}\label{b0gen}\\
B_1(0,x,y)={1\over2}B_0(0,x,y)-{1\over4}{x+y\over x-y}+{1\over2}
{xy\over(x-y)^2}\log{x\over y}={1\over4}+{1\over2}C_2(x,y,y)
\label{b1gen}
\end{eqnarray}
The function $H_2$ of eq.~(\ref{eqn:hfun}) is related to $C_0$ as follows:
\begin{eqnarray}
H_2(x,y) = -m^2 C_0(m^2,xm^2,ym^2) \equiv -C_0(1,x,y).
\end{eqnarray}


\newpage

\begin{thebibliography}{99}
  
\bibitem{MIPORO} M. Misiak, S. Pokorski and J. Rosiek,in 
                 {\sl Heavy Flavours II}, eds. A.J. Buras and M. Lindner, 
                 World Scientific, Singapore, 1998, p. 795 (hep-ph/9703442).
  
\bibitem{GRNIRA} Y. Grossman, Y. Nir and R. Rattazzi, in 
                 {\sl Heavy Flavours II}, eds. A.J. Buras and M. Lindner, 
                 World Scientific, Singapore, 1998, p. 755 (hep-ph/9701231). 
  
\bibitem{LS02} L. Silvestrini, hep-ph/0210031.

\bibitem{BUGAGOJASI} A.J. Buras, P. Gambino, M. Gorbahn, S. J\"ager and L. 
                     Silvestrini, {\sl Nucl. Phys.} {\bf B592} (2001) 55.
  
\bibitem{GAGAMASI} F. Gabbiani, E. Gabrielli, A. Masiero and L. Silvestrini, 
                   {\sl Nucl. Phys.} {\bf B477} (1996) 321.

\bibitem{CHSL} P.H. Chankowski and \L . S\l awianowska, {\sl Phys. Rev.}  
               {\bf D63} (2001) 054012; {\sl Acta Phys. Polon.} 
               {\bf B32} (2001) 1895.

\bibitem{CHRO} P.H. Chankowski and J. Rosiek, {\sl Acta Phys. Polon.}  
               {\bf B33} (2002) 2329.

\bibitem{ISRE2} G. Isidori and A. Retico, {\sl JHEP} {\bf 0209} (2001) 063. 

\bibitem{BRBUGE} G.C. Branco, A.J. Buras and J.-M. G\'erard, 
                 {\sl Phys. Lett.} {\bf B155} (1985) 192; {\sl Nucl. Phys.}  
                 {\bf B259} (1985) 306.
  
\bibitem{OLPO} M. Olechowski and S. Pokorski, {\sl Phys. Lett.}
               {\bf B214} (1988) 393.
  
\bibitem{HARASA} L.J. Hall, R. Rattazzi and U. Sarid, {\sl Phys. Rev.}  
                 {\bf D50} (1994) 7048; 
                 M. Carena, M. Olechowski, S. Pokorski and C.E.M. Wagner, 
                 {\sl Nucl. Phys.} {\bf B426} (1994) 269;
                 R. Rattazzi and U. Sarid, {\sl Nucl. Phys.} 
                 {\bf B501} (1997) 297; 
                 H. Baer, M. Brhlik, D. Castano and X. Tata, {\sl Phys. Rev.} 
                 {\bf D58} (1998) 015007; 
                 T. Bla\v zek and S. Raby, {\sl Phys. Rev.} {\bf D59} 
                 (1999) 095002.
  
\bibitem{COSO} J.A. Coarasa and J. Sol\`a, {\sl Phys. Lett.} 
               {\bf B389} (1996) 53; 
               J.A. Coarasa, D. Garcia, J. Guasch, R.A. Jim\'enez and 
               J. Sol\`a, {\sl Eur. Phys. J.} {\bf C2} (1998) 373; 
               J.A. Coarasa, R.A. Jim\'enez and J. Sol\`a, {\sl Phys. Lett.} 
               {\bf B406} (1997) 337.

\bibitem{BAKO2} K.S. Babu and C. Kolda, {\sl Phys. Lett.} {\bf B451} 
                (1999) 77.
  
\bibitem{GUHOPE} J. Guasch, W.F.L. Hollik and S. Penaranda, 
                 {\sl Phys. Lett.} {\bf B515} (2001) 367.



\bibitem{HAHE} H.E. Haber et al., {\sl Phys. Rev.} {\bf D63} (2001) 055004; 
               M.J. Herrero, S. Pennaranda and D. Temes, 
               {\sl Phys. Rev.} {\bf D64} (2001) 115003.

\bibitem{CUHE} A.M. Curiel, M.J. Herrero, D. Temes and J.F. De Troconiz, 
               {\bf D65} (2002) 075006;
               A. Dobado, M.J. Herrero, D. Temes, {\bf D65} (2002) 075023;
               A.M. Curiel, M.J. Herrero and D. Temes, preprint 
               FTUAM-02-21, hep-ph/0210335. 


\bibitem{CAGANIWA1} M. Carena, D. Garcia, U. Nierste and C.E.M. Wagner, 
                    {\sl Nucl. Phys.} {\bf B577} (2000) 88.
  
\bibitem{HUYA} C.-S. Huang and Q.-S. Yan, {\sl Phys. Lett.} 
             {\bf B442} (1998) 209;
             C.-S. Huang, W. Liao and Q.-S. Yan, {\sl Phys. Rev.} {\bf D59}
             (1999) 011701. 

\bibitem{HAPOTO} C. Hamzaoui, M. Pospelov and M. Toharia, {\sl Phys. Rev.}  
                 {\bf D59} (1999) 095005.

\bibitem{BAKO} K.S. Babu and C. Kolda, {\sl Phys. Rev. Lett.} {\bf 84} 
               (2000) 228.

\bibitem{HULIYAZH} C.-S. Huang, W. Liao, Q.-S. Yan and S.-H. Zhu, 
                   {\sl Phys. Rev.} {\bf D63} (2001) 114021, Erratum: 
                   {\sl Phys. Rev.} {\bf D64} (2001) 059902.
  
\bibitem{BOEWKRUR} C. Bobeth, T. Ewerth, F. Kr\"uger and J. Urban,
                   {\sl Phys. Rev.} {\bf D64} (2001) 074014 and 
                   {\bf D66} (2002) 074021. 
  
\bibitem{BUCHROSL1} A.J. Buras, P.H. Chankowski, J. Rosiek and 
                   \L . S\l awianowska, {\sl Nucl. Phys.} {\bf B619} 
                   (2001) 434.

\bibitem{BLPORA} T. Bla\v zek, S. Raby and S. Pokorski, {\sl Phys. Rev.}  
                 {\bf D52} (1995) 4151.
  
\bibitem{DEGAGI} G. Degrassi, P. Gambino and G.-F. Giudice, {\sl JHEP} 
                 {\bf 0012} (2000) 009.

\bibitem{CAGANIWA2} M. Carena, D. Garcia, U. Nierste and C.E.M. Wagner, 
                    {\sl Phys. Lett.} {\bf B499} (2001) 141.

\bibitem{ISRE1} G. Isidori and A. Retico, {\sl JHEP} {\bf 0111} (2001) 001 
                and hep-ph/0110121 v3, July 2002.

\bibitem{AMGIISST} G. D'Ambrosio, G.-F.  Giudice, G. Isidori and A. Strumia,
                   {\sl Nucl. Phys.} {\bf B645} (2002) 155.

\bibitem{BUCHROSL2} A.J. Buras, P.H. Chankowski, J. Rosiek and \L .
                    S\l awianowska, {\sl Phys. Lett.} {\bf B546} (2002) 96.

\bibitem{ROS} J. Rosiek, {\sl Phys. Rev.} {\bf D41} (1990) 3464,
              {\sl Erratum} hep-ph/9511250.
  
\bibitem{CHPO} P.H. Chankowski and S. Pokorski in {\sl Perspectives
               on supersymmetry}, G.L. Kane ed., World Scientific Publishing
               Co., Singapore 1998 (hep-ph/9707497); 
               M. Carena, U. Nierste and C.E.M. Wagner, {\sl Nucl. Phys.} 
               {\bf B577} (2000) 88.

\bibitem{CHELOLPO} P.H. Chankowski, J. Ellis, M. Olechowski and S. Pokorski, 
                   {\sl Phys. Lett.} {\bf B214} (1988) 393.

\bibitem{CIDEGAGI2} M. Ciuchini, G. Degrassi, P. Gambino and G.-F. Giudice, 
                    {\sl Nucl. Phys.} {\bf B534} (1998) 3.
  
\bibitem{ALEXAN} B. Aubert et al., BaBar Collaboration, hep-ex/0207083.
  
\bibitem{CDF} F. Abe et al., The CDF Collaboration, {\sl Phys. Rev.}
              {\bf D57} (1998) 3811.    

\bibitem{DEDRNI} A. Dedes, H.K. Dreiner and U. Nierste, 
                 {\sl Phys. Rev. Lett.} {\bf 87} (2001) 251804.

\bibitem{BUER} A.J. Buras, hep-ph/0101336, lectures at the
               International Erice School, August, 2000.

\bibitem{BUJAWE} A.J. Buras, M. Jamin, and P.H. Weisz, {\sl Nucl. Phys.}  
                 {\bf B347} (1990) 491.  
  
\bibitem{CET0} M. Ciuchini, E. Franco, V. Lubicz, G. Martinelli,
               I. Scimemi and L. Silvestrini, {\sl Nucl. Phys.} {\bf B523}
               (1998) 501; 
               M. Ciuchini, et al., {\sl JHEP} {\bf 9810} (1998) 008.
  
\bibitem{BUMIUR} A.J. Buras, M. Misiak and J. Urban, {\sl Nucl. Phys.}  
                 {\bf B586} (2000) 397.

\bibitem{BUJAUR} A.J. Buras, S. J\"ager and J. Urban, {\sl Nucl. Phys.}  
                 {\bf B605} (2001) 600.
  
\bibitem{BECIREVIC} D. Becirevic, V. Gimenez, G. Martinelli, M. Papinutto 
                    and J. Reyes, hep-lat/0110091, hep-lat/0110117.

\bibitem{BOBUKRUR} C. Bobeth, A.J. Buras, F. Kr\"uger and J. Urban, 
                   {\sl Nucl. Phys.}  {\bf B630} (2002) 87.

\bibitem{BUPAST} A.J. Buras, F. Parodi and A. Stocchi, {\sl JHEP} 
                 {\bf 0209} (2003) 0301.
 
\bibitem{GAMI} P. Gambino and M. Misiak, {\sl Nucl. Phys.} {\bf B611} 
               (2001) 338.
  
\bibitem{CIDEGAGI1} M. Ciuchini, G. Degrassi, P. Gambino and G.-F. Giudice, 
                   {\sl Nucl. Phys.} {\bf B527} (1998) 21.

\bibitem{CIROST} P. Ciafaloni, A. Romanino and A. Strumia, {\sl Nucl. Phys.}
                 {\bf B524} (1998) 361; 
                 F. Borzumati and C. Greub, {\sl Phys. Rev.} {\bf D58} 
                 (1998) 074004.
  
\bibitem{BAGI} R. Barbieri and G.-F.  Giudice, {\sl Phys.  Lett.}
               {\bf B309} (1993) 86.

\bibitem{DEPI} A. Dedes and A. Pilaftsis, {\sl Phys. Rev.} {\bf D67}
               (2003) 015012.

\end{thebibliography}
\end{document}